\documentclass[12pt]{iopart}
\usepackage{iopams}
\usepackage{graphicx}% Include figure files
\usepackage{dcolumn}% Align table columns on decimal point
\usepackage{bm, amssymb, color}% bold math
\usepackage{hyperref}% add hypertext capabilities
\usepackage{booktabs,tabulary}
\usepackage{soul}
\usepackage[caption=false]{subfig}
\usepackage{array}
\usepackage{cases}
\usepackage{multirow}

\usepackage[textwidth=16cm, top=1in, bottom=1in]{geometry}

\usepackage{enumitem}

\newcommand{\fn}[2]{\mathinner{#1\mathopen{\left(#2\right)}}}

\newcommand{\spD}[1]{\fn{\tilde{\chi}_{_V}}{#1}}

\renewcommand{\Im}{\mathrm{Im}}
\newcommand{\tens}[1]{\boldsymbol{#1}}

\newcommand{\uvect}[1]{\hat{\vect{#1}}}

\newcommand{\R}{\mathbb{R}}
\newcommand{\dd}[1]{\mathinner{\mathrm{d}#1}}
\newcommand{\vect}[1]{{\bf #1}}%{\bm{#1}}
\newcommand{\E}[1]{\left\langle#1\right\rangle}
\newcommand{\F}[2]{\fn{F^\mathrm{(#1D)}}{#2}}
\newcommand{\J}[2]{\fn{\mathcal{J}^{(#1)}}{#2}}
\newcommand{\FTJ}[2]{\fn{\tilde{\mathcal{J}}^{(#1)}}{#2}}
\newcommand{\BETA}[2]{\beta^\mathrm{(#1D)}_{#2}}

\newcommand{\Ctm}[1]{C_{#1}^{TM}}
\newcommand{\CL}[1]{C_{#1}^{\perp}}
\newcommand{\abs}[1]{\left\vert #1 \right\vert}

\newcommand{\ATM}[2]{\fn{A_{#1}^{TM}}{#2}}
\newcommand{\ATE}[2]{\fn{A_{#1}^{TE}}{#2}}
\newcommand{\ETM}[1]{\fn{\varepsilon_e^{TM}}{#1}}
\newcommand{\ETE}[1]{\fn{\varepsilon_e^{TE}}{#1}}

\usepackage[colorinlistoftodos,shadow,textwidth=18mm]{todonotes}

\setstcolor{red}

\newcommand{\eqref}[1]{(\ref{#1})}

\begin{document}

\title{Extraordinary Optical and Transport Properties of Disordered Stealthy Hyperuniform Two-Phase Media}

\author{Jaeuk  Kim$^{1,3}$}
\vspace{-0.15in}

\author{Salvatore Torquato$^{1,2,3,4}$}

$^1$  Department of Chemistry,
Princeton University,
Princeton, NJ 08544, USA  

$^2$  Department of Physics,
Princeton University,
Princeton, NJ 08544, USA  

$^3$  Princeton Materials Institute,
Princeton, NJ 08544, USA

$^4$ Program in Applied and Computational Mathematics,
Princeton University,
Princeton, NJ 08544, USA

\noindent {Corresponding author:  Salvatore Torquato;\, E-mail: torquato$@$princeton.edu}
%\begin{tabbing} 
 %\hspace{0.25in}
 %\= Salvatore Torquato \\
% \>  Tel.: 609-258-3341 \\
% \>  Fax: 609-258-6746 \\
% \>  E-mail: torquato$@$princeton.edu
%\end{tabbing}

%\noindent \textbf{Short title}: Disordered Hyperuniform Heterogeneous Materials  \newline

%\noindent \mboxbf{Classification numbers}: ?????

\begin{abstract}

    Disordered stealthy hyperuniform two-phase media are a special subset of hyperuniform structures 
    with novel physical properties due to their hybrid crystal-liquid nature.
    We have previously shown that the rapidly converging strong-contrast expansion of a linear fractional form
    of the effective dynamic dielectric constant $\fn{\tens{\varepsilon}_e}{\vect{k}_1,\omega}$ [Phys. Rev. X {\bf 11}, 296 021002 (2021)] leads to 
    accurate approximations for both hyperuniform and nonhyperuniform two-phase composite media when truncated at the two-point level
    for distinctly different types of microstructural symmetries in three dimensions.
    In this paper, we further elucidate the extraordinary optical and transport properties of disordered stealthy hyperuniform media.
    Among other results, we provide detailed proofs that stealthy hyperuniform layered and transversely isotropic media are perfectly transparent (i.e., no Anderson localization, in principle) within finite wavenumber intervals through the third-order terms. Remarkably, these results imply that there can be no Anderson localization within the predicted perfect transparency interval in stealthy hyperuniform layered and transversely isotropic media in practice because the localization length (associated with only possibly negligibly small higher-order contributions) should be very large compared to any practically large sample size.
    We further contrast and compare the extraordinary physical properties of stealthy hyperuniform two-phase
    layered, transversely isotropic media, and fully 3D isotropic media to other model nonstealthy microstructures, including their attenuation characteristics,
    as measured by the imaginary part of $\fn{\tens{\varepsilon}_e}{\vect{k}_1,\omega}$, and transport properties, as measured by the time-dependent diffusion spreadability ${\cal S}(t)$.
    We demonstrate that there are cross-property relations between them, namely,  we quantify how the imaginary parts of $\fn{\tens{\varepsilon}_e}{\vect{k}_1,\omega}$ and the spreadability at long times
    are positively correlated as the structures span from nonhyperuniform, nonstealthy hyperuniform, and stealthy hyperuniform media.
    It will also be useful to establish cross-property relations for stealthy hyperuniform media for other wave phenomena (e.g., elastodynamics) as well as other transport properties.
    Cross-property relations are generally useful because they enable one to estimate one property, given a measurement of another property.

    % Our study provides useful guidance for studying the cross-property relations among attenuation and spreadability characteristics in general situations (e.g., general incidence directions and metallic phases) and other physical properties associated with the two-point statistics, such as fluid permeability.

\end{abstract}

%Uncomment for PACS numbers title message
%\pacs{%05.20.-y,05.40.-a,61.20.Gy,61.50.Ah
%}
% Keywords required only for MST, PB, PMB, PM, JOA, JOB?
\vspace{2pc}
%\noindent{\it Keywords}: hyperuniformity, fluctuations,
%heterogeneous media, disordered materials
% Uncomment for Submitted to journal title message
%\submitto{\JPA}
% Comment out if separate title page not required
%\maketitle
%\ioptwocol
\section{Introduction}

Hyperuniform many-body systems and materials are characterized by an anomalous suppression of density fluctuations at large length scales compared to typical disordered systems \cite{To03a}. The broad importance of the hyperuniformity concept for statistical physics, condensed matter physics and
materials science was introduced two decades ago in an investigation that focused on fundamental theoretical issues pertaining to local density
fluctuations, including how it provides a unified means to classify and categorize crystals, quasicrystals, and special disordered point configurations \cite{To03a}. A hyperuniform (or superhomogeneous \cite{Ga02}) many-particle system in $d$-dimensional Euclidean space $\mathbb{R}^d$ possesses a  structure factor $S({\bf k})$ 
(proportional to the scattering intensity) that vanishes  as the wavenumber $k\equiv |\bf k|$ tends to zero,
i.e.,
\begin{equation}
\lim_{|{\bf k}| \rightarrow 0} S({\bf k}) = 0.
\label{hyper}
\end{equation}
Equivalently, a hyperuniform system is one in which the number variance of particles within a
spherical observation window of radius $R$, denoted by $\sigma^2_{_N}(R)$,  grows, for large $R$, more slowly than the window volume, i.e., $R^d$ \cite{To03a}. Typical disordered systems, such as ordinary gases and liquids, have the expected asymptotic volume scaling,  i.e., $\sigma^2_{_N}(R) \sim R^d$. 
On the other hand, all perfect crystals and many perfect quasicrystals are hyperuniform with the surface-area 
scaling $\sigma^2_{_N}(R)\sim R^{d-1}$. 
%Surprisingly, there are a special class of disordered particle configurations that have the same asymptotic behavior as crystals.
Disordered hyperuniform systems are exotic amorphous states of matter that lie between a crystal and liquid: they can behave like perfect crystals in the way they suppress large-scale density fluctuations and yet have characteristics of liquids or glasses at small length scales such that they are statistically isotropic with no Bragg peaks. Thus, disordered hyperuniform systems
can be thought of as having a   {\it hidden order} (see Fig. 2 of Ref. \cite{To18a} for a vivid example), and their hybrid crystal-liquid nature can endow them with novel physical properties, as described below. 
It is noteworthy that the hyperuniformity concept extends our
traditional notions of long-range order
to not only include crystals and quasicrystals but exotic disordered states of matter \cite{To18a}.

Two decades ago, only a few examples of disordered hyperuniform systems and their manifestations were known \cite{To03a,Ga02}.
It has come to be discovered that
these special disordered systems arise
in a variety of contexts, across the physical, materials, mathematical, and biological sciences,
including disordered hard-sphere plasmas  \cite{Lo18a,To18a,Fl22}, classical disordered
(noncrystalline) ground states \cite{Uc04b,Zh15a,Zh17a},
maximally random jammed (MRJ) hard-particle packings \cite{Do05d,Ma23},
jammed bidisperse emulsions \cite{Ri17}, 
jammed thermal colloidal packings~\cite{Ku11,Dr15},
jammed athermal soft-sphere models of granular media~\cite{Si09,Be11},
nonequilibrium phase transitions  \cite{Ja15,He15,We15,Tj15,He17a,He17b,We17,Kw17,Wi18}, 
dynamical processes in ultracold atoms~\cite{Le14},
avian photoreceptor patterns \cite{Ji14},
%geometry of neuronal tracts \cite{Bur15}, 
receptor organization in the immune system \cite{Ma15}, vegetation patterns in arid ecosystems \cite{Ge23}, 
certain quantum ground states (both fermionic and bosonic) \cite{To08b, Fe56}, vortex structures in superconductors \cite{sanchez_disordered_2023},
``perfect" glasses \cite{Zh16a},
the distribution of the nontrivial zeros of the Riemann zeta function \cite{To08b, Mon73}, and the eigenvalues of various random matrices \cite{Dy70, Me91, To08b, To18a, La19}.
Thus, it is apparent that disordered hyperuniform states of matter can exist as both equilibrium
and nonequilibrium phases and come in classical and quantum-mechanical varieties.

\begin{figure}
\begin{center}
\includegraphics*[  width=2.5in,clip=keepaspectratio]{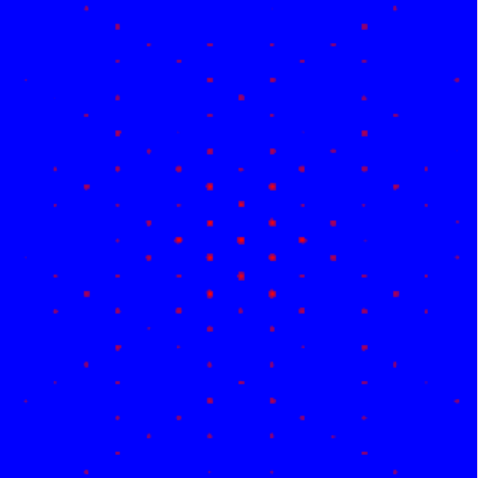}\hspace{0.3in}
\includegraphics*[  width=2.5in,clip=keepaspectratio]{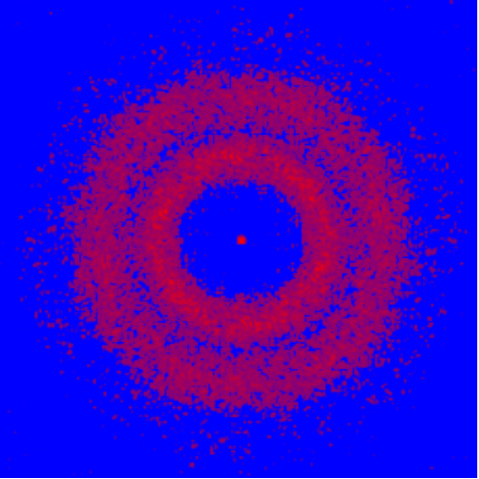}
\caption{Left: Scattering pattern for a crystal. Right: Scattering pattern for a  disordered ``stealthy" hyperuniform material
defined by relation (\ref{stealthy-points}).
Scattering intensity increases from zero (blue color) to the highest value (red color). Notice that apart from forward scattering, there is a circular region around the origin
in which there is no scattering, a singularly exotic situation for an amorphous state of matter.
Reprinted (figure) with permission from \cite{To16a}, Copyright (2016) by the American Physical Society.
%This figure is adapted from Fig. 1 of Ref. .
}
\label{pattern}
\end{center}
\end{figure}

The fundamental and practical importance of the hyperuniformity concept in the context of condensed matter physics began to emerge when it was demonstrated that classical many-particle systems with certain long-ranged
pair potentials could counterintuitively freeze into disordered hyperuniform states
at absolute zero with singular scattering patterns, such as the stealthy hyperuniform one depicted 
in the right panel of Fig. \ref{pattern} \cite{Uc04b,Zh15a}. 
%This exotic situation runs counter to our everyday experience, where we expect liquids to freeze into crystal structures (like ice). 
{\it Stealthy hyperuniform} many-particle systems possess a
structure factor that is zero not only at infinite wavelengths but also vanishes for a range of wavenumbers around the origin, i.e.,
\begin{equation}
S({\bf k}) = 0 \qquad \mbox{for}\; 0 < |{\bf k}| \le  K,
\label{stealthy-points}
\end{equation}
implying that there is no single scattering down to  intermediate wavelengths of the order of $2\pi/K$ \cite{Uc04b, Zh15a}.
Mapping such particle configurations to networks enabled the discovery of the first disordered
dielectric networks to have large isotropic photonic band gaps comparable in
size to photonic crystals \cite{Fl09b}.
This computational study led to the design and  fabrication of disordered cellular solids with the predicted
photonic band-gap characteristics for the microwave regime, enabling unprecedented
free-form waveguide geometries that are robust to defects
not possible with crystalline structures \cite{Ae22}.
Afterward, stealthy hyperuniform materials were demonstrated to possess singular wave propagation, transport, and elasticity characteristics, including wave transparency
\cite{Le16,Fr17,Ki20a,To21a,Ki23,Fr23,Kl22,alhaitz_experimental_2023},
tunable localization and diffusive regimes \cite{Fr17,Sg22, Sc22},
enhanced absorption of waves \cite{Bi19}, enhanced solar cell efficiency \cite{merkel_stealthy_2023}, phononic properties \cite{Gk17,Ro19,Roh20}, Luneberg lenses
with reduced backscattering \cite{Zh19}, extraordinary phased arrays \cite{Ch21,tang_hyperuniform_2023}, optimal sampling array of 3D ultrasound imaging \cite{tamraoui_hyperuniform_2023}, high quality factor optical cavity \cite{granchi_nearfield_2023},
and network materials with nearly optimal effective electrical conductivities
and elastic moduli \cite{To18c}.
It has been noted that the novel physical properties of disordered isotropic
stealthy hyperuniform materials is due to their hybrid liquid-crystal
nature, including the fact that they cannot tolerate arbitrarily
large holes in the infinite-volume limit, which is a property that is also possessed
by systems with long-range order, such as crystals and quasicrystals \cite{To18a}.

The hyperuniformity concept was generalized to describe other contexts, including two-phase heterogeneous media \cite{Za09, To16b},
and random scalar and vector fields \cite{To16a}. Two-phase media, of central interest in this paper, are ubiquitous; examples include porous media, composites, cellular solids, biological media, colloids, granular media, foams, and polymer blends \cite{To02a, Sa03}. Here, the {\it phase volume fraction}
fluctuates within a spherical window of radius $R$, which is useful to describe  by the
volume-fraction variance $\sigma_{_V}^2(R)$. For ordinary disordered two-phase media, the variance  $\sigma_{_V}^2(R)$ goes to zero 
like $R^{-d}$ for large $R$. 
By contrast, for hyperuniform disordered two-phase media, $\sigma_{_V}^2(R)$  goes to zero faster than the inverse of the window volume in the large-$R$ asymptotic limit, i.e., faster than $R^{-d}$, which is equivalent to the vanishing of the spectral density $\spD{\vect{k}}$ (defined in Sec. \ref{sec:back}) in the infinite-wavelength limit, i.e.,
\begin{eqnarray}
\lim_{|\mathbf{k}|\rightarrow 0}\tilde{\chi}_{_V}(\mathbf{k}) = 0.
\label{hyper-2}
\end{eqnarray}
Similar to the instance of hyperuniform point configurations \cite{To03a, To18a, Za09}, three different 
large-$R$ scaling regimes 
arise when the spectral density goes to zero with the power-law form
\begin{equation}
{\tilde \chi}_{_V}({\bf k})\sim |{\bf k}|^\alpha;
\label{hyper-3}
\end{equation}
specifically,
\begin{eqnarray}  
    \sigma^2_{_V}(R) \sim 
    \cases{
        R^{-(d+1)}, \qquad \qquad \alpha >1 & \mbox{(Class I)}\\
        R^{-(d+1)} \ln R, \qquad \alpha = 1 & \mbox{(Class II)}\\
        R^{-(d+\alpha)}, \qquad 0 < \alpha < 1 & \mbox{(Class III)},
    }
\label{sigma-V-asy}
\end{eqnarray}
where the exponent $\alpha$ is a positive constant. 
Classes I and III are the strongest and weakest forms of hyperuniformity, respectively.
Stealthy hyperuniform  media, the major focus of this paper, are also of class I and are defined to be those that possess
zero-scattering intensity for a range of wavevectors in the vicinity of the origin \cite{To16b}, i.e.,
\begin{equation}
{\tilde \chi}_{_V}({\bf k})=0 \qquad \mbox{for}\; 0 < |{\bf k}| \le K.
\label{stealthy}
\end{equation}
Examples of stealthy hyperuniform media are periodic packings of spheres
as well as unusual disordered sphere packings derived from stealthy point patterns \cite{To16b, Zh16b}.
The reader is referred to Ref. \cite{To22b} for a review of the extraordinary multifunctional
characteristics of disordered hyperuniform media.

On the other hand, for any nonhyperuniform two-phase medium, it is simple to demonstrate that
the local volume-fraction variance has the following large-$R$ scaling behaviors \cite{To21c}:
\begin{eqnarray}
\sigma^2_{_V}(R) \sim 
\cases{
    R^{-d}, \qquad \qquad \alpha =0 & \mbox{(typical nonhyperuniform)}\\
    R^{-(d+\alpha)}, \quad -d <\alpha < 0 & \mbox{(antihyperuniform)}.
}
\label{sigma-nonhyper}
\end{eqnarray}
For a ``typical" nonhyperuniform system, ${\tilde \chi}_{_V}(0)$ is bounded \cite{To18a}. In {\it antihyperuniform} systems,
${\tilde \chi}_{_V}(0)$ is unbounded, i.e.,
\begin{equation}
\lim_{|{\bf k}| \to 0} {\tilde \chi}_{_V}({\bf k})=+\infty,
\label{antihyper}
\end{equation}
and hence are diametrically opposite to hyperuniform systems.
Antihyperuniform systems include  systems at thermal critical points (e.g., liquid-vapor and magnetic critical points) \cite{St87b, Bi92}, fractals \cite{Ma82}, disordered non-fractals \cite{To18a},
and certain substitution tilings \cite{Og19}.

In this paper, we further contrast and compare the extraordinary physical properties of stealthy hyperuniform two-phase
layered, transversely isotropic media and fully three-dimensional (3D) isotropic media to other model microstructures, including the optical characteristics
as measured by the effective dynamic dielectric constant, and transport properties, as measured by the time-dependent diffusion spreadability. 
We also show, for the first time, that there are cross-property relations between the attenuation characteristics
(imaginary part of the dielectric constant) and the spreadability. (Cross-property relations have been profitably used to link seemingly disparate
physical properties to one another \cite{Be78a, Av88, To90e, To91f, Av91b, Gi95b, Gi96b, Gi97c, To02a, To04b, Se09, Ki20a}.)
For these purposes, it is instructive to briefly review the strong-contrast formalism
for the effective dielectric constant and the spreadability concept  \cite{To21d}, as described immediately below.

    \begin{figure}
    \centering
        \includegraphics[width=0.7\textwidth]{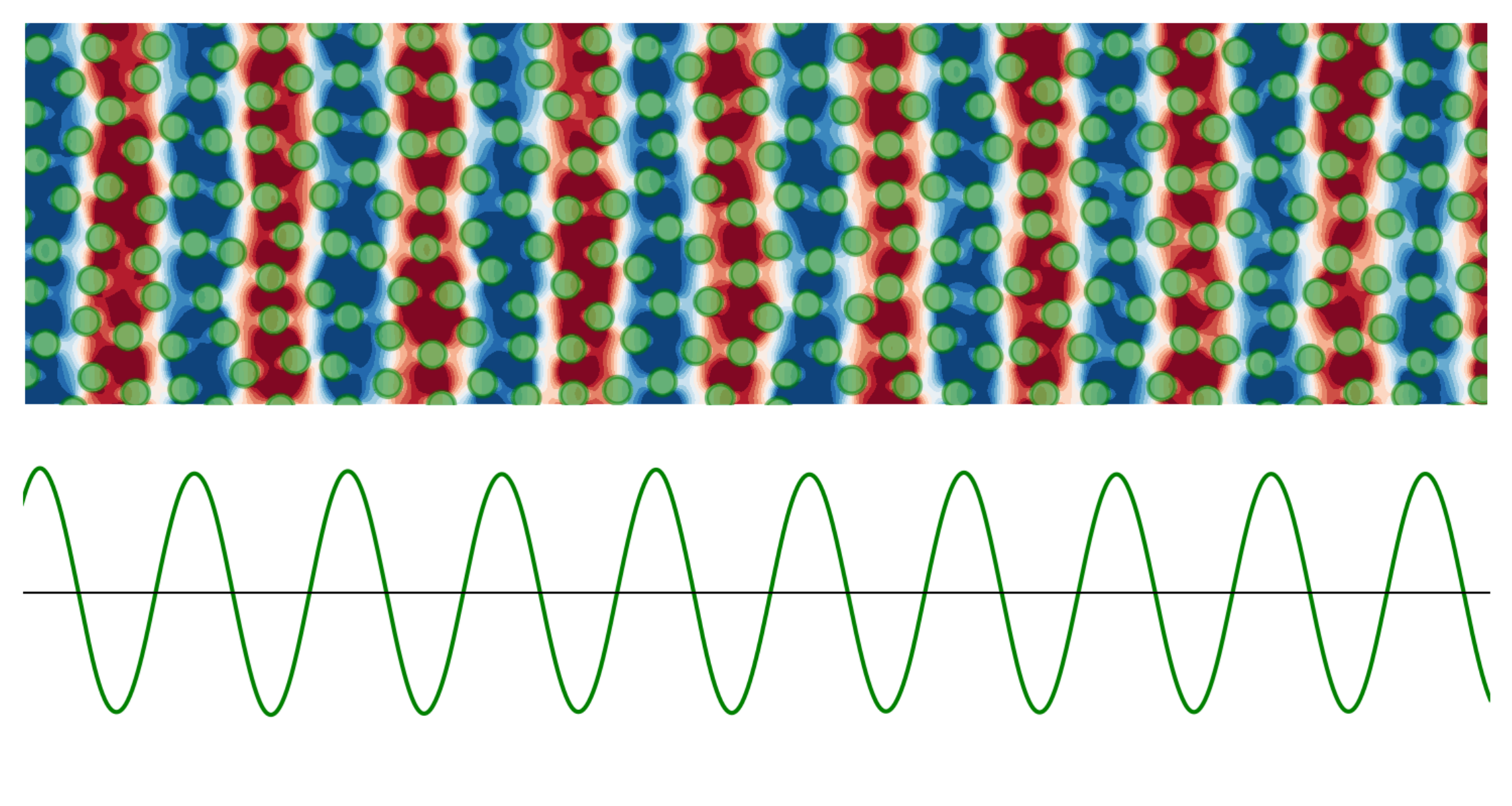}
        \caption{
        This schematic illustrates how electric waves propagate through a 2D stealthy hyperuniform medium consisting
 of identical disks (green circles) of dielectric constant $\varepsilon_2$ embedded in the matrix phase  of dielectric constant $\varepsilon_1$
 within a perfect transparency interval.
The upper panel depicts the spatial distribution of the electric
field inside the sample as the wave travels from the left to the right, assuming $\varepsilon_2 > \varepsilon_1$.
Blue and red colors represent the regions where the electric field becomes positive and negative, respectively.
The lower panel depicts the average electric field as a function of the propagation distance.
Importantly, the electric field is a perfect sine wave, implying that the wave can propagate indefinitely without loss.
            \label{fig:wave-SHU}
        }
    \end{figure}

We have derived exact nonlocal strong-contrast expansions of the effective dynamic dielectric constant tensor $\fn{\tens{\varepsilon}_e}{\vect{k}_1,\omega}$ that treat general three-dimensional (3D) two-phase composite microstructures \cite{To21a}.
These expansions are rational functions of the effective dielectric constant tensor $\fn{\tens{\varepsilon}_e}{\vect{k}_1, \omega}$, 
the terms of which depend on the microstructure via functionals of the $n$-point correlation functions $\fn{S_n ^{(i)}}{\vect{x}_1,\ldots,\vect{x}_n}$ for all $n$ (see Sec. \ref{sec:back}) and exactly treat multiple scattering to all orders beyond the long-wavelength regime (i.e., $0\leq \abs{\vect{k}_1}\xi \lesssim 1$). Due to the rapid convergence of the strong-contrast expansions \cite{To21a, Ki23}, 
their truncations at low orders yield highly accurate estimates of $\fn{\tens{\varepsilon}_e}{\vect{k}_1, \omega}$ suited for various microstructural symmetries, implying that this resulting formula very accurately approximates multiple scattering through all orders.
Specifically, we previously showed that second-order truncations, which depend on the microstructure via
functionals involving the spectral density $\spD{\vect{k}}$, already provide accurate approximations of  $\fn{\tens{\varepsilon}_e}{\vect{k}_1,\omega}$
beyond the long-wavelength regime for transverse electric (TE) polarization in transversely isotropic media \cite{To21a} and transverse polarizations in layered \cite{Ki23} and fully 3D isotropic media \cite{To21a}. Among other results, we predicted that
stealthy hyperuniform two-phase dielectric composites possess the perfect transparency intervals \cite{To21a, Ki23, kim_accurate_2024}.
As illustrated in Fig. \ref{fig:wave-SHU}, within such transparency intervals, the sinusoidal electric waves can propagate indefinitely without decaying (i.e., no Anderson localization \cite{anderson_absence_1958,mcgurn_anderson_1993, sheng_introduction_2006,aegerter_coherent_2009, izrailev_anomalous_2012,wiersma_disordered_2013}, in principle).

%For layered, transversely isotropic, and statistically isotropic media, these second-order truncations of the strong-contrast expansions already provide accurate approximations \cite{To21a, Ki23,kim_accurate_2024}; see section \ref{sec:theory}.

%Stealthy hyperuniform two-phase dielectric composites possess the perfect transparency intervals \cite{To21a,Ki23,kim_accurate_2024}.
%    As illustrated in Fig. \ref{fig:wave-SHU}, within such transparency intervals, the electric waves have a perfect sinusoidal form without decaying, implying that the %wave can propagate indefinitely (i.e., no Anderson localization \cite{anderson_absence_1958,mcgurn_anderson_1993, sheng_introduction_2006,aegerter_coherent_2009, %izrailev_anomalous_2012,wiersma_disordered_2013}, in principle).

\begin{figure*}[bt]
    \subfloat[]{\includegraphics[width=2.1in,keepaspectratio,clip=]{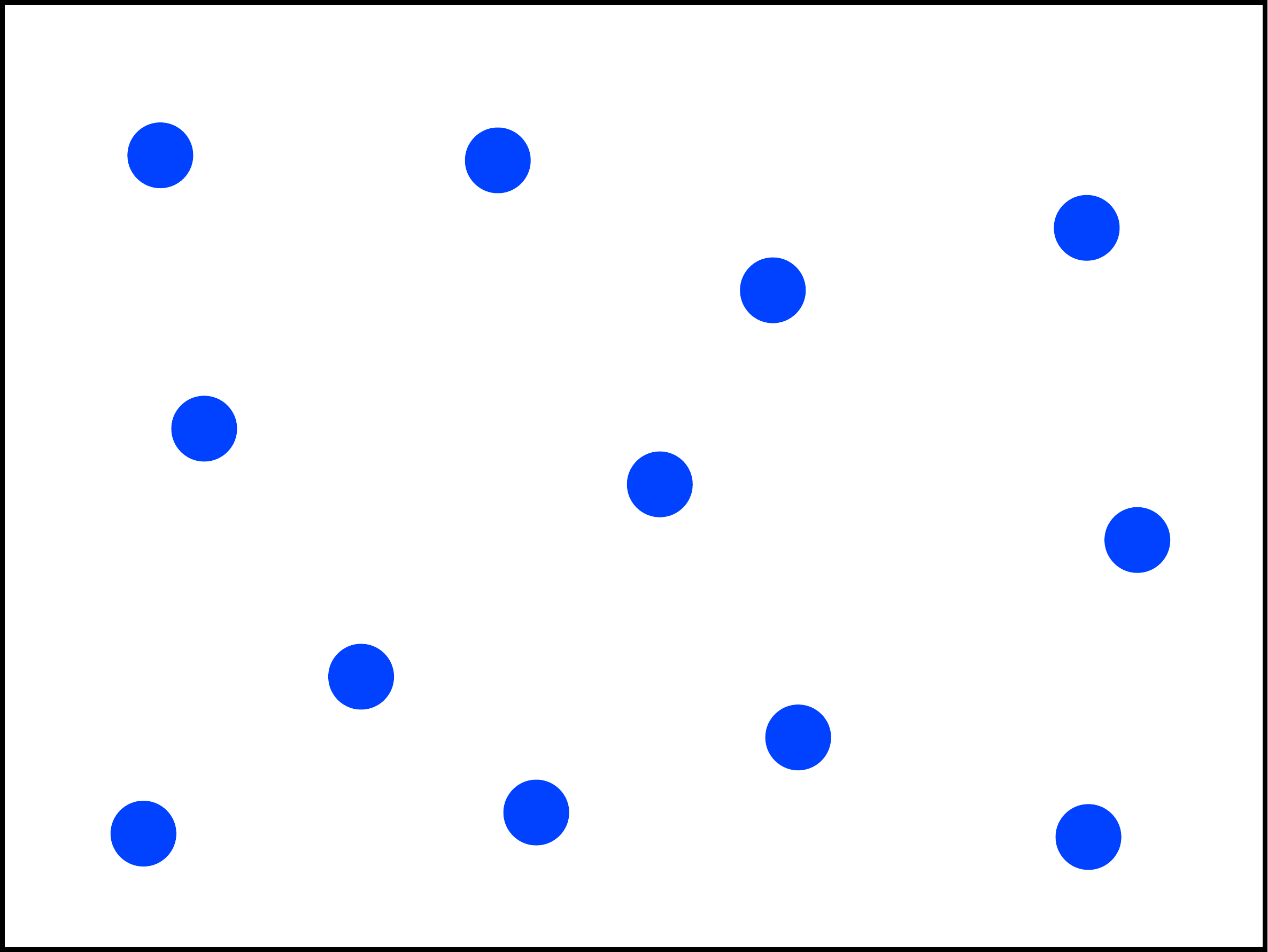}}
    \hspace{0.2in}\subfloat[]{\includegraphics[width=2.1in,keepaspectratio,clip=]{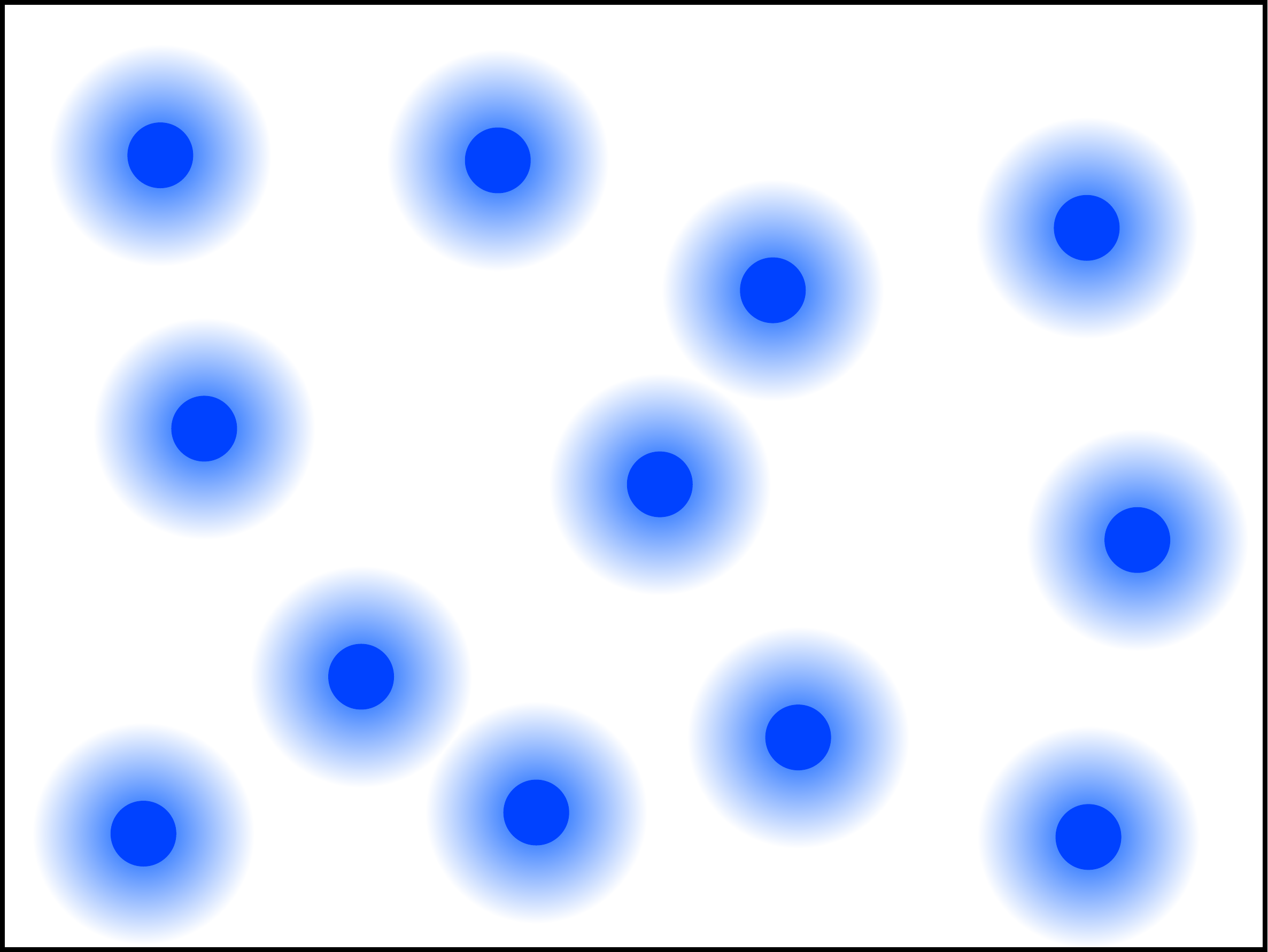}}
    \hspace{0.2in}\subfloat[]{\includegraphics[width=2.1in,keepaspectratio,clip=]{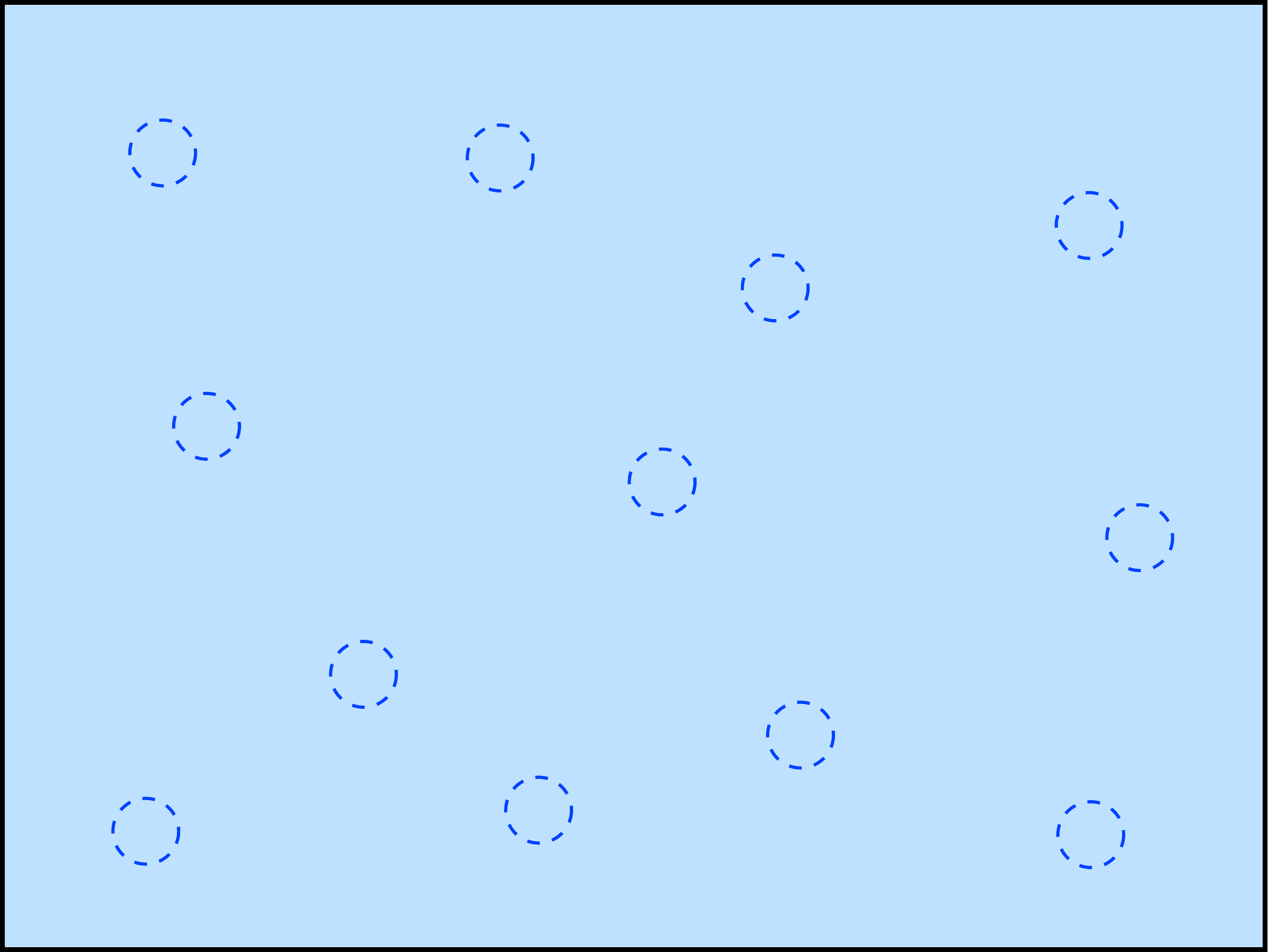}}
    \caption{For purposes of illustration, this schematic shows diffusion
    spreadability at different times for the special case
    in which phase 2 is comprised of a spatial distribution of particles. The
    left panel (a) depicts
    the uniform concentration of the solute species within phase 2 (dark blue
    regions) at time $t=0$. The middle panel (b)
    depicts the spreading of diffusion information at short times. The right
    panel (c) depicts the uniform concentration of the solute
    species throughout both phases (light blue region) in the infinite-time
    limit. The behavior of the spreadability
    ${\cal S}(t)$ as a function of time is intimately related to the
    underlying microstructure. 
    Reprinted (figure) with permission from \cite{To21d}, Copyright (2021) by the American Physical Society.}
    \label{cartoon}
    \end{figure*}

The diffusion spreadability, developed recently by Torquato \cite{To21d}, is a dynamical probe that directly links the time-dependent diffusive transport with the microstructure of heterogeneous media across length scales. 
Here, one investigates the time-dependent problem of mass transfer of a solute in a two-phase medium where all of the solute is initially contained in phase 2, and it is assumed that the solute has the same diffusion coefficient $\cal D$ in each phase. The spreadability $\mathcal{S}(t)$ is defined as the total solute present in phase 1 at time $t$; see
Fig. \ref{cartoon} for a  schematic that shows diffusion
    spreadability at different times for the very special case, for purposes of illustration,
    in which phase 2 consists of a spatial distribution of particles. For two different microstructures at some time $t$, the one with the larger value of $\mathcal{S}(t)$ spreads diffusion information more rapidly \cite{To21d}. Torquato showed that the spreadability is exactly determined
by the spectral density ${\tilde \chi}_{_V}({\bf k})$ and its
small-, intermediate-, and long-time behaviors of $\mathcal{S}(t)$ are
directly determined by the small-, intermediate-, and large-scale structural characteristics of the two-phase medium.
Notably, the spreadability of hyperuniform media always approaches their long-time asymptotic behaviors substantially faster than those in nonhyperuniform ones.
Remarkably, disordered stealthy hyperuniform media approach their long-time asymptotic behaviors exponentially faster among any hyperuniform variety \cite{To21d}; see Sec. \ref{spreadability} for relevant
formulas and more detail.

Among other results, we further study their transparency characteristics for different microstructural symmetries (Sec. \ref{sec:results}) as well as provide detailed proofs that stealthy hyperuniform layered and transversely isotropic media are perfectly transparent (i.e., no Anderson localization, in principle) within finite wavenumber intervals through the third-order terms;  see Sec. \ref{sec:trans} and \ref{sec:proofs}. Remarkably, these results imply that there can be no Anderson localization within the predicted perfect transparency interval in stealthy hyperuniform layered and transversely isotropic media in practice because the localization length (associated with only possibly negligibly small higher-order contributions) should be very large compared to any practically large sample size.

We further contrast and compare the extraordinary physical properties of 3D stealthy hyperuniform two-phase media with various symmetries
(layered, transversely isotropic media, and fully 3D isotropic media) to other model nonstealthy microstructures, including  their attenuation characteristics,
as measured by the imaginary part of $\fn{\tens{\varepsilon}_e}{\vect{k}_1,\omega}$, and transport properties, as measured by the time-dependent diffusion spreadability ${\cal S}(t)$.
Specifically, for layered media, we consider two prototypical nonhyperuniform models: Debye random media and equilibrium hard rods in a matrix; and two hyperuniform ones: nonstealthy hyperuniform polydisperse packings in a matrix and stealthy hyperuniform packings in a matrix (see Sec. \ref{sec:models}).
\footnote{ Transport of waves and mass in layered and transversely isotropic media can be
rigorously considered to be those transporting in one-dimensional ($d = 1$) and two-dimensional ($d=2$) systems, 
and so we sometimes refer to them as 1D and 2D media, respectively.}
For transversely isotropic and fully 3D isotropic media, we study the two models
that attenuate the most and least, i.e.,  Debye random media and stealthy hyperuniform ones, respectively  (see Sec. \ref{sec:models}).
\footnote{In the present work, we focus on the imaginary parts of the effective dielectric constants since their behaviors, as opposed to the real
parts, are highly sensitive to the microstructures. The reader is referred to Ref. \cite{To21a} for estimates
of the real parts as a function of the wavenumber of various hyperuniform and nonhyperuniform models.}

For all models, we estimate the imaginary part of $\fn{\tens{\varepsilon}_e}{\vect{k}_1,\omega}$ using the second-order strong-contrast formulas presented
in Sec. \ref{sec:theory}. In Sec. \ref{sec:trans}, we elaborate on the perfect transparency intervals of stealthy hyperuniform media predicted
by the second-order formulas
for all symmetries considered here, including proofs for certain cases of the same transparency intervals
through the third-order terms in the strong-contrast expansion.
We present results for the attenuation characteristics for all of our models considered here in Sec. \ref{sec:attenuation}.
We also provide in Sec. \ref{sec:spreadability} comparisons of the spreadability
as a function of time for all of our models.
In Sec. \ref{sec:cross},  we demonstrate that there are cross-property relations between $\Im[\fn{\varepsilon_e}{k_1}]$ and ${\cal S}(t)$, namely,  we quantify how the imaginary parts of the dielectric and the spreadability at long times are positively correlated as the structures span from nonhyperuniform, nonstealthy hyperuniform and stealthy hyperuniform media.
We utilize the inverse of specific surface (i.e., the mean interface area per volume) $1/s$, as a natural characteristic inhomogeneity length scale $\xi$  to scale distance or wavenumbers in order to compare properties of different models of two-phase media, as discussed in Ref.  \cite{Ki21}.
In Sec. \ref{sec:discussion}, we discuss our results and future avenues of research.

\section{Definitions and Background}
\label{sec:back}

\subsection{Two-Phase Media}

A two-phase  random medium is a domain of space $\mathcal{V} \subseteq \mathbb{R}^d$ of volume $V$
that is partitioned into two disjoint regions that make up  $\mathcal{V}$:
a phase 1 region $\mathcal{V}_1$ with volume fraction $\phi_1$ and a phase 2 region $\mathcal{V}_2$ with volume fraction $\phi_2$ \cite{To02a}.

\subsubsection{$n$-Point Correlation Functions}
\label{n-point}

The phase indicator function ${\cal I}^{(i)}({\bf x})$ for a given realization is defined as
\begin{equation}
{\cal I}^{(i)}({\bf x}) = 
\cases{
    1, \qquad {\bf x} \in {\cal V}_i,\\
    0, \qquad {\bf x} \notin {\cal V}_i.
}
\label{phase-char}
\end{equation}
The two-phase medium is fully statistically characterized by the $n$-point correlation $S^{(i)}_n({\bf x}_1,{\bf x}_2,\ldots,{\bf x}_n)$ 
associated with phase $i$ defined by \cite{To02a}
\begin{equation}
S^{(i)}_n({\bf x}_1,{\bf x}_2,\ldots,{\bf x}_n) \equiv \left\langle{{\cal I}^{(i)}({\bf x}_1) {\cal I}^{(i)}({\bf x}_2)\cdots 
{\cal I}^{(i)}({\bf x}_n) }\right\rangle
\end{equation}
for all $n \ge 1$,  where angular brackets denote
an ensemble average and ${\bf x}_1,{\bf x}_2,\ldots, {\bf x}_n$ are position vectors. 
The function $S^{(i)}_n({\bf x}_1,{\bf x}_2,\ldots,{\bf x}_n)$
can also be interpreted to be the
probability of simultaneously finding $n$ points with positions ${\bf x}_1,{\bf x}_2,\ldots,{\bf x}_n$
in phase $i$. 

For \textit{statistically homogeneous} media, there is no preferred origin, and hence
the $n$-point correlation function depends on their relative displacement vectors between the $n$ points, i.e.,
\begin{equation}
\label{eq1002}
S^{(i)}_n({\bf x}_1,{\bf x}_2,\cdots,{\bf x}_n) = S^{(i)}_n({\bf x}_{12},\cdots,{\bf x}_{1n}),
\end{equation}
where ${\bf x}_{ij}={\bf x}_j-{\bf x}_i$, where we have chosen the origin 
be at position ${\bf x}_1$.
In such instances, the one-point correlation function is a constant,  namely, the phase volume fraction $\phi_i$, i.e.,
\begin{equation}
\phi_i = \langle {\cal I}^{(i)}({\bf x}) \rangle,
\end{equation}
such that $\phi_1+\phi_2=1$, and two-point correlation function  depends on the relative displacement vector ${\bf r} \equiv {\bf x}_2-{\bf x}_1$ 
and hence $S_2^{(i)}({\bf x}_1,{\bf x}_2)=S_2^{(i)}({\bf r})$ \cite{To02a,To16a}. 
The autocovariance function $\chi_{_V}({\bf r})$ is defined as 
\begin{eqnarray}    \label{eq108}
    \chi_{_V}({\bf r}) \equiv 
    \E{\J{i}{\vect{x}+\vect{r}}\J{i}{\vect{x}}} = S^{(i)}_2({\bf r}) - {\phi_i}^2,
\end{eqnarray}
where the fluctuating part of Eq. \eqref{phase-char} is defined as 
    \begin{eqnarray}   \label{eq:J}
        \J{i}{\vect{x}} \equiv \fn{\mathcal{I}^{(i)}}{\vect{x}}-\phi_i,
    \end{eqnarray}
and thus is identical for each phase $i=1,2$.
At the extreme limits of its argument, $\chi_{_V}$ has the following asymptotic behavior
\begin{equation}
\chi_{_V}({\bf r}=0)=\phi_1\phi_2, \qquad \lim_{|{\bf r}| \rightarrow \infty} \chi_{_V}({\bf r})=0,
\label{limits}
\end{equation}
the latter limit applying when the medium possesses no long-range order. If the medium is statistically homogeneous and isotropic, then the  autocovariance
function ${\chi}_{_V}({\bf r})$ depends only on the magnitude of its argument $r=|\bf r|$,
and hence is a radial function. In such instances, its slope at the origin is directly related 
to the {\it specific surface} $s$ (interface area per unit volume); specifically, we have in any space
dimension $d$, the asymptotic form \cite{To02a},
\begin{equation}
\chi_{_V}({\bf r})= \phi_1\phi_2 - \beta(d) s \;|{\bf r}| + {\cal O}(|{\bf r}|^2),
\label{specific}
\end{equation}
where 
\begin{equation}
\beta(d)= \frac{\Gamma(d/2)}{2\sqrt{\pi} \Gamma((d+1)/2)}.
\label{beta}
\end{equation}

\subsubsection{Spectral Density}

The nonnegative spectral density ${\tilde \chi}_{_V}({\bf k})$, which can be obtained from  scattering experiments \cite{De49,De57},
is  the Fourier transform of $\chi_{_V}({\bf r})$, i.e.,
\begin{equation}
{\tilde \chi}_{_V}({\bf k}) = \int_{\mathbb{R}^d} \chi_{_V}({\bf r}) e^{-i{\bf k \cdot r}} {\rm d} {\bf r} \ge 0, \qquad \mbox{for all} \; {\bf k}.
\label{1}
\end{equation}
For isotropic media, the spectral density only depends
on $k=|{\bf k}|$ and, as a consequence of  (\ref{specific}), its decay in the large-$k$ limit is controlled
by the exact following power-law form \cite{To02a}:
\begin{equation}
{\tilde \chi}_{_V}({\bf k}) \sim \frac{\gamma(d)\,s}{k^{d+1}}, \qquad k \rightarrow \infty,
\label{decay}
\end{equation}
where 
\begin{equation}
\gamma(d)=2^d\,\pi^{(d-1)/2} \,\Gamma((d+1)/2)
\end{equation}
is a $d$-dimensional  constant.

For the special case of packing of identical spheres of radius $a$ that comprises phase 2 with packing fraction $\phi_2$, the spectral
density is simply related to the structure factor \cite{To85b,To02a}:
\begin{eqnarray}
{\tilde \chi}_{_V}({\bf k})&=& \phi_2{\tilde \alpha}(k;a) S({\bf k}).
\label{chi_V-S}
\end{eqnarray}
where
\begin{equation}
{\tilde \alpha}(k;a)= \frac{1}{v_1(a)} \left(\frac{2\pi a}{k}\right)^{d} J_{d/2}^2(ka),
\end{equation}
where $v_1(a)=\pi^{d/2} a^d/\Gamma(1+d/2)$ is the volume of a $d$-dimensional sphere, and $\fn{J_\nu}{x}$ is the Bessel function of the first kind of order $\nu$.
Since ${\tilde \alpha}(k;a)$ is a positive, bounded well-behaved function in the vicinity of   the origin,
it immediately follows from expression (\ref{chi_V-S}) that if the underlying
point process is hyperuniform with a power-law structure factor $S({\bf k}) \sim |{\bf k}|^{\alpha}$ in the limit $|{\bf k}|\to 0$, then
the spectral density ${\tilde \chi}_{_V}({\bf k})$ inherits the same power-law form [cf. (\ref{hyper-3})]
only through the structure factor, not ${\tilde \alpha}(k;a)$ \cite{To16b}.
Moreover, it is clear that for stealthy hyperuniform packings, relation (\ref{chi_V-S})
dictates that ${\tilde \chi}_{_V}({\bf k})$ is zero for the same wavenumbers at which $S({\bf k})$ is zero [i.e., Eqs. \eqref{stealthy-points} and \eqref{chi_V-S}
yield Eq. \eqref{stealthy}], but the spectral density also vanishes at the zeros
of the function ${\tilde \alpha}(k;a)$, which is determined by the zeros of $J_{d/2}(ka)$ \cite{To16b}.

\subsection{Diffusion Spreadability}
\label{spreadability}

Torquato demonstrated that the time-dependent diffusion spreadability
$\mathcal{S}(t)$ in any $d$-dimensional Euclidean space $\mathbb{R}^d$ is \textit{exactly} related to the microstructure via the autocovariance function $\chi_{_V}(\mathbf{r})$ in direct space, or equivalently, via the spectral density $\tilde{\chi}_{_V}(\mathbf{k})$ in Fourier space \cite{To21d}:
\begin{equation}
    \hspace{-1cm}\mathcal{S}(\infty) - \mathcal{S}(t)  =\frac{1}{(2\pi)^d\phi_2}\int_{\mathbb{R}^d} \tilde{\chi}_{_V}(\mathbf{k})\exp[-k^2{\cal D}t]d\mathbf{k}.
\label{spread}
\end{equation}
Here, $\phi_2$ is the volume fraction of phase 2, $\mathcal{S}(\infty)=\phi_1$ is the infinite-time limit of $\mathcal{S}(t)$, and $\mathcal{S}(\infty)-\mathcal{S}(t)$ is called the {\it excess spreadability} \footnote{Larger (smaller) spreadability $\mathcal{S}(t)$ implies smaller (larger) excess spreadabilty $\mathcal{S}(\infty)-\mathcal{S}(t)$.}.
This situation is a rare example of transport in two-phase media that is exactly described by only the first two correlation functions, namely, $\phi_1$ and two-point statistics via either $\chi_{_V}(\mathbf{r})$ or $\tilde{\chi}_{_V}(\mathbf{k})$.
The reader is referred to Ref. \cite{To21d} for a description of remarkable
links between      the spreadability ${\cal S}(t)$, covering problem of discrete geometry,
and nuclear magnetic resonance (NMR) measurements.

For statistically homogeneous microstructures whose spectral densities exhibit the following power-law form
\begin{equation}
\tilde{\chi}_{_V}(\mathbf{k})\sim
B|\mathbf{k}|^{\alpha}\label{eqn:smallk}
\end{equation}
in the limit $|\mathbf{k}|\to0$, Torquato \cite{To21d} showed that the
long-time excess spreadability for two-phase materials in $\mathbb{R}^d$
is given by the inverse power-law
\begin{equation}
{\cal S}(\infty)- {\cal S}(t) = \frac{ C}{
({\cal D}t/a^2)^{(d+\alpha)/2}}
+{o}\left(({\cal D}t/a^2)^{-(d+\alpha)/2}\right) \quad ({\cal D}t/a^2 \gg 1),
\label{long-time}
\end{equation}
where $a$ is a characteristic heterogeneity length scale, ${o}(x)$ signifies all terms of
order less than $x$, and
\begin{equation} \label{eq:C}
C= B\,
\Gamma((d+\alpha)/2)\,\phi_2/[2^d\,{\pi}^{d/2}\Gamma(d/2)] 
\end{equation}
is a  microstructure-dependent coefficient; see Fig. \ref{fig:cartoon-2}.
Thus, it is seen that the long-time asymptotic behavior of ${\cal S}(t)$ is determined by the exponent $\alpha$
and the space dimension $d$, i.e., at long times, ${\cal S}(t)$ approaches the value $\phi_1$ with a power-law decay
$1/t^{(d+\alpha)/2}$, implying a faster decay as $\alpha$ increases for some dimension $d$.
Thus, compared to a nonhyperuniform medium with a power-law decay $t^{-d/2}$, a hyperuniform medium with a decay rate $1/t^{(d+\alpha)/2}$ can be viewed as having an effective dimension that is higher than the space dimension, namely, $d+\alpha$.
When $\alpha$ is bounded and positive, this result means that class I hyperuniform media has the fastest decay, followed by class II and then class III, which has the slowest decay
among hyperuniform media. Of course, antihyperuniform media with $\alpha \to -d$
has the slowest decay among all translationally invariant media.
In the {\it  stealthy limit} in which $\alpha =+ \infty$, the predicted infinitely-fast
inverse-power-law decay rate implies that the infinite-time asymptote is approached exponentially
fast. Indeed, for disordered stealthy media defined by (\ref{stealthy}), it was shown \cite{To21d} that
long-time behavior is exactly given by
\begin{equation}
{\cal S}(\infty)- {\cal S}(t) \sim \frac{\exp(-K^2 {\cal D}t)}{K^2 {\cal D}t} \qquad ({\cal D}t/a^2 \gg 1).
\label{S-stealthy}
\end{equation}

\begin{figure*}[bthp]
\includegraphics[width=5in,keepaspectratio,clip=]{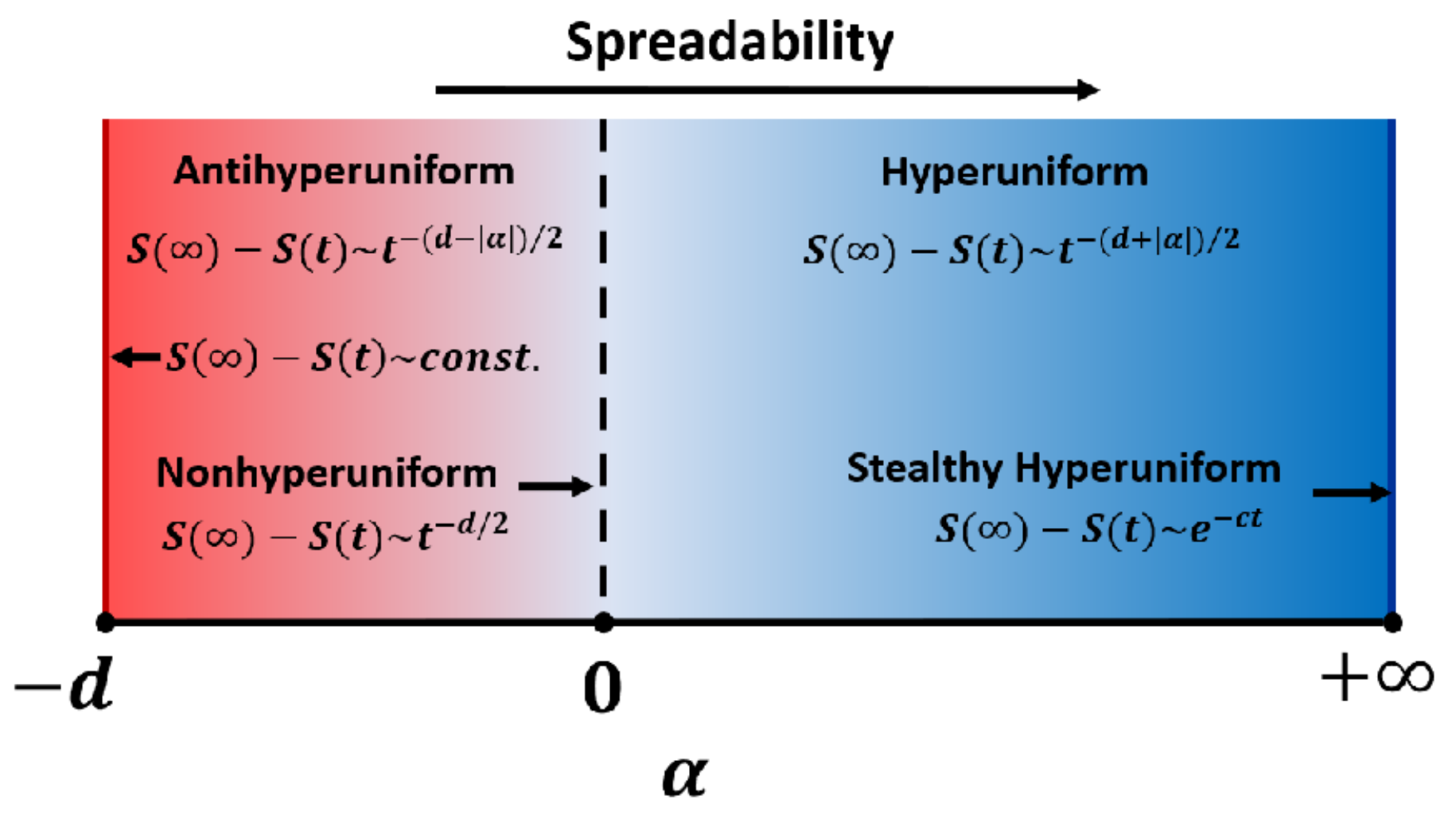}
\caption{``Phase diagram" that schematically shows the spectrum of spreadability regimes in terms of the exponent $\alpha$  (taken from Ref. \cite{To21d}).
As $\alpha$ increases from the extreme antihyperuniform limit of $\alpha \to -d$, the spreadability decay rate gets faster, i.e.,
the excess spreadability follows the inverse power law $1/t^{(d+\alpha)/2}$, except when $\alpha = +\infty$, which
corresponds to stealthy hyperuniform media with a  decay rate that is exponentially fast.
Reprinted (figure) with permission from \cite{To21d}, Copyright (2021) by the American Physical Society.
}
\label{cartoon-2}
\end{figure*}

\section{Model Microstructures}
\label{sec:models}

\begin{figure}
    \centering
    \includegraphics[width=\textwidth]{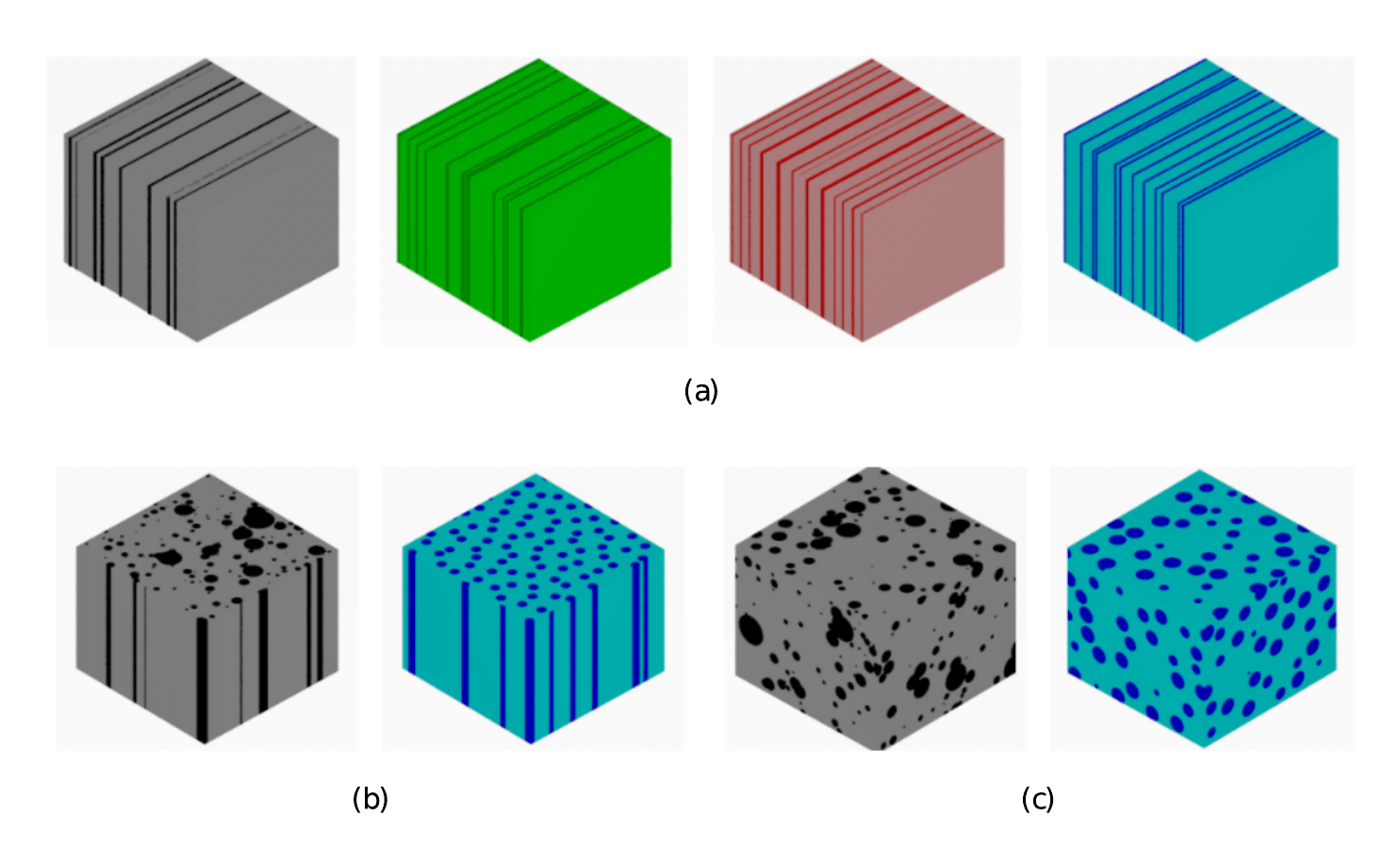}

    \caption{
        Representative images of 3D two-phase media  
        with $\phi_2=0.20$ possessing three different microstructural symmetries that are considered in the present study.
        (a) Four models of layered media, consisting of infinite parallel slabs of phases 1 and 2 whose thicknesses are derived from 1D models.
        From the leftmost to the rightmost, we present Debye random media (black), equilibrium hard rods in a matrix (green), nonstealthy hyperuniform polydisperse sphere packing in a matrix (red), and stealthy hyperuniform sphere packing in a matrix (blue), respectively.
        (b) Two models of transversely isotropic media with cylindrical symmetry obtained from 2D models: Debye media (black) and stealthy hyperuniform sphere packings in a matrix (blue).
        (c) Two models of fully isotropic media: Debye media (black) and stealthy hyperuniform sphere packings (blue).
        In each subfigure, the bright and dark colors represent phases 1 and 2, respectively.
%        In each symmetry, all models are scaled to the same specific surface. 
        \label{fig:schm-models}
    }
\end{figure}

In what follows, we describe nonhyperuniform and hyperuniform models of 3D  two-phase media with certain microstructural symmetries for which
we investigate the aforementioned optical and transport properties. Specifically, we describe 3D
model microstructures of layered, transversely isotropic media, and fully isotropic media, , including a description of their spectral densities, which are used to compute their effective dynamic dielectric constants and spreadabilities. Representative images of the models are shown in Fig. \ref{fig:schm-models}.
In Fig. \ref{fig:schm-models}(a), we depict representative images of the four models of layered media: Debye random media, equilibrium hard rods in a matrix, nonstealthy hyperuniform polydisperse sphere packing in a matrix, and stealthy hyperuniform sphere packing in a matrix.
For transversely isotropic and fully isotropic media, we depict representative images of the two models for each symmetry: Debye media and stealthy hyperuniform packings in a matrix; see Fig. \ref{fig:schm-models}(b) and (c).
Among four models, three, except for Debye random media, are derived from packings of nonoverlapping particles such that the particles of dielectric constant $\varepsilon_2$ are distributed throughout a matrix of dielectric constant $\varepsilon_1$, and thus $\phi_2$ indicates the packing fraction.
We take $\phi_2=0.2$ for all models considered here.
In what follows, we describe all of the models.

\subsection{Models} \label{sec:models-def}

\subsubsection{Debye Random Media} \label{sec:Debye}

Debye {\it et al.} \cite{De57} hypothesized
that the following autocovariance function characterizes  isotropic random media
in which the  phases form domains of ``random shape and size":  
\begin{equation}
\chi_{_V}(r)=\phi_1\phi_2 \exp(-r/a),
\label{Debye}
\end{equation}
where $a$ is a characteristic length scale. This functional form for $\chi_{_V}(r)$
was later shown to be realizable by two-phase media across space dimensions
\cite{Ye98a,To20,Sk21} and was dubbed {\it Debye random media} \cite{Ye98a,To02a}.
%Figure \ref{fig:schm-models}(a) shows the realized Debye media for $d=1,2,3$ by using overlapping polydisperse spherse \cite{Sk21}.
The Taylor expansion of (\ref{Debye}) about $r=0$ and comparison to (\ref{specific}) reveals
that the specific surface $s$ of a Debye random medium in any space dimension is given by
\begin{equation}
s= \frac{\phi_1\,\phi_2}{\kappa(d)\,a},
\end{equation}
The spectral density of Debye random media in any space dimension is given by \cite{To20}
\begin{equation}    \label{eq:chi-Debye}
{\tilde \chi}_{_V}(k) = \frac{ \phi_1\phi_2\, c_d\, a^d}{[1+(ka)^2]^{(d+1)/2}},
\end{equation}
where $c_d=2^d \pi^{(d-1)/2} \Gamma((d+1)/2)$. The representative images of this model shown in Fig. \ref{fig:schm-models} are obtained using the construction techniques of Ref. \cite{Sk21}.

\subsubsection{Equilibrium Hard Rods in a Matrix}
\label{sec:equilibrium}

%\textcolor{red}{Make the ensuing words consistent with the subsubsection heading.}

We also examine equilibrium (Gibbs) ensembles of identical hard spheres of radius $a$ with packing fraction $\phi_2$ \cite{Ha86} in a matrix. 
In particular, we consider such disordered packings along the stable disordered fluid branch in the phase diagram \cite{To02a}.
All such states are nonhyperuniform. In the case of 1D equilibrium hard rods in a matrix, pair statistics
are known exactly \cite{Pe64}. In particular, using the exact solution of the 
direct correlation function \cite{Ze27, Pe64} and the Ornstein-Zernike integral equation, one can obtain the exact structure factor $\fn{S}{k}$. 
This solution for $\fn{S}{k}$, together with Eq. \eqref{chi_V-S}, yields the corresponding spectral density $\spD{k}$ \cite{To02a, To21d}:
\begin{equation}    \label{eq:chi-EHS}
    \fl
    \spD{k} = \phi_2 \left[ \frac{2\sin(ka)^2}{k^2a} \right]\left[ 1 - \frac{2\phi_2\left\{ \phi_2\left[\cos(2ak)-1\right] +2ak\sin(2ak)(\phi_2-1) \right\}}{(1-\phi_2)^2(2ak)^2} \right]^{-1}.
\end{equation}
The representative image of this model in Fig. \ref{fig:schm-models}(a) is generated via the Monte Carlo method \cite{To02a}.
Henceforth, we call this model equilibrium (Equil.) media. 
%A representative image of 1D equilibrium medium is shown in Fig. \ref{fig:schm-models}(b).

%we can express the exact structure factor. 
% as
%     \begin{equation*}
%         S(k) = \left[ 1 - \frac{2\phi_2\left\{ \phi_2\left[\cos(2ak)-1\right] +2ak\sin(2ak)(\phi_2-1) \right\}}{(1-\phi_2)^2(2ak)^2} \right]^{-1}. \\
%     \end{equation*}
%For $d=3$, we utilize the Percus-Yevick approximation of the structure factor $S(k)$ \cite{Ha86}:
% \begin{align}
% \fn{S}{k} = 
% &\Big(1-\rho \frac{16 \pi  a^3 }{q^6} 
% \Big\{\big[24 a_1 \phi_2 - 12 (a_1 + 2 a_2) \phi_2 q^2 
%     \nonumber \\
%     &\quad+ (12 a_2 \phi_2 + 2 a_1 + a_2\phi_2) q^4] \cos(q) 
%     \nonumber \\
% &+ [24 a_1 \phi_2 q - 2 (a_1 + 2 a_1 \phi_2 + 12 a_2 \phi_2) q^3\big] \sin(q)
%     \nonumber \\
% & -24 \phi_2 (a_1 - a_2 q^2) \Big\}\Big)^{-1},
% \end{align}
% where $q = 2ka$, $a_1 = (1+2\phi_2)^2/(1-\phi_2)^4$, and $a_2 = -(1+\phi_2/2)^2 /(1-\phi_2)^4$.
%Using this solution for the structure factor in conjunction with (\ref{chi-packing}) yields the corresponding spectral density ${\tilde \chi}_{_V}(k)$.
 
\subsubsection{Nonstealthy Hyperuniform Polydisperse Sphere Packings in a Matrix}
\label{sec:poly}

%\textcolor{red}{Make the ensuing words consistent with the subsubsection heading.}
Nonstealthy hyperuniform (NSHU) packings of spheres with a polydispersity in size in a matrix can be constructed from nonhyperuniform progenitor point patterns via a tessellation-based procedure \cite{kim_new_2019}.
Specifically, we begin with the Voronoi tessellation \cite{To02a} of the progenitor point patterns, which are the centers of 1D equilibrium media (see section \ref{sec:equilibrium}) with packing fraction $0.20$ in this work.
We then move the particle center in a Voronoi cell to its centroid and rescale the particle such that the packing fraction inside this cell is identical to a prescribed value $\phi_2<1$.
The same process is repeated over all cells.
The final packing fraction is $\phi_2 = \sum_{j=1}^N \fn{v_1}{a_j}/V_\mathfrak{F} = \rho \fn{v_1}{a}$, where $\rho$ is the number density of particle centers and $a$ represents the mean sphere radius.
In the thermodynamic limit, the spectral densities of the resulting particulate composites, which are NSHU media, exhibit a power-law scaling $\spD{\vect{k}} \sim \abs{\vect{k}}^4$ for small wavenumbers \cite{kim_new_2019}, which are of class I.

\subsubsection{Stealthy Hyperuniform Sphere Packings in a Matrix}
\label{sec:SHU}

%\textcolor{red}{Make the ensuing words consistent with the subsubsection heading.}

Stealthy hyperuniform (SHU) two-phase media have $\spD{\vect{k}}=0$ for the finite range $0<\abs{\vect{k}}\leq K$, called the {\it exclusion-region radius}.
Specifically, we consider $d$-dimensional SHU sphere packings in a matrix of packing fraction $\phi_2$ that can be numerically generated in the following two-step procedure. 
First, we generate point configurations consisting of $N$ particles in a fundamental cell $\mathfrak{F}$ under periodic boundary conditions via the collective-coordinate optimization technique \cite{Uc04b,  Zh15a}, which finds numerically the ground-state configurations of the following potential energy;
\begin{equation*}\label{eq:CC_potential}
\fn{\Phi}{\vect{r}^N} =\frac{N}{2V_\mathfrak{F}} \sum_{\vect{k}} \fn{\tilde{v}}{\vect{k}}\fn{S}{\vect{k}} +  \sum_{i <j} \fn{u}{r_{ij}},
\end{equation*}
where $V_\mathfrak{F}$ is the volume of $\mathfrak{F}$, $\fn{\tilde{v}}{\vect{k}}=\fn{\Theta}{K-\abs{\vect{k}}}$, $\Theta(x)$ (equal to 1 for $x>0$ and zero otherwise) is the Heaviside step function, soft-core repulsion $\fn{u}{r}=(1-r/\sigma)^2\fn{\Theta}{\sigma-r}$ \cite{Zh17a}. 
The consequent ground-state configurations, if they exist, are still stealthy and hyperuniform, and their nearest-neighbor distances are larger than the length scale $\sigma$.
Finally, to create packings in a matrix, we follow Ref. \cite{Zh16b} by circumscribing the points by identical spheres of radius $a<\sigma/2$ under the constraint that they cannot overlap.
Importantly, one cannot obtain disordered SHU packings in a matrix with $\phi_2\geq 0.2$ for $d=1,2$ without the soft-core repulsion $\fn{u}{r}$ at low $\chi$ values \cite{Zh15a}.

For such SHU media, the degree of stealthiness $\chi$ is measured by the ratio of the number of the wave vectors within the exclusion-region radius in the Fourier space to the total degrees of freedom \cite{Uc04b,  Zh15a}, i.e., $\chi = K / (2\pi \rho)$ in one dimension, and $\chi=K^2 / (16\pi \rho)$ in two dimensions.
These SHU media are highly degenerate and disordered if $\chi <1/3$ in one dimension or $\chi<1/2$ in two and three dimensions \cite{Zh15a}.
In the present work, we consider the cases of $\chi=0.3$ in the first three space dimensions.

\subsection{Spectral Densities}
\label{sec:spectral}

\begin{figure}
    \includegraphics[width=0.32\textwidth]{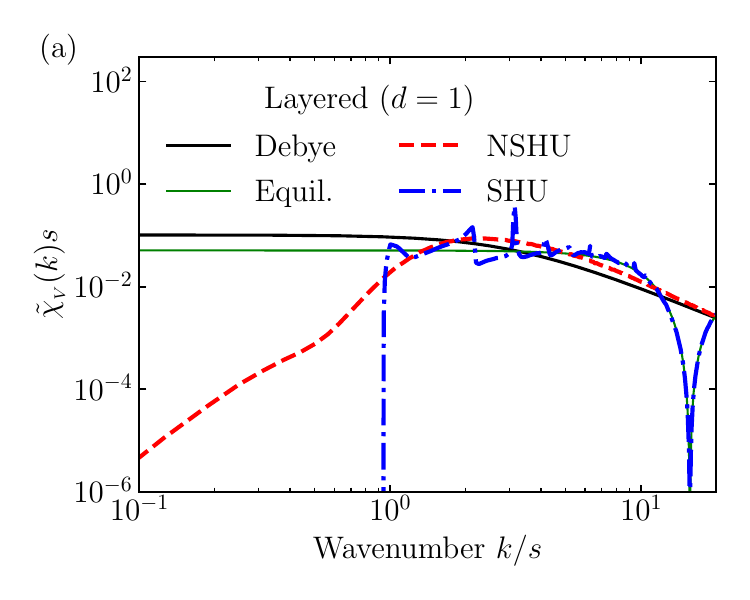}
    \includegraphics[width=0.32\textwidth]{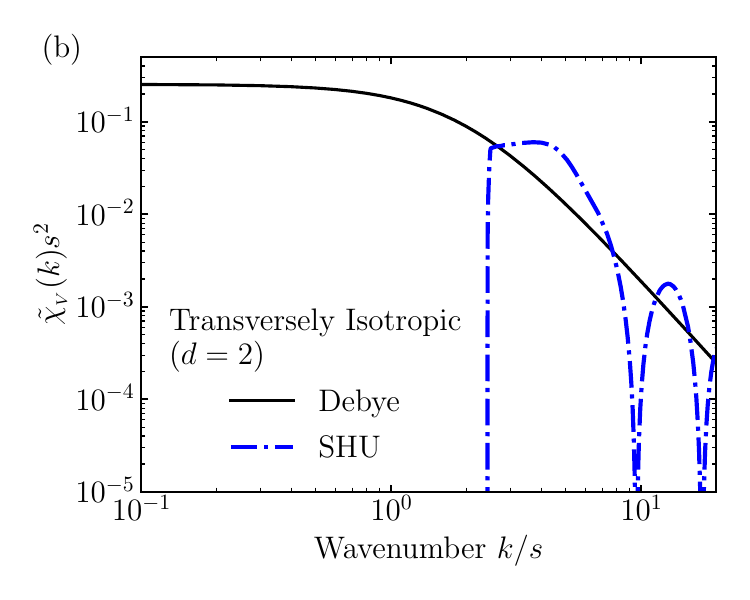}
    \includegraphics[width=0.32\textwidth]{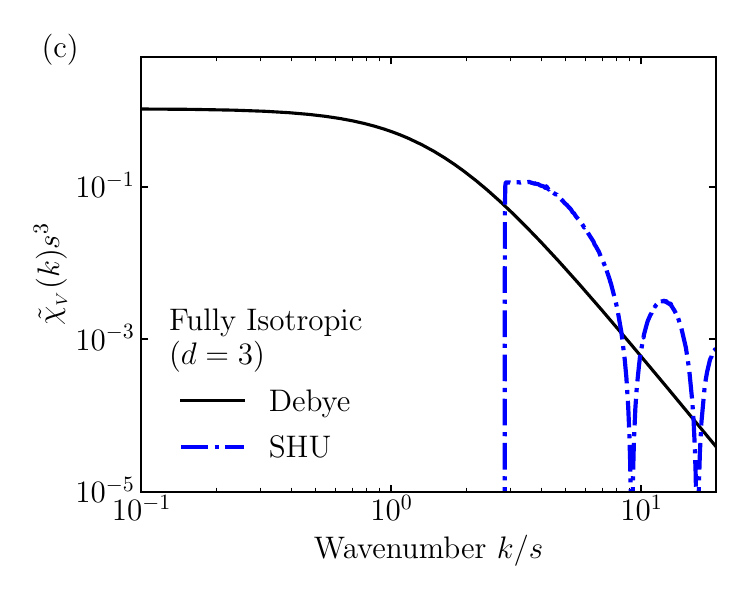}

    \caption{
        Log-log plots of dimensionless radial spectral densities $\spD{k}s^d$ as functions of dimensionless wavenumber $k/s$ for all two-phase media considered here with three structural symmetries, each with 
        $\phi_2=0.20$: (a) layered, (b) transversely isotropic, and (c) fully isotropic media.
        Representative images of each model are shown in Fig. \ref{fig:schm-models}.  
        %(a) layered, i.e., $d=1$, (b) transversely isotropic, i.e., $d=2$, and (c) fully isotropic media, i.e., $d=3$. 
        %We consider four models (Debye, Equil., NSHU, SHU media) in (a), whereas we consider two (Debye and SHU media) in (b,c).
        \label{fig:chik}
    }
\end{figure}

Estimation of the attenuation and spreadability behaviors reported in section \ref{sec:results} require the computation of the spectral density for the models.
Figure \ref{fig:chik} shows the dimensionless radial spectral density $\spD{k}s^d$ for all models considered here with three microstructural symmetries, each with $\phi_2=0.20$.
For all nonhyperuniform media (Debye for $d=1,2,3$ and equilibrium for $d=1$), $\spD{k}$ is evaluated by using the analytic expressions given in Eqs. \eqref{eq:chi-Debye} and \eqref{eq:chi-EHS}.
For all hyperuniform media (NSHU for $d=1$ and SHU for $d=1,2,3$), we obtain $\spD{k}$ from numerically generated configurations.

%We begin by considering the four models of layered media, i.e., Debye, equilibrium, NSHU, and SHU media; see Fig. \ref{fig:chik}(a).
Across all length scales, the spectral densities for all four models of layered media are considerably different from one another; see Fig. \ref{fig:chik}(a).
In the long-wavelength regime ($k/s \ll 1$), the SHU medium suppresses volume-fraction fluctuations [i.e., $\spD{k}=0$] to a greater degree than the NSHU medium over a wider range of wavelengths.
By contrast, Debye media depart the most from hyperuniformity.     
In the small-wavelength regime ($k/s \gg 10$), $\spD{k}$ of SHU and equilibrium media commonly show sharp oscillations because these two media consist of particles of identical size.
By contrast, for the same wavelength regime, Debye and NSHU media do not exhibit any oscillations because they consist of phase domains with a broad range of sizes. 
%For transversely isotropic ($d=1$) and fully isotropic ($d=3$) media, we consider two models (i.e., Debye and SHU media); see Fig. \ref{fig:chik} (b) and (c), respectively.
For the other two symmetries ($d=2,3$), the behaviors of $\spD{k}$ are qualitatively similar to those observed in layered media, as seen in Fig. \ref{fig:chik}(b) and (c).

\section{Strong-Contrast Formulas for the Effective Dynamic Dielectric Constant}    
\label{sec:theory}

In what follows, we present the key formulas for the effective dynamic
dielectric constant tensor of layered and transversely
isotropic media, which were extracted from the general strong-contrast
expansion in Refs. \cite{To21a,Ki23,kim_accurate_2024}.
We present relevant formulas at the three-point level to prove that SHU layered and transversely isotropic media are perfectly transparent within finite wavenumber intervals through the third-order terms in section \ref{sec:trans}.
At the two-point level, these formulas
involve $\spD{k}$, which are applied in section \ref{sec:results} to estimate the effective attenuation behaviors of the various models due to their simplicity and great accuracy.
We also employ the corresponding
formula for 3D isotropic media at the two-point level, which was previously derived in Ref.
\cite{To21a}, and apply it in the present work.
The reader is referred to \ref{sec:details} for the problem setup and assumptions from which the formulas below are obtained.

\subsection{Multiple-Scattering Approximations for Layered Media}   \label{sec:layered}

We consider our multiple-scattering approximations for the effective dynamic dielectric constant tensor $\fn{\tens{\varepsilon}_e}{\vect{k}_1}$ of layered media [Fig. \ref{fig:schm-models}(a)] when waves are normally incident, i.e., $\vect{k}_1 = k_1 \uvect{z}$, where $\uvect{z}$ is a unit vector along the $z$-direction, which is the symmetry axis of the layered media.
(The reader is referred to Supplement1 of Ref. \cite{kim_accurate_2024} for the formulas of obliquely incident waves.)
Thus, these formulas depend on the wavenumber $k_1$.
%They are extracted from exact strong-contrast series \eqref{eq:str-exp-trunc} by choosing a disk-like exclusion volume normal to $\uvect{z}$ that involves the singularity of the dyadic Green's function.
%The reader is referred to Refs. \cite{Ki23,kim_accurate_2024} for derivations.
%Sec. \ref{secS:SM-Layered} of \supp.
Due to the symmetries of layered media, one can decompose $\fn{\tens{\varepsilon}_e}{k_1}$ into two orthogonal components $\fn{\varepsilon_e ^\perp}{k_1}$ and $\fn{\varepsilon_e ^z}{k_1}$ for the transverse and longitudinal polarizations, respectively, as follows:
$\fn{\tens{\varepsilon}_e}{k_1} = \fn{\varepsilon_e ^\perp}{k_1} \left(\tens{I}-\uvect{z}\uvect{z} \right) + \fn{\varepsilon_e^z}{k_1}\uvect{z}\uvect{z}$. 
%Thus, we derive two independent approximations by truncating the strong-contrast expansion and solving for $\fn{\tens{\varepsilon}_e}{k_1}$ from this linear fractional form.
%Renormalization of the resulting formulas with the reference phase for the optimal convergence, equivalent to using the effective Green's function in Ref. \cite{To21a}, yields 
% {\color{red}Among them, $\fn{\varepsilon_e ^\perp}{k_1}$ is of central interest in this paper.

Here, we focus on the formulas for $\fn{\varepsilon_e ^\perp}{k_1}$. (The reader is referred to \ref{sec:formula-layered} for the formula for the longitudinal polarization.)
We first present the formula at the three-point level, which is used to prove that SHU layered media are perfectly transparent at the three-point level; see section \ref{sec:trans}.
%Then, the second-order formula is used to estimate the effective attenuation behaviors for the four different models in section \ref{sec:results} because of its simplicity and great accuracy, as demonstrated in Refs. \cite{Ki23,kim_accurate_2024}.
The \emph{scaled} strong-contrast approximation for $\fn{\varepsilon_e ^\perp}{k_1}$ at the three-point level is given as
\begin{eqnarray}
    \fn{\varepsilon_e^\perp}{k_1} =& \varepsilon_1 \left[ 
        1 + \frac{{\phi_1}^2 \varepsilon_2/\varepsilon_1\BETA{1}{21}}{\phi_2 - (\varepsilon_2 \BETA{1}{21}) \fn{A_2^\perp}{k_*; \E{\varepsilon}} - (\varepsilon_2 \BETA{1}{21})^2 \fn{A_3^\perp}{k_*; \E{\varepsilon}} }
    \right],
    \label{eq:scaled_eps_layered_trans}
    % \\
    % \fn{\varepsilon_e^z}{k_1} =& \varepsilon_1 (1 - \phi_2\BETA{1}{21})^{-1},
    % \label{eq:scaled_eps_layered_long}
\end{eqnarray}
where $k_*\equiv k_1 \sqrt{\E{\varepsilon}/\varepsilon_1}$, $\BETA{1}{pq}\equiv  1-\varepsilon_q/\varepsilon_p$ is the 1D counterpart of the \emph{dielectric polarizability}, and the second- and third-order terms are defined, respectively, as 
\begin{eqnarray}
    \fl \fn{A_2^\perp}{k;\varepsilon} 
    \equiv & 
    %\frac{1}{\varepsilon} \F{1}{k}
    %= 
    \frac{1}{\varepsilon}\left\{\frac{{k}^2}{\pi} \mathrm{p.v.}\int_{0}^{\infty} \dd{q} 
    \frac{\spD{q}}{{q}^2 - {(2k)}^2}
    +
    \frac{i k}{4} [
    \spD{0} +  \spD{2k}
    ]\right\}   \label{eq:A2-strat-Fourier},    \\
    \fl \fn{A_3^\perp}{k;\varepsilon} 
    \equiv & 
    \frac{-1}{\phi_2} \left(\frac{{k}^2}{2\pi \varepsilon} \right)^2
    \int_{-\infty}^\infty \dd{q_{1}} \frac{1}{(k+q_1)^2-{k}^2} \int_{-\infty}^\infty \dd{q_2} \frac{1}{(k+q_2)^2-{k}^2} \fn{\tilde{\Delta}_3^{(p)}}{q_1,q_2}
    \label{eq:A3-strat-Fourier},
\end{eqnarray}
where $\mathrm{p.v.}$ stands for the Cauchy principal value, and $\fn{\tilde{\Delta}_3^{(2)}}{q_1,q_2}$ is the Fourier transform of $\fn{\Delta_3^{(2)}}{z_{21},z_{31}}\equiv \fn{S_2^{(2)}}{z_{21}} \fn{S_2^{(2)}}{z_{32}} - \phi_2 \fn{S_3^{(2)}}{z_{21}, z_{32}+z_{21}}$.
%The second-order counterpart of \eqref{eq:scaled_eps_layered_trans}, where $A_3^\perp = 0$, was derived in Ref. \cite{Ki23}, and its accuracy was numerically demonstrated for hyperuniform layered media \cite{Ki23,kim_accurate_2024}.
% {\color{red}We mainly rely on the second-order formula of Eq. \eqref{eq:scaled_eps_layered_trans}, where $A_3^\perp = 0$, because it is already very accurate for hyperuniform layered media \cite{Ki23,kim_accurate_2024}.
% {\color{red}We then use full expression of Eq. \eqref{eq:scaled_eps_layered_trans} to further investigate the perfect transparency of SHU media in section \ref{sec:trans}.

Note that $\fn{\varepsilon_e^\perp}{k_1}$ given in \eqref{eq:scaled_eps_layered_trans} is complex-valued, implying that the media can be lossy due to forward scattering and backscattering from fluctuations in the local dielectric constant.
%By contrast, $\fn{\varepsilon_e^z}{k_1}$ is independent of $k_1$, since a traveling longitudinal wave cannot exist under the aforementioned conditions.
%Hence, we focus on $\fn{\varepsilon_e^\perp}{k_1}$ in the rest of this work.   
% In the static limit, Eqs. \eqref{eq:scaled_eps_layered_trans} and \eqref{eq:scaled_eps_layered_long} reduce to  the arithmetic and harmonic means of the local dielectric constants, respectively:
% $
%     \fn{\varepsilon_e^\perp}{0} = \E{\varepsilon}\equiv \phi_1 \varepsilon_1 + \phi_2 \varepsilon_2
% $
% and
% $
%     \fn{\varepsilon_e^z}{0} = (\phi_2/\varepsilon_2 + \phi_1 /\varepsilon_1)^{-1},
% $
% which are exact for any microstructure  \cite{To02a}.
The static limit of Eq. \eqref{eq:scaled_eps_layered_trans} is the arithmetic mean of the local dielectric constants, i.e.,
$
    \fn{\varepsilon_e^\perp}{0} = \E{\varepsilon}\equiv \phi_1 \varepsilon_1 + \phi_2 \varepsilon_2
$,
which is exact for any 1D microstructure  \cite{To02a}.
%The corresponding formulas for longitudinal polarization are given in \ref{sec:formula-layered}.

In the long-wavelength regime ($k_1 /s \ll 1$), assuming a power-law scaling of $\spD{k}$ for small $k$ as specified by Eq. \eqref{eqn:smallk}, the imaginary part of Eq. \eqref{eq:scaled_eps_layered_trans} has the following asymptotic behaviors:
\begin{eqnarray}
\fl    \Im[\fn{\varepsilon_e^\perp}{k_1}] 
    \sim (\varepsilon_2-\varepsilon_1)^2\Im[\fn{A_2^\perp}{k_*; \E{\varepsilon}}]
    \sim
    \cases{
        k_1 ,   \quad \alpha = 0 
        ~~\mbox{(typical nonhyperuniform)} \\
        {k_1}^{1+\alpha}, ~ \alpha > 0
        ~~\mbox{(hyperuniform)}
    }.
    \label{eq:asymptotic-layered}
\end{eqnarray}
It is seen that hyperuniform media attenuate less than nonhyperuniform ones in the long-wavelength regime.
Furthermore, %while all typical nonhyperuniform systems exhibit a similar attenuation behavior [i.e., $\Im[\varepsilon_e]\propto {k_1}$ for small $k_1$], 
hyperuniform systems can exhibit a wide range of behaviors by tuning the exponent $\alpha$.
The attenuation behavior of the SHU medium is elaborated in section \ref{sec:trans}.
    
\subsection{Multiple-Scattering Approximations for Transversely Isotropic Media}    \label{sec:iso}

    % \outl{
    %     Explain the polarizations and incident direction in words.
    % }

    We obtain our multiple-scattering approximations for the effective dynamic dielectric constant tensor $\fn{\tens{\varepsilon}_e}{\vect{k}_1}$ of transversely isotropic media [see Fig. \ref{fig:schm-models}(b)] for the situation in which waves are normally incident to the symmetry axis of the media, i.e., $\vect{k}_1 = k_1 \uvect{y}$, where $\uvect{y}$ is a unit vector along the $y$-direction.
    (The reader is referred to Supplement1 of Ref. \cite{kim_accurate_2024} for the formulas obliquely incident waves.)
    These formulas depend on the wavenumber $k_1$ in the reference phase $q~(=1)$.
 %   We extract them from exact strong-contrast series \eqref{eq:str-exp-trunc} by choosing a needle-like exclusion volume aligned along $\uvect{y}$ that involves the singularity of the dyadic Green's function.
 %  
 Due to the symmetries of the problems, one can decompose $\fn{\tens{\varepsilon}_e}{\vect{k}_1}$ into two orthogonal components $\fn{\varepsilon_e ^{TM}}{k_1}$ and $\fn{\varepsilon_e ^{TE}}{k_1}$ for transverse magnetic (TM) and TE polarizations \footnote{The electric field for TM (TE) polarization is parallel (normal) to the symmetry axis.}, respectively, as follows:
    $\fn{\tens{\varepsilon}_e}{k_1} = \fn{\varepsilon_e ^{TM}}{k_1} \uvect{z}\uvect{z} + \fn{\varepsilon_e^{TE}}{k_1}(\tens{I}-\uvect{z}\uvect{z})$.
    %Thus, we extract two independent approximations of $\ETE{k_1}$ and $\ETM{k_1}$ by truncating the strong-contrast series at third-order terms.
%    see \eqref{eqS:appr-trans-TE} and \eqref{eqS:appr-trans-TM} of \supp.
    %Solving for the effective dielectric constants from these linear fractional forms and then renormalizing them with the optimal reference phase \cite{tsang_scattering_1981,mackay_strongpropertyfluctuation_2000, To21a}, we finally obtain the \emph{scaled} strong-contrast approximations at the three-point level for disordered transversely isotropic media.

    In this paper, we also consider $\ETM{k_1}$. (Formulas for TE polarization are given in \ref{sec:formula-trans-iso}.)
    We begin by presenting the three-point formula, which is used to prove that SHU transversely isotropic media are also perfectly transparent at the three-point level; see section \ref{sec:trans}.
    %Then, the second-order formula is used to estimate the effective attenuation behaviors for the two different models.
%   We mainly use the second-order formula to estimate the attenuation behaviors because of its simplicity and great accuracy \cite{kim_accurate_2024}.
    The \emph{scaled} strong-contrast approximation for $\ETM{k_1}$ at the three-point level is given as
    \begin{eqnarray}
        \fl \frac{\fn{\varepsilon_e^{TM}}{k_1}}{\varepsilon_1} 
        =& 
        1+ \frac{ 
            {\phi_2}^2 \left[(\varepsilon_2+\varepsilon_1) \BETA{2}{21} \right]
        } 
        {\phi_2 - \ATM{2}{k_*^{TM}, \E{\varepsilon}} \left[(\varepsilon_2+\varepsilon_1) \BETA{2}{21} \right]
        - \ATM{3}{k_*^{TM}, \E{\varepsilon}} \left[(\varepsilon_2+\varepsilon_1) \BETA{2}{21} \right]^2 
        },    \label{eq:scaled_eps_2D_TM}
        % \\
        % \fl \frac{\ETE{k_1}}{\varepsilon_1}
        % =& 
        % 1+ \frac{ 
        %     2  {\phi_2}^2 \BETA{2}{21}
        % }
        % { 
        %     \phi_2(1-\phi_2 \BETA{2}{21}) 
        %     -  
        %     \ATE{2}{k_*^{TE}; \varepsilon_{BG}^{(2D)}} \left[2\varepsilon_1 \BETA{2}{21} \right]
        %     -  
        %     \ATE{3}{k_*^{TE}; \varepsilon_{BG}^{(2D)}} \left[2\varepsilon_1 \BETA{2}{21} \right]^2
        % }
        %     \label{eq:scaled_eps_2D_TE} , 
    \end{eqnarray}
    where $\BETA{2}{pq}\equiv (\varepsilon_p - \varepsilon_q)/(\varepsilon_p + \varepsilon_q)$ is the 2D counterpart of the dielectric polarizability, %and $k_*^{TE} \equiv k_1 \sqrt{\varepsilon_{BG}^{(2D)}/\varepsilon_1}$ and $k_*^{TM} \equiv k_1 \sqrt{\E{\varepsilon}/\varepsilon_1}$ are wavenumbers in the optimal reference phase for TE and TM polarizations, respectively, and $\varepsilon_{BG}^{(2D)}$ is the Bruggeman approximation for 2D two-phase media \cite{bruggeman_berechnung_1935, To02a}.
    and $k_*^{TM} \equiv k_1 \sqrt{\E{\varepsilon}/\varepsilon_1}$ is wavenumber in the optimal reference phase for TM polarization.
%     We note that Eq. \eqref{eq:scaled_eps_2D_TE} is identical to the 2D formula derived in Ref. \cite{To21a}, except for the reference phase.
%     The current choice for the reference dielectric constant $\varepsilon_{BG}^{(2D)}$ offers slightly better renormalization than 2D Maxwell-Garnett approximation employed in Ref. \cite{To21a} because $\varepsilon_{BG}^{(2D)}$ makes the mean of the depolarization tensors exactly zero, i.e., $\phi_1 \tens{L}_1 ^{(*)} + \phi_2 \tens{L}_2 ^{(*)}=\tens{0}$. % (see Eq. (S105) of \supp).
% %    (see \eqref{eqS:optimal-reference} of \supp).
%     We henceforth focus on the approximation for the TM polarization given in Eq. \eqref{eq:scaled_eps_2D_TM}.
    The second- and third-order coefficients are defined as
    \begin{eqnarray}
        \fl &\ATM{2}{k; \varepsilon}
        =
        \frac{1}{\varepsilon}
        \left\{
            \frac{{k}^2}{\pi^2}
            \int_{0}^{\pi/2} \dd{\phi} 
            \left[ \mathrm{p.v.} 
                \int_0^\infty \dd{q} \frac{2q\spD{q}}{q^2-(2k\cos\phi)^2}
            +i\pi \spD{2k \cos\phi}
            \right]
        \right\},
        \label{eq:F-trans-Fourier}
        \\
        \fl &\ATM{3}{k; \varepsilon}
        =
        \frac{-1}{\phi_2}\left[\frac{{k}^2}{\varepsilon (2\pi)^2 } \right]^2
        \int_{\R^2} \dd{\vect{q}_1}
        \int_{\R^2} \dd{\vect{q}_2}
            \frac{\fn{\tilde{\Delta}_3^{(2)}}{\vect{q}_1,\vect{q}_2}}{[\abs{\vect{q}_1 + k\uvect{y}}^2 - {k}^2] [\abs{\vect{q}_2 + k\uvect{y}}^2 - {k}^2]} ,
        \label{eq:A3-TM}
        % \\
        % \fl &\ATE{3}{k; \varepsilon}
        % = 
        % \frac{-1}{\phi_2}
        % \left[\frac{1}{2\varepsilon (2\pi)^2} \right]^2
        % \int_{\R^2} \dd{\vect{q}_1}
        % \int_{\R^2} \dd{\vect{q}_2} 
        % \fn{\tilde{\Delta}_3^{(2)}}{\vect{q}_1,\vect{q}_2}
        % \nonumber \\
        % \fl&\qquad  \times \frac{\Big[
        %     ({k}^2)^2 - \abs{k\uvect{y}+\vect{q}_1}^2 \abs{k\uvect{y}+\vect{q}_2}^2
        %     + 2 [(k\uvect{y}+\vect{q}_1)\cdot (k\uvect{y}+\vect{q}_2)]^2
        % \Big]}
        % {\left(\abs{k\uvect{y}+\vect{q}_1}^2 - {k}^2 \right) \left(\abs{k\uvect{y} + \vect{q}_2}^2 - {k}^2 \right)}
        % ,
        % \label{eq:A3-TE}
    \end{eqnarray}
    where %$\F{2}{k}$ is the nonlocal attenuation function for 2D statistically isotropic two-phase media \cite{To21a}, and 
    $\fn{\tilde{\Delta}_3^{(2)}}{\vect{q}_1,\vect{q}_2}$ is the Fourier transform of $\fn{\Delta_3^{(2)}}{\vect{x}_{21},\vect{x}_{31}}\equiv \fn{S_2^{(2)}}{\vect{x}_{21}} \fn{S_2^{(2)}}{\vect{x}_{32}} - \phi_p \fn{S_3^{(2)}}{\vect{x}_{21}, \vect{x}_{32}+\vect{x}_{21}}$.
    % In the static limit ($k_1 \to 0^+$), Eqs. \eqref{eq:scaled_eps_2D_TE} and \eqref{eq:scaled_eps_2D_TM} reduce to the arithmetic means of the local dielectric constant and the third-order static strong-contrast approximation for $d=2$ \cite{To02a}, respectively:
    The static limit of Eq. \eqref{eq:scaled_eps_2D_TM} is the arithmetic mean of the local dielectric constant:
    \begin{equation}
        \ETM{0}  = \E{\varepsilon} =\varepsilon_1\phi_1 + \varepsilon_2 \phi_2.
        % \quad 
        % \ETE{0} 
        % =
        % \varepsilon_1
        % \frac{
        %     \phi_2 (1+\phi_2\BETA{2}{21}) - (1-\phi_2) \zeta_2 [\BETA{2}{21}]^2}
        % {
        %     \phi_2 (1-\phi_2\BETA{2}{21}) - (1-\phi_2) \zeta_2 [\BETA{2}{21}]^2
        % },    
        \label{eq:eps_2D_static} 
    \end{equation}
    
%    where $\zeta_2$ is the three-point parameter that lies in the closed interval $[0,1]$ \cite{To02a}.
    
  %  \outline{
  %      \item Eq. \eqref{eq:eps_2D_static} and the sentence around it can be removed. }

    In the long-wavelength regime ($k_1 /s \ll 1$), assuming a power-law scaling of $\spD{k}$ for small $k$ as specified by Eq. \eqref{eqn:smallk}, the imaginary part of Eq. \eqref{eq:scaled_eps_2D_TM} has the following asymptotic behaviors:
    \begin{equation}
        \fl \Im[\ETM{k_1}] 
        \sim (\varepsilon_2-\varepsilon_1)^2\Im[\ATM{2}{k_*^{TM}; \E{\varepsilon}}]
        \sim
        \cases{
            {k_1}^2 , \quad \alpha = 0 
            ~\mbox{(typical nonhyperuniform)} \\
            {k_1}^{2+\alpha},  \alpha > 0
             ~\mbox{(hyperuniform)} 
        },
        \label{eq:asymptotic-trans-TM}
    \end{equation}
    which are identical to those for TE polarization \cite{To21a}.
    Thus, in the long-wavelength regime, hyperuniform media attenuate less than nonhyperuniform ones and can exhibit a wide range of behaviors by tuning the exponent $\alpha$. 
    The attenuation behavior of the SHU medium is elaborated in section \ref{sec:trans}.

\section{Perfect Transparency Intervals of Stealthy Hyperuniform Media}
\label{sec:trans}
    
We have previously shown that second-order strong-contrast formulas for SHU two-phase composites, i.e., those in which the spectral density obeys the condition (\ref{stealthy}), predict perfect transparency or, equivalently, a zero imaginary part of the
effective dielectric constant, in a finite range of wavenumbers that is determined by the exclusion-region radius $K$.
This was done specifically for TM \cite{kim_accurate_2024} and TE \cite{To21a} polarization in transversely isotropic media and transverse polarization in fully isotropic media \cite{To21a} and layered media \cite{Ki23}.
Across space dimensions, the predictions of such perfect transparency intervals have a similar form \footnote{For layered media, the perfect transparency
interval \eqref{eq:trans-interval} at the two-point level was derived from the condition that the second-order formula \eqref{eq:scaled_eps_layered_trans}
has $\Im[\fn{A_2^\perp}{k_*;\E{\varepsilon}}]=0$. The corresponding intervals for TM and TE polarizations in transversely isotropic media were derived by finding the intervals in which 
$\ATM{2}{k_*^{TM};\E{\varepsilon}}$ [Eq. \eqref{eq:scaled_eps_2D_TM}] and 
$\ATE{2}{k_*^{TE};\varepsilon_{BG}^{(2D)}}$ [Eq. \eqref{eq:scaled_eps_2D_TE}] have
zero imaginary parts, respectively.}:
    \begin{equation}
    0 \leq k_1 < K_T \equiv \frac{K}{2 \sqrt{\varepsilon_* /\varepsilon_1}} , \label{eq:trans-interval}
\end{equation}
where the dielectric constant of the optimal reference phase $\varepsilon_*$ is given as
\begin{numcases}{\varepsilon_*=}
    \E{\varepsilon},    \qquad ~~ \mbox{Transverse polarization in layered media}, \label{eq:eps_layered}\\
    \E{\varepsilon}, \qquad ~~\mbox{TM polarization in transversely isotropic media }, \label{eq:eps_2D_TM}\\
    \varepsilon_{BG}^{(2D)}, \qquad \mbox{TE polarization in transversely isotropic media}, \label{eq:eps_2D_TE} \\
    \varepsilon_{BG}^{(3D)}, \qquad \mbox{Transverse polarization in fully isotropic media}  ,\label{eq:eps_3D}
    \end{numcases}
and $\varepsilon_{BG}^{(dD)}$ is the Bruggeman approximation \cite{bruggeman_berechnung_1935, To02a} of the effective static dielectric constant of $d$-dimensional two-phase composites \footnote{For a contrast ratio $\varepsilon_2/\varepsilon_1 \leq 4$, the Bruggeman formulas are very close to the Maxwell-Garnett formulas employed in Ref. \cite{To21a}.
The former offers a slightly better renormalization, since it uses the fact that the mean of the depolarization tensors is exactly zero; see Ref. \cite{kim_accurate_2024}. 
%See the discussion in section \ref{sec:theory}.
}.
Within the finite perfect transparency interval \eqref{eq:trans-interval}, the rapid convergence of strong-contrast expansions and high
predictive power of their second-order truncations, as validated in Refs. \cite{To21a, Ki23, kim_accurate_2024}, imply that third- and higher-order contributions to the imaginary part $\Im[\fn{\varepsilon_e}{k_1}]$ are negligibly small for relatively large contrast ratios ($\varepsilon_2/\varepsilon_1\lesssim 10$).

In the present work, we now also prove that the interval (\ref{eq:trans-interval}) for layered media and transversely isotropic media (with TM polarization) applies not only at the two-point but the three-point level as well.
We begin by proving that the perfect transparency interval specified by Eq. \eqref{eq:trans-interval} in conjunction with Eq. \eqref{eq:eps_layered} for
layered media is exact through the third-order terms.
Accounting for the perfect transparency at the two-point level, from the third-order formula \eqref{eq:scaled_eps_layered_trans}, we obtain
$\Im[\fn{\varepsilon_e^\perp}{k_1}]\propto (\varepsilon_2-\varepsilon_1)^3 \Im[\fn{A_3^\perp}{k_*, \E{\varepsilon}}]$ for $0\leq k_* < K/2 $.
Thus, it is sufficient to show that
        \begin{equation}   \label{eq:im-layered-3pt}
            \Im[\fn{A_3^\perp}{k_*, \E{\varepsilon}}] = 0,  \quad\mbox{for }k_*<K/2.
        \end{equation}
The reader is referred to \ref{sec:PT-layered-3pt} for a  detailed proof of Eq. \eqref{eq:im-layered-3pt}.

We now show that the perfect transparency interval for TM polarization in transversely isotropic media is specified by Eq. \eqref{eq:trans-interval}
together with Eq. \eqref{eq:eps_2D_TM} through the third-order terms.
Due to the perfect transparency at the two-point level, the third-order formula of Eq. \eqref{eq:scaled_eps_2D_TM} yields $\Im[\ETM{k_1}] \propto
(\varepsilon_2-\varepsilon_1)^3 \Im[\ATM{3}{k_*^{TM};\E{\varepsilon}}]$ for $0\leq k_*^{TM}<K/2$.
    Therefore, it is sufficient to show that
    \begin{eqnarray}   \label{eq:im-TM-3pt}
        \Im[\ATM{3}{k_*^{TM},\E{\varepsilon}}] = 0,  \quad\mbox{for }k_*^{TM}<K/2.
    \end{eqnarray}
The detailed proof of \eqref{eq:im-TM-3pt} is presented in \ref{sec:PT-trans-3pt}.

The fact that the above results show perfect transparency through third-order terms in the strong-contrast expansion implies that the localization length for such stealthy hyperuniform media should be very large compared to any practically large sample size, and thus, there can be no Anderson localization within the predicted perfect transparency intervals in stealthy hyperuniform layered and transversely isotropic media in practice, as noted in Refs. \cite{kim_accurate_2024}.
This prediction for layered media is especially remarkable because the traditional understanding is that localization must occur for any type of disorder in 1D systems \cite{sheng_introduction_2006, aegerter_coherent_2009, izrailev_anomalous_2012, wiersma_disordered_2013, yu_engineered_2021}.

\begin{figure}
\centering
\includegraphics[width=0.9\textwidth]{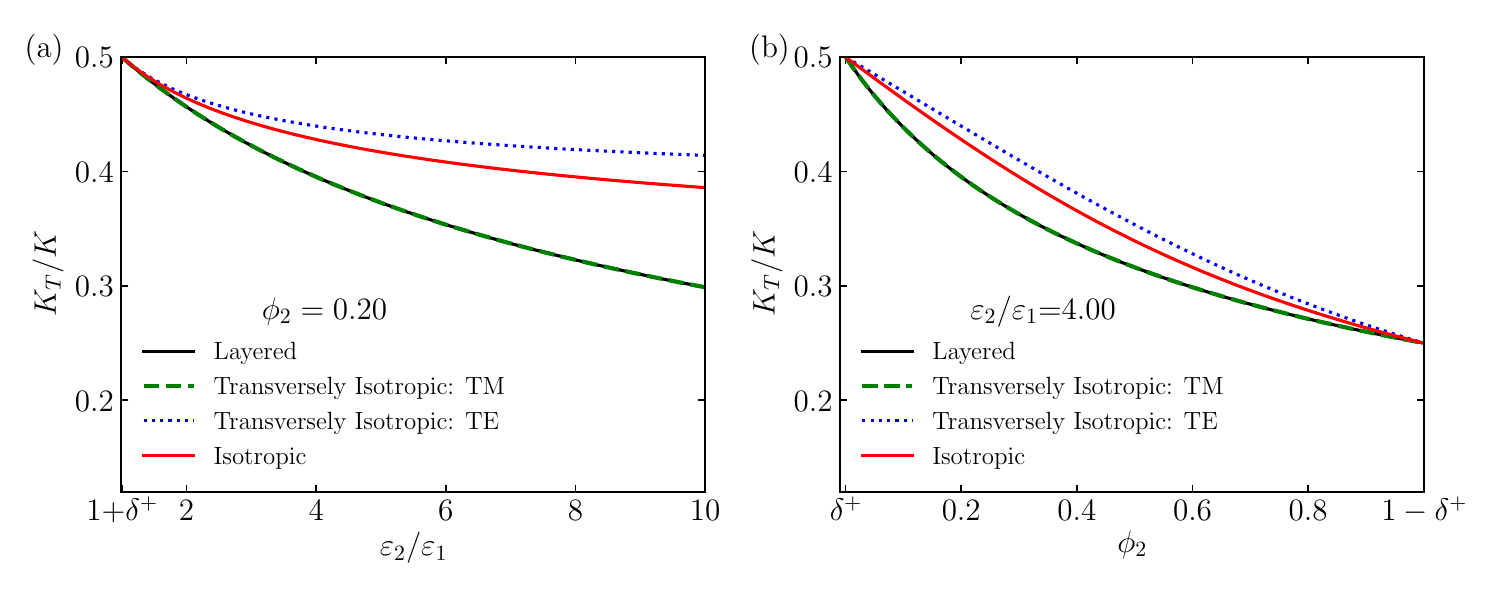}
\caption{
Prediction \eqref{eq:trans-interval} of the dimensionless sizes $K_T/K$ of the perfect transparency intervals for stealthy hyperuniform media in the four cases: transverse polarization in layered media, TM and TE polarizations in transversely isotropic media, and transverse polarization in fully isotropic media.
(a) The sizes as functions of the contrast ratios $\varepsilon_2/\varepsilon_1$ with the volume fraction $\phi_2=0.20$ of the higher dielectric constant.
Here, $\delta^+$ denotes an infinitesimally small positive number.
(b) The sizes as functions of $\phi_2$ with $\varepsilon_2/\varepsilon_1=4.0$.
\label{fig:interval}
}
\end{figure}    
    
It is noteworthy that the dimensionless size of the perfect transparency interval, $K_T/K$, defined by Eq. \eqref{eq:trans-interval}, tends to decrease as the wave propagation is disturbed
further for the four cases considered in Eqs. \eqref{eq:eps_layered}-\eqref{eq:eps_3D}; see Fig. \ref{fig:interval}.
Specifically, Fig. \ref{fig:interval}(a) shows that the dimensionless interval size $K_T/K$ decreases with the contrast ratio $\varepsilon_2/\varepsilon_1$, since a higher contrast
ratio strengthens scattering.
Similarly, in Fig. \ref{fig:interval}(b), we see that $K_T/K$ decreases with the volume fraction $\phi_2$ at a fixed contrast ratio $\varepsilon_2/\varepsilon_1=4$, 
since the frequency of scattering events is proportional to $\phi_2$.
Figure \ref{fig:interval} also reveals the relative ranking of the four cases for $K_T/K$: $K_T/K$ is the lowest for TM polarization in transversely
isotropic media (and layered media), followed by fully isotropic ones, and is the highest for TE polarization in transversely isotropic media.
This ranking is again due to the strength of reflectance (i.e.,
scattering) on an interface between two dielectric materials.
Specifically, the Fresnel equations \cite{jackson_classical_1999} dictate that among $s$/$p$-polarized lights, in which the electric field is normal/parallel to the
incident plane, the $s$-polarized one is always reflected more than the $p$-polarized one.
Thus, for transversely isotropic media, the dimensionless size of $K_T$ for TM is lower than TE because the TM/TE wave is always $s$/$p$-polarized.
For layered media, the ratio $K_T/K$ is identical to $K_T/K$ for the TM polarization because the transverse waves are always $s$-polarized.
Similarly, the dimensionless interval size $K_T/K$ for isotropic media is in between those for TE and TM polarizations, since the reflected waves in a 3D isotropic media have both $s$ and $p$ polarizations due to interfaces that are isotropically oriented with respect
to the incident radiation.

\section{Results for Attenuation and Spreadability Characteristics and Their Correlations}
\label{sec:results}

\subsection{Attenuation Characteristics}
\label{sec:attenuation}

\begin{figure}[h]
    \includegraphics[width=0.5\textwidth]{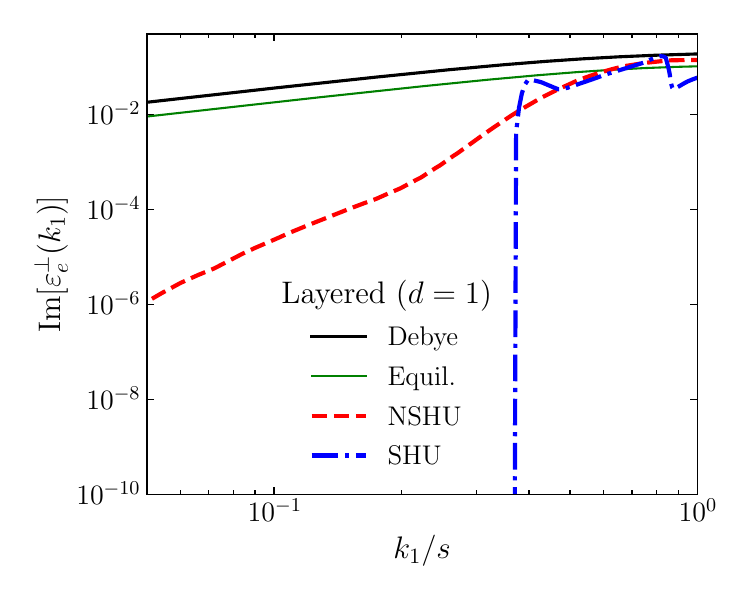}
    \caption{
        Predictions of the scaled strong-contrast approximation [Eq. \eqref{eq:scaled_eps_layered_trans}]  of the imaginary part of the effective dynamic dielectric constant $\Im[\fn{\varepsilon_e^\perp}{k_1}]$ as a function of the dimensionless incident wavenumber $k_1/s$ for disordered layered media with $\phi_2=0.20$ and $\varepsilon_2/\varepsilon_1=4.0$, where  $s$ is the specific surface.  We consider the four models: Debye, Equil., NSHU, and SHU media. The plot is shown on a log-log scale.
        \label{fig:1D-imag}
    }
\end{figure}

Here, we compare the extraordinary effective attenuation characteristics of SHU media to those of their nonstealthy counterparts. 
For this purpose, we compute the imaginary part of the effective dielectric constant $\Im[\fn{\varepsilon_e}{k_1}]$ (which is related to the imaginary part of the effective wavenumber $k_e\equiv k_1 \sqrt{\varepsilon_e/\varepsilon_1}$), as predicted from the scaled second-order strong-contrast approximations. %\eqref{eq:scaled_eps_layered_trans}.  
We begin by showing in Fig. \ref{fig:1D-imag}, the imaginary part $\Im[\fn{\varepsilon_e^\perp}{k_1}]$ for the four models of layered media ($d=1$)  (see Sec. \ref{sec:models}) with $\phi_2=0.20$ and $\varepsilon_2/\varepsilon_1=4.0$.
As noted in the previous section,  the stealthy hyperuniform medium exhibits perfect transparency (i.e., $\Im[\fn{\varepsilon_e^\perp}{k_1}]=0$) in the finite interval specified by Eq. \eqref{eq:trans-interval}, whereas the three nonstealthy media, whether hyperuniform or not, always exhibit nonzero attenuation with power-law scalings [i.e., $\Im[\fn{\varepsilon_e^\perp}{k_1}]\sim {k_1}^{1+ \alpha}$] for small wavenumbers  ($k_1/s < 0.3$).
Furthermore, for $k_1/s < 0.3$, the relative ranking of their degree of attenuation is identical to their corresponding rankings of 
the large-scale volume-fraction fluctuations, which are proportional to the scattering intensity [i.e., $\spD{k}$ for small $k$], because the attenuation comes solely from scattering due to dielectric inhomogeneity (not absorption).
Thus, the Debye medium exhibits the strongest attenuation, followed by the equilibrium and then the NSHU media, down to the SHU one, in which the attenuation is zero.

\begin{figure}
    \subfloat[]{
    \includegraphics[width=0.49\textwidth]{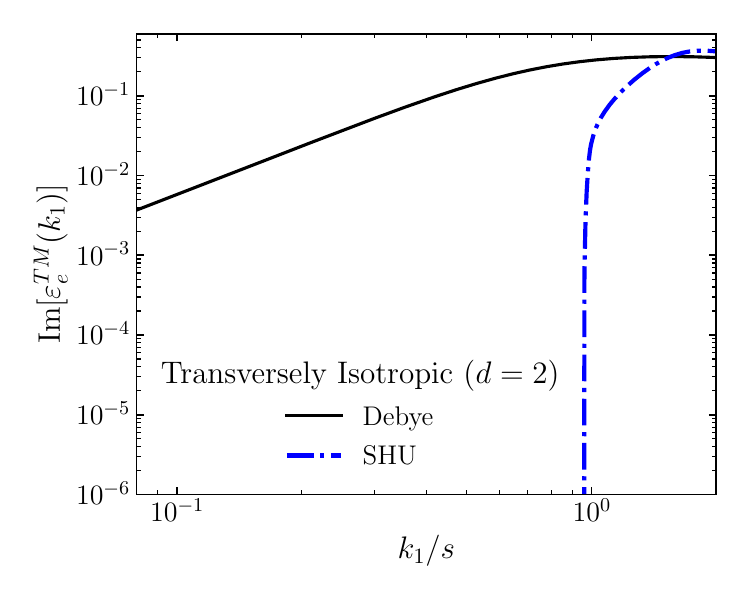}}
    \subfloat[]{
    \includegraphics[width=0.49\textwidth]{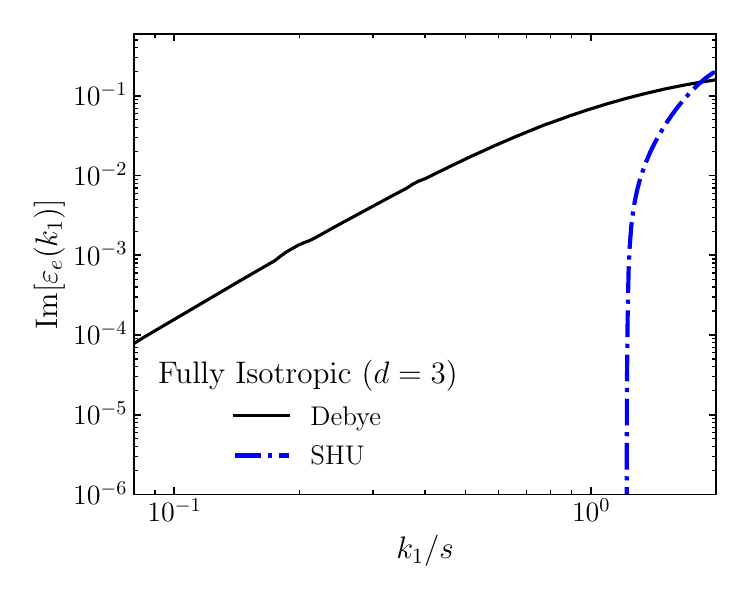}
    }
    \caption{
        Predictions of the scaled strong-contrast approximation of the imaginary part of the effective dynamic dielectric constant as a function of the dimensionless incident wavenumber $k_1/s$ for disordered two-phase media in three dimensions with $\phi_2=0.20$ and $\varepsilon_2/\varepsilon_1=4.0$ with two distinct symmetries: (a) TM polarization in transversely isotropic media ($d=2$) and (b) transverse polarization in fully isotropic media ($d=3$), where $s$ is the specific surface.
%        In each case, we consider two models: Debye medium and stealthy hyperuniform (SHU) one.
        For each symmetry, we consider Debye and SHU media.
        The plots shown in  (a) are obtained from Eq. \eqref{eq:scaled_eps_2D_TM}, whereas the plots shown in (b) are obtained from the scaled approximation for $d=3$ in Ref. \cite{To21a}.
        Both plots are shown on log-log scales.
        \label{fig:high-D-imag}
    }
\end{figure}

For transversely isotropic and fully 3D isotropic media, we study the imaginary parts of the effective dielectric constant for two of the aforementioned models that 
exhibit the greatest and least attenuation, namely,
Debye  and SHU media (see Sec. \ref{sec:models}), respectively, with $\phi_2=0.20$ and $\varepsilon_2/\varepsilon_1=4$.
We do not report corresponding results for the equilibrium and
the NSHN media for $d=2$ and $d=3$, since qualitatively they are similar to the layered media
models. 
The predictions shown in Fig. \ref{fig:high-D-imag}(a) are obtained from the second-order formula of Eq. \eqref{eq:scaled_eps_2D_TM} for TM polarization in transversely isotropic media ($d=2$), whereas those in Fig. \ref{fig:high-D-imag}(b) are obtained from the scaled approximation derived in Ref. \cite{To21a}; see Eq. (73) for $d=3$ therein.
For these two symmetries, the perfect transparency interval \eqref{eq:trans-interval} of the SHU medium is singularly different
from the nonhyperuniform Debye medium with a substantial degree of attenuation, i.e.,  with a power-law scaling $\Im[\fn{\varepsilon_e}{k_1}\sim {k_1}^{d}]$ for small wavenumbers ($k_1/s <0.3$). As for layered-media cases for $k_1/s <0.3$, the relative ranking of the
attenuation behaviors of these two models is consistent with the corresponding rankings of their large-scale volume-fraction fluctuations.

%Specifically, while Debye medium for $d=2,3$ exhibits a notable degree of attenuation ($\Im[\varepsilon_e(k_1)]\sim {k_1}^d$), stealthy hyperuniform one is perfectly transparent within a finite interval.
%Similar to the case of layered media, for small wavenumbers ($k_1/s <0.3$), the relative ranking of these two models for $\Im[\varepsilon_e(k_1)]$ is consistent with the ranking for the large-scale density fluctuations.

\subsection{Spreadability Characteristics}
\label{sec:spreadability}

\begin{figure}
    \includegraphics[width=0.5\textwidth]{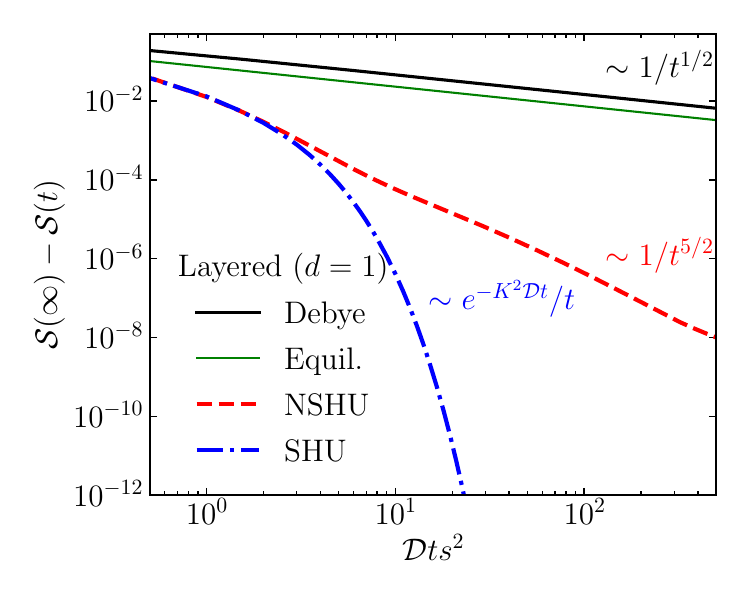}
    \caption{
        Comparison of the excess spreadabilities for disordered layered media ($d=1$) with $\phi_2=0.20$ as functions of the dimensionless time ${\cal D}ts^2$, where $s$ is the specific surface. We consider four models: Debye, Equil., NSHU, and SHU media.
        %random media, equilibrium hard rods (EHR), nonstealthy hyperuniform polydisperse packings (HUP), and stealthy hyperuniform packings (SHU) with $\chi=0.3$.
        The long-time behaviors of $\fn{\mathcal{S}}{\infty}-\fn{\mathcal{S}}{t}$ for each of these models is indicated.
        \label{fig:1D-S}
    }
\end{figure}

\begin{table}
    \caption{
        Values of the coefficients $B$ and $C$ for all models considered here in three distinct structural symmetries with $\phi_2 = 0.20$.
        We consider four models for layered media: Debye, Equil., NSHU, and SHU media.
        For the other two symmetries, we focus on Debye and SHU media.
        Here, the exponent $\alpha$ and $B$ are specified by Eq. \eqref{eqn:smallk} in units of the specific surface $s$, and $C$ is specified by Eq. \eqref{eq:C}.
        For SHU, we denote $\alpha=+\infty$ because its spectral density is smaller than any power-law scaling.
        \label{tab:coefficients}
    }
    \begin{tabular}{c|c|c|c}
        \hline
        Symmetry& Model& $B$ & $C$ \\
        \hline
        \multirow{4}{*}{Layered ($d=1$)}
        & Debye ($\alpha=0$) &1.01$\times10^{-1}$   &1.44$\times10^{-1}$   \\  
        & Equil. ($\alpha=0$) &5.12$\times10^{-2}$ &7.22$\times10^{-2}$   \\
        & NSHU ($\alpha=4$) & 5.13$\times10^{-2}$ & 4.37$\times10^{-2}$   \\    
        & SHU ($\alpha=+\infty$) & 0.0 & 0.0  \\
        \hline      
        \multirow{2}{*}{Transversely Isotropic ($d=2$)}
        & Debye ($\alpha=0$) &4.94$\times10^{-1}$   &2.01$\times10^{-1}$\\  
        & SHU ($\alpha=+\infty$) & 0.0 & 0.0  \\
        \hline      
        \multirow{2}{*}{Fully Isotropic ($d=3$)}
        & Debye ($\alpha=0$) &2.50  &2.86$\times10^{-1}$   \\  
        & SHU ($\alpha=+\infty$) & 0.0 & 0.0  \\
        \hline      
    \end{tabular}
\end{table}

We begin by considering the four models of layered media ($d=1$) with $\phi_2=0.2$ to demonstrate the extraordinary property of diffusion spreadability for the SHU medium; see Fig. \ref{fig:1D-S}.
At the intermediate times ($0.01 < {\cal D}ts^2 < 1$), we find that the excess spreadability for the Debye medium shows the slowest decay, followed by the equilibrium one, and those for two hyperuniform media commonly shows the fastest decay, as predicted in Ref. \cite{To21d}.
Consistent with previous theoretical
results \cite{To21d}, we also find that the excess spreadability for nonhyperuniform Debye medium has the slowest long-time decay to its infinite-time behavior
among the four models, followed by the equilibrium media with the same decay rate, i.e., $1/t^{1/2}$. The NSHU medium has the second fastest
decay rate ($1/t^{5/2}$). Of course, the SHU medium is distinguished among all of the nonstealthy models, since
its excess spreadability decays exponentially fast, as predicted by
the asymptotic formula \eqref{S-stealthy}.
More specifically, referring to the asymptotic formula \eqref{long-time}, the two nonhyperuniform layered media exhibit a common power-law decay with a coefficient $C$, specified by  Eq. \eqref{eq:C}.
Table \ref{tab:coefficients} summarizes the values of the coefficients $B$, defined in Eq. \eqref{eqn:smallk}, and $C$ for all models of layered media
considered here, along with their values of the exponent $\alpha$.
% considered and confirms the predicted proportionality between $C$ and $B$ for $d=1$.
We take SHU medium to be one with $\alpha=+\infty$, roughly speaking, because its decay rate of
$\mathcal{S}(t)$ is faster than any inverse power law.
The long-time decay rate  of the excess spreadability  $\fn{\mathcal{S}}{\infty}-\fn{\mathcal{S}}{t}$ is faster for media with smaller large-scale
    volume-fraction fluctuations (i.e., small values of $\spD{k}$ for small $k$ around the origin) \cite{To22a}, i.e.,
 it takes less time to reach the state of uniform concentration of the solute in
    the infinite-time limit in media when such fluctuations are smaller; see Fig. \ref{cartoon}.

\begin{figure}
    \subfloat[]{ 
    \includegraphics[width=0.49\textwidth]{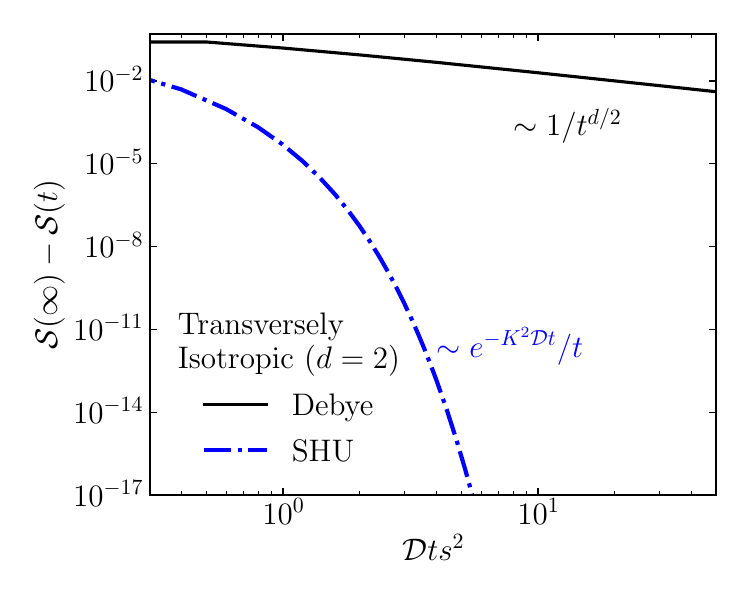}
    }
    \subfloat[]{ 
    \includegraphics[width=0.49\textwidth]{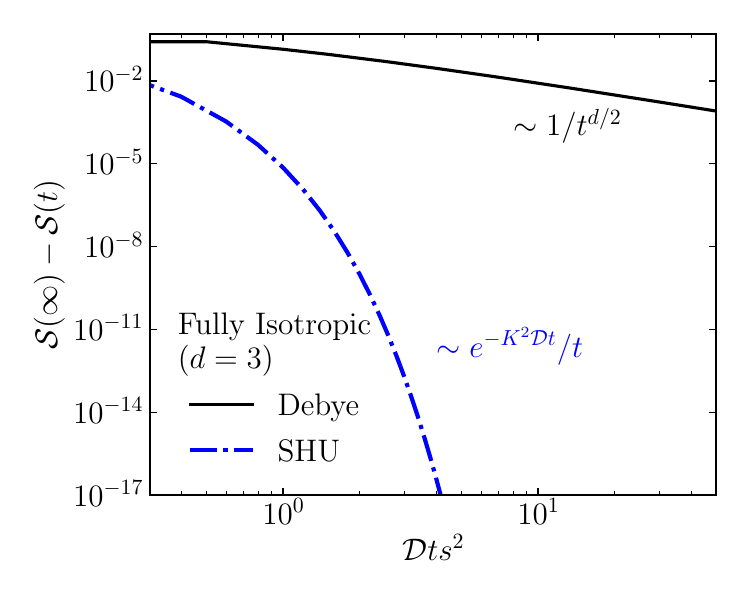}
    }
    \caption{
        Comparison of the excess spreadabilities as functions of the dimensionless time ${\cal D}ts^2$ for (a) disordered transversely isotropic media ($d=2$) and (b) disordered fully isotropic media ($d=3$) with $\phi_2=0.20$, where  $s$ is the specific surface.
        We consider two models for each symmetry: Debye and SHU media.
        The long-time behavior of $\fn{\mathcal{S}}{\infty}-\fn{\mathcal{S}}{t}$ for each of these models is indicated.
        \label{fig:high-D-S}
    }
\end{figure}

For both transversely isotropic ($d=2$) and fully isotropic ($d=3$) media, we study the two models with $\phi_2=0.20$
whose $\fn{\mathcal{S}}{\infty}-\fn{\mathcal{S}}{t}$ at long times decay slowest and fastest, i.e., Debye media and SHU ones, respectively; see Fig. \ref{fig:high-D-S}.
Even at the intermediate times, $\fn{\mathcal{S}}{\infty}-\fn{\mathcal{S}}{t}$ for SHU media decays much faster than those for the Debye ones.
While the long-time $\fn{\mathcal{S}}{\infty}-\fn{\mathcal{S}}{t}$ for Debye media decays like
$1/t^{d/2}$, as predicted by Eq. \eqref{long-time}, those for the SHU media 
decay exponentially fast, as
predicted by Eq. \eqref{S-stealthy}.
The corresponding coefficients $B$ and $C$ are also listed in Table \ref{tab:coefficients}.
For the same reasons given for layered media, the relative rankings of these two models for the asymptotic decay rate of
    the long-time behavior of $\fn{\mathcal{S}}{\infty}-\fn{\mathcal{S}}{t}$ is also identical to that of their 
large-scale volume-fraction fluctuations \cite{To22a}.

\subsection{Cross-Property Relations}
\label{sec:cross}

    \begin{figure}
        \centering
        \includegraphics[width=0.5\textwidth]{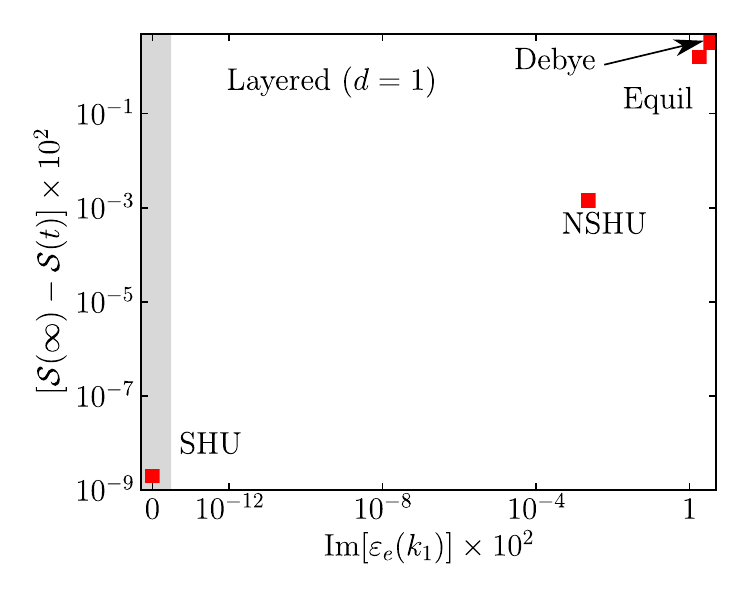}
        \caption{
           Cross-property relation between the imaginary part of the effective dielectric constant $\Im[\varepsilon_e]$ at a small dimensionless wavenumber $k_1/s=0.1$ within the transparency interval and the excess spreadability $\fn{\mathcal{S}}{\infty}-\fn{\mathcal{S}}{t}$ at a large dimensionless time ${\cal D}ts^2 = 20$ for the four models of layered media with $\phi_2 = 0.20$.
We consider Debye, Equil., NSHU, and SHU media.
The data point for the SHU medium is shown in the gray shaded region, since it has a zero imaginary part, which cannot be shown in the log scale. The Debye medium has the largest attenuation and spreadability, while the SHU has zero attenuation and the fastest (exponentially fast) spreadability decay rate.
            \label{fig:cross-property}
        }
    \end{figure}
    
    \begin{table}
        \caption{
            Values of the imaginary part of the effective dielectric constant $\Im[\varepsilon_e]$ at a small dimensionless wavenumber $k_1/s=0.1$ within the transparency interval and the excess spreadability $\fn{\mathcal{S}}{\infty}-\fn{\mathcal{S}}{t}$ at a large dimensionless time ${\cal D}ts^2 = 20$ for all models considered here for three distinct structural symmetries with $\phi_2 = 0.20$.
            For layered media ($d=1$), we consider Debye, Equil., NSHU, and SHU media.
            For other symmetries, we consider Debye and SHU media.
            While we arbitrarily choose the values for $k_1/s$ and ${\cal D}ts^2$, the relative rankings for the models are independent of the specific values if $k_1/s<0.3$ and ${\cal D}ts^2 > 10$.
            \label{tab:cross}
        }
        \begin{tabular}{c|c|c|c}
            \hline
            Symmetry& Model& $\Im[\varepsilon_e(k_1)]$ & $\mathcal{S}(\infty)-\mathcal{S}(t)$ \\
            \hline
            \multirow{4}{*}{Layered ($d=1$)}
            &   Debye&  3.60$\times10^{-2}$&    3.22$\times10^{-2}$ \\ 
            &   Equil.&    1.81$\times10^{-2}$&    1.61$\times10^{-2}$ \\ 
            &   NSHU&    2.33$\times10^{-5}$&    1.41$\times10^{-5}$ \\ 
            &   SHU&    0.0&    1.97$\times10^{-11}$ \\ 
            \hline      
            \multirow{2}{*}{Transversely Isotropic ($d=2$)}
            &   Debye&  5.77$\times10^{-3}$&    9.87$\times10^{-3}$ \\ 
            &   SHU&    0.0&    7.74$\times10^{-22}$ \\ 
            \hline      
            \multirow{2}{*}{Fully Isotropic ($d=3$)}
            &   Debye&  1.58$\times10^{-4}$&    3.05$\times10^{-3}$ \\ 
            &   SHU&    0.0 &   2.75$\times10^{-23}$ \\ 
            \hline      
        \end{tabular}
    \end{table}

    We now show  that the attenuation behaviors of electromagnetic waves for
    sufficiently small wavenumbers and the long-time decaying behaviors of the excess
    spreadabilities are positively correlated with one another, i.e., we
    demonstrate that there are cross-property
    relations between them. We begin by considering our four models of
    layered media (i.e., $d=1$): Debye,
    equilibrium, NSHU, and SHU media.
    Figure \ref{fig:cross-property}
    depicts the values of the imaginary part of the effective dielectric
    constant $\Im[\varepsilon_e(k_1)]$ at a dimensionless wavenumber $k_1/s=0.1$
    and the excess spreadability
    $\fn{\mathcal{S}}{\infty}-\fn{\mathcal{S}}{t}$ at a large dimensionless time ${\cal D}ts^2 = 20$.
    This plot clearly demonstrates that as the degree of attenuation increases from zero attenuation and the associated smallest  long-time excess
spreadability (fastest decay) for the SHU media, long-time excess spreadability increases (slower decay rate) until the largest attenuation and long-time excess
spreadability (smallest decay rate) is achieved for the Debye media.
    We see that the NSHU medium
has a substantially smaller attenuation and excess spreadability than that of the equilibrium media.
     Table \ref{tab:cross} lists the values of the imaginary part of the effective dielectric constant $\Im[\fn{\varepsilon_e}{k_1}]$ at a  dimensionless wavenumber $k_1/s=0.1$ and the excess spreadability $\fn{\mathcal{S}}{\infty}-\fn{\mathcal{S}}{t}$ at a large dimensionless time ${\cal D}ts^2 = 20$ corresponding to the plotted values in Fig. \ref{fig:cross-property}. 
      While our choices for the values of the small dimensionless wavenumber $k_1/s$ and the dimensionless time ${\cal D}ts^2$ are arbitrary, the relative rankings of the models are independent of the specific values if $k_1 / s < 0.3$ and $10< {\cal D}ts^2 $.

Table \ref{tab:cross} lists corresponding values the imaginary part of the effective dielectric constant $\Im[\fn{\varepsilon_e}{k_1}]$ at a  dimensionless wavenumber $k_1/s=0.1$ and the excess spreadability $\fn{\mathcal{S}}{\infty}-\fn{\mathcal{S}}{t}$ 
    for Debye and SHU media in both the transversely isotropic ($d=2$) and fully isotropic case ($d=3$).
The analogous positive correlations between the attenuation behaviors and the long-time spreadability behaviors
seen in for $d=1$ are observed in these instances, even if we do not report corresponding results of the equilibrium and  NSHN media for $d=2$ and $d=3$, since qualitatively, they are similar to the layered media models.
    Specifically, the SHU media achieve both zero attenuation (i.e., perfect transparency) and the smallest long-time excess spreadability (i.e., exponential decay), while Debye media exhibit the strongest attenuation and largest long-time excess spreadability among all four models.
    Furthermore, for a $d$-dimensional NSHU model, the power-law scaling of its small-wavenumber attenuation behavior [i.e., $\Im[\fn{\varepsilon_e}{k_1}] \sim {k_1}^{d+\alpha}$, as predicted by Eqs. \eqref{eq:asymptotic-layered} and \eqref{eq:asymptotic-trans-TM}] determines that of its long-time excess spreadability [i.e., $\fn{\mathcal{S}}{\infty}-\fn{\mathcal{S}}{t} \sim (\mathcal{D}ts^2)^{-(d+\alpha)/2}$, as predicted by Eq. \eqref{long-time}], where the exponent $\alpha$ is specified by Eq. \eqref{eqn:smallk}.

    \section{Discussion}
    \label{sec:discussion}
   
   %Among the possible hyperuniform states of composite media, stealthy hyperuniform ones are emerging as stand outs.
   %A critical fundamental and practical question is whether their unique physical
   %properties persist in the infinite-volume or thermodynamic limit. Describe \cite{Kl22}
   
   In this paper, we have further elucidated the extraordinary optical and transport properties of 3D
   disordered SHU media with three distinct microstructural symmetries: layered, transversely isotropic, and fully isotropic cases.
   We had previously shown that the accurate truncations of the strong-contrast expansions of the
    effective dynamic dielectric constant $\fn{\tens{\varepsilon}_e}{\vect{k}_1}$\cite{To21a}  predicted the perfect transparency interval defined by Eq. (\ref{eq:trans-interval})
   for SHU layered \cite{Ki23}, transversely isotropic \cite{To21a,kim_accurate_2024} and fully isotropic media \cite{To21a}.
   In \ref{sec:proofs}, we provided detailed proofs that the same perfect transparency interval exactly applies
   through the third-order truncations for such layered media and transversely isotropic in the case
   of TM polarization. Given the rapid convergence of the strong contrast expansion, the previously demonstrated accuracy of the second-order approximations \cite{To21a, Ki23,kim_accurate_2024}  leads to the conclusion that there can be no Anderson localization within the predicted perfect transparency interval in such SHU media in practice because the localization length (associated with only possibly negligibly small higher-order contributions) should be very large compared to any practically large sample size, as noted in Refs. \cite{kim_accurate_2024}.
   This prediction for layered media is especially remarkable because the traditional understanding is that localization must occur for any type of disorder in 1D media.
   
   Additionally, we further revealed the singular physical properties of 3D SHU two-phase media with three distinct symmetries (layered, transversely isotropic media, and fully isotropic media) by comparing and contrasting them to those of other model nonstealthy microstructures. Specifically, we studied their attenuation characteristics, as measured by the imaginary part of $\fn{\tens{\varepsilon}_e}{\vect{k}_1,\omega}$, and transport properties, as measured by the time-dependent diffusion spreadability ${\cal S}(t)$.
   Among the possible hyperuniform states of composite media, the SHU ones stand out as having singular physical properties, such as exponentially fast decay of excess spreadability and finite perfect transparency intervals for all symmetries.
   We demonstrated that there are cross-property relations between the attenuation and spreadability behaviors
   by quantifying how the imaginary parts of the effective dielectric constant and the excess spreadability at long times
   are positively correlated as the structures span from nonhyperuniform, NSHU, and SHU media.
   Such cross-property relations are useful because they enable one to estimate one of the properties, given a measurement of the other \cite{To02a}.

    There are a variety of outstanding open problems for future research.
    For example, while our results imply that there can be no Anderson localization for 1D disordered SHU medium in practice, its rigorous and direct proof in the thermodynamic limit remains an open problem.
    To resolve this problem, we will systematically study the localization length of 1D SHU media as a function of system size \cite{Kl_local_2023}.
    Another interesting direction is to extend the cross-property relations between diffusion spreadability and attenuation behaviors for general cases, such as arbitrary incidence direction or two-phase media with absorbing phases.
    It will also be useful to establish cross-property relations for SHU media for other
    wave phenomena, such as elastodynamics characteristics \cite{Ki20a}, as well 
    as other transport properties.
    Concerning the latter possibility, it has recently been shown that the fluid permeability associated with fluid transport in porous media is bounded from above in terms of a functional involving the spectral density $\spD{k}$\cite{To20}. Since the second-formulas for $\fn{\tens{\varepsilon}_e}{\vect{k}_1}$
    and the formulas for the spreadability ${\cal S}(t)$ also depend on the spectral density, it would be relatively straightforward to determine
    whether useful cross-property relations exist between these three different physical properties for nonhyperuniform
    and hyperuniform media.
   
       %Cross-property relations for such situations could provide useful insight for designing metastructures that require effective cooling from the heating due to radiations because heat conduction is physically equivalent to mass diffusion.
    %[@Sal:The previous sentence needs some justification from literature.]
   % due to the rapid convergence properties of strong-contrast expansions and the high predictive power of their second-order truncations shown in Ref. \cite{To21a,kim_accurate_2024}, their higher-order contributions in the imaginary part are negligibly small for relatively large contrast ratios.
   % Therefore, the localization
   % length associated with only possibly negligibly small higher-order contributions should be very large compared to any practically large
   % sample size, and thus, there can be no Anderson localization within the predicted perfect transparency interval in 1D and 2D stealthy
   % hyperuniform media in practice.

   \appendix

    \section{Scaled Strong-Contrast Formulas at Three-Point Levels}
    \label{sec:details}

    Here, we provide detailed descriptions, including the problem setup and assumptions, of the strong-contrast formulas for the effective dynamic dielectric constant, omitted in section \ref{sec:theory}.    
    We consider a plane electric wave of an angular frequency $\omega$
    and a wavevector $\vect{k}_q$ in the reference phase $q$, which is taken as the matrix phase (i.e., $q = 1$) of the composite unless otherwise stated.
    We make three assumptions about both phases: (a) they have
    constant
    real-valued dielectric constants, (b) they are dielectrically isotropic,
    and (c) they are nonmagnetic.
    Thus, such a dielectric composite cannot absorb waves but attenuate them solely due to multiple scattering from fluctuations in the
    local dielectric constant.
    Since these assumptions give a simple relation between the wavenumber
    $k_1$ in the reference phase $q=1$ and angular frequency $\omega$ [i.e.,
    $\fn{k_1}{\omega}\equiv\abs{\vect{k}_1(\omega)} =
    \sqrt{\varepsilon_1}\omega/c$], where $c$ is the speed of light in vacuum,
    we henceforth do not explicitly indicate the $\omega$ dependence, i.e.,
    $\fn{\tens{\varepsilon}_e}{\vect{k}_1, \omega
    }=\fn{\tens{\varepsilon}_e}{\vect{k}_1}$.

    Under these three assumptions, we derived the following  general strong-contrast
    expansion for the linear fractional form of $\fn{\tens{\varepsilon}_e}{\vect{k}_1}$ in Ref. \cite{To21a}:
    \begin{eqnarray}
        &\phi_2 \tens{L}_2^{(1)}\cdot 
        \left(
        \left\{\tens{I} + \tens{D}^{(1)}  \cdot \left[\fn{\tens{\varepsilon}_e}{\vect{k}_1} - {\varepsilon}_1 \tens{I} \right] \right\} \cdot \left[\fn{\tens{\varepsilon}_e}{\vect{k}_1} - {\varepsilon}_1 \tens{I} \right]^{-1} 
        \right) 
        \cdot \phi_2 \tens{L}_2^{(1)}
        \nonumber \\
        & \qquad =
        \phi_2 \tens{L}_2^{(1)} 
        - \fn{\tens{\mathcal{A}}_2^{(2)}}{\vect{k}_1} - \fn{\tens{\mathcal{A}}_3^{(2)}}{\vect{k}_1} - \ldots ,
        \label{eq:str-exp-trunc}
    \end{eqnarray}
    where $p~(= 2)$ indicates the polarized phase, $\fn{\tens{\mathcal{A}}_n^{(2)}}{\vect{k}_1}$ is a wavevector-dependent second-rank tensor that is a functional involving the set of correlation functions $S_1^{(2)}, S_2^{(2)}, \ldots, S_n^{(2)}$, defined in Sec.~\ref{n-point}, and products of the principal part of the dyadic Green's function, and the two constant second-rank tensors $\tens{L}_2^{(1)}$ and $\tens{D}^{(1)}$ are associated with an infinitesimal region around the singularity in the Green's function \cite{To21a, yaghjian_electric_1980}.
    The linear fractional form of the strong-contrast expansion \eqref{eq:str-exp-trunc} results in a
    rapidly converging series whose lower-order truncations lead to accurate
    approximation formulas for $\fn{\tens{\varepsilon}_e}{\vect{k}_1}$, even for large contrast ratios.
    This is to be contrasted with standard weak-contrast expansions that do
    not converge rapidly for large contrast ratios; see Refs.
    \cite{To21a,kim_accurate_2024} for such a quantitative explanation.
    Indeed, its truncation at the two-point level yields a highly accurate
    estimation of $\fn{\tens{\varepsilon}_e}{\vect{k}_1}$ due to the fact that higher-order contributions are negligibly small, implying that
    this resulting formula very accurately approximates multiple scattering
    through all orders \cite{To21a, Ki23}.
    Furthermore, this
    series takes into account the singularity of the dyadic Green's function
    and thus is valid even for large values of phase contrast ratio
    $\varepsilon_2/\varepsilon_1$ and various microstructural symmetries.
    In what follows, we briefly outline the derivations of the key formulas for the effective dynamic
    dielectric constant tensor of layered and transversely
    isotropic media, which were extracted from the general strong-contrast
    expansion \eqref{eq:str-exp-trunc} in Refs. \cite{To21a,Ki23,kim_accurate_2024}.

    \subsection{Layered Media} \label{sec:formula-layered}

    Under the condition specified in section \ref{sec:layered}, the scaled strong-contrast approximations associated with layered media are extracted from exact strong-contrast series \eqref{eq:str-exp-trunc} by choosing a disk-like exclusion volume normal to the symmetry axis $\uvect{z}$ that involves the singularity of the dyadic Green's function.
    Due to the symmetries of layered media, one can decompose $\fn{\tens{\varepsilon}_e}{\vect{k}_1}$ into two orthogonal components $\fn{\varepsilon_e ^\perp}{k_1}$ and $\fn{\varepsilon_e ^z}{k_1}$ for the transverse and longitudinal polarizations, respectively. %, as follows:
    % $\fn{\tens{\varepsilon}_e}{k_1} = \fn{\varepsilon_e ^\perp}{k_1} \left(\tens{I}-\uvect{z}\uvect{z} \right) + \fn{\varepsilon_e^z}{k_1}\uvect{z}\uvect{z}$. 
    Thus, we derive two independent approximations by truncating the strong-contrast expansion and solving for $\fn{\tens{\varepsilon}_e}{k_1}$ from this linear fractional form.
    Renormalization of the resulting formulas with the reference phase for the optimal convergence, equivalent to using the effective Green's function in Ref. \cite{To21a}, yields scaled strong-contrast approximations for disordered layered media at the three-point level.
    The reader is referred to Refs. \cite{Ki23,kim_accurate_2024} for derivations.
    %Sec. \ref{secS:SM-Layered} of \supp.
    
    Such a formula for the transverse polarization is given in Eq. \eqref{eq:scaled_eps_layered_trans}.
    The analogous formula for longitudinal polarization is given as 
    \begin{eqnarray}
        \fn{\varepsilon_e^z}{k_1} =& \varepsilon_1 (1 - \phi_2\BETA{1}{21})^{-1}.
        \label{eq:scaled_eps_layered_long}
    \end{eqnarray}       
    Note that $\fn{\varepsilon_e^z}{k_1}$ is independent of $k_1$, since a traveling longitudinal wave cannot exist under the aforementioned conditions.
    The static limit of Eq. \eqref{eq:scaled_eps_layered_long} is the harmonic mean of the local dielectric constants, i.e., 
    $
        \fn{\varepsilon_e^z}{0} = (\phi_2/\varepsilon_2 + \phi_1 /\varepsilon_1)^{-1},
    $
    which is exact for any 1D microstructure \cite{To02a}.

    \subsection{Transversely Isotropic Media}   \label{sec:formula-trans-iso}

    Under the condition specified in section \ref{sec:iso}, the scaled strong-contrast approximations associated with transversely isotropic media are extracted from exact strong-contrast series \eqref{eq:str-exp-trunc} by choosing a needle-like exclusion volume aligned along $\uvect{y}$ that involves the singularity of the dyadic Green's function.
    Due to the symmetries of the problems, one can decompose $\fn{\tens{\varepsilon}_e}{k_1}$ into two orthogonal components $\fn{\varepsilon_e ^{TM}}{k_1}$ and $\fn{\varepsilon_e ^{TE}}{k_1}$ for TM and TE polarizations, respectively.
    Thus, we extract two independent approximations of $\ETE{k_1}$ and $\ETM{k_1}$ by truncating the strong-contrast series at third-order terms.
%    see \eqref{eqS:appr-trans-TE} and \eqref{eqS:appr-trans-TM} of \supp.
    Solving for the effective dielectric constants from these linear fractional forms and then renormalizing them with the optimal reference phase \cite{tsang_scattering_1981,mackay_strongpropertyfluctuation_2000, To21a}, we finally obtain the scaled strong-contrast approximations at the three-point level for disordered transversely isotropic media.

    Such a formula for the TM polarization is given in Eq. \eqref{eq:scaled_eps_2D_TM}.
    Its TE analog is given as 
    \begin{eqnarray}
        \fl \frac{\ETE{k_1}}{\varepsilon_1}
        =& 
        1+ \frac{ 
            2  {\phi_2}^2 \BETA{2}{21}
        }
        { 
            \phi_2(1-\phi_2 \BETA{2}{21}) 
            -  
            \ATE{2}{k_*^{TE}; \varepsilon_{BG}^{(2D)}} \left[2\varepsilon_1 \BETA{2}{21} \right]
            -  
            \ATE{3}{k_*^{TE}; \varepsilon_{BG}^{(2D)}} \left[2\varepsilon_1 \BETA{2}{21} \right]^2
        }
            \label{eq:scaled_eps_2D_TE} , 
    \end{eqnarray}
    where $k_*^{TE} \equiv k_1 \sqrt{\varepsilon_{BG}^{(2D)}/\varepsilon_1}$ is wavenumber in the optimal reference phase for TE polarization, and $\varepsilon_{BG}^{(2D)}$ is the Bruggeman approximation for 2D two-phase media \cite{bruggeman_berechnung_1935, To02a}.
    Here, the second-order coefficient is given as $\ATE{2}{k; \varepsilon}
    =
    \ATM{2}{k; \varepsilon}/2$; see Eq. \eqref{eq:F-trans-Fourier}, and the third-order coefficient is defined as 
    \begin{eqnarray}
        \fl &\ATE{3}{k; \varepsilon}
        = 
        \frac{-1}{\phi_2}
        \left[\frac{1}{2\varepsilon (2\pi)^2} \right]^2
        \int_{\R^2} \dd{\vect{q}_1}
        \int_{\R^2} \dd{\vect{q}_2} 
        \fn{\tilde{\Delta}_3^{(2)}}{\vect{q}_1,\vect{q}_2}
        \nonumber \\
        \fl&\qquad  \times \frac{\Big[
            ({k}^2)^2 - \abs{k\uvect{y}+\vect{q}_1}^2 \abs{k\uvect{y}+\vect{q}_2}^2
            + 2 [(k\uvect{y}+\vect{q}_1)\cdot (k\uvect{y}+\vect{q}_2)]^2
        \Big]}
        {\left(\abs{k\uvect{y}+\vect{q}_1}^2 - {k}^2 \right) \left(\abs{k\uvect{y} + \vect{q}_2}^2 - {k}^2 \right)}
        .
        \label{eq:A3-TE}
    \end{eqnarray}
 %   where %$\F{2}{k}$ is the nonlocal attenuation function for 2D statistically isotropic two-phase media \cite{To21a}, and 
 %   $\fn{\tilde{\Delta}_3^{(2)}}{\vect{q}_1,\vect{q}_2}$ is the Fourier transform of $\fn{\Delta_3^{(2)}}{\vect{x}_{21},\vect{x}_{31}}\equiv \fn{S_2^{(2)}}{\vect{x}_{21}} \fn{S_2^{(2)}}{\vect{x}_{32}} - \phi_p \fn{S_3^{(2)}}{\vect{x}_{21}, \vect{x}_{32}+\vect{x}_{21}}$.
    We note that Eq. \eqref{eq:scaled_eps_2D_TE} is identical to the 2D formula derived in Ref. \cite{To21a}, except for the reference phase.
    The current choice for the reference dielectric constant $\varepsilon_{BG}^{(2D)}$ offers slightly better renormalization than 2D Maxwell-Garnett approximation employed in Ref. \cite{To21a} because $\varepsilon_{BG}^{(2D)}$ makes the mean of the depolarization tensors exactly zero, i.e., $\phi_1 \tens{L}_1 ^{(*)} + \phi_2 \tens{L}_2 ^{(*)}=\tens{0}$. % (see Eq. (S105) of \supp).
%    (see \eqref{eqS:optimal-reference} of \supp).
    The static limit ($k_1 \to 0^+$) of Eq. \eqref{eq:scaled_eps_2D_TE} is the third-order static strong-contrast approximation for $d=2$ \cite{To02a}:
    \begin{equation}
        \ETE{0} 
        =
        \varepsilon_1
        \frac{
            \phi_2 (1+\phi_2\BETA{2}{21}) - (1-\phi_2) \zeta_2 [\BETA{2}{21}]^2}
        {
            \phi_2 (1-\phi_2\BETA{2}{21}) - (1-\phi_2) \zeta_2 [\BETA{2}{21}]^2
        },    
        \label{eq:eps_2D_static-TE} 
    \end{equation}
    where $\zeta_2$ is the three-point parameter that lies in the closed interval $[0,1]$ \cite{To02a}.

    \section{Perfect Transparency at the Three-Point Level} 
\label{sec:proofs}

Here, we analytically show that the finite perfect transparency interval given in \eqref{eq:trans-interval} for SHU two-phase media with $\spD{k}=0$ for $k<K$ is exact through the third-order terms.
The detailed proofs presented here were only very briefly sketched in Ref. \cite{kim_accurate_2024}.

We first show the proof for transverse polarization in the SHU layered medium (or, equivalently, 1D SHU medium) in \ref{sec:PT-layered-3pt}.
We then show the proof for TM polarization in the SHU transversely isotropic medium (or, equivalently, 2D SHU medium) in \ref{sec:PT-trans-3pt}.
For these proofs, we utilize the three general properties of $\FTJ{p}{\vect{k}}$ for disordered SHU two-phase media in $\R^d$:
\begin{eqnarray}
    \fl \mbox{Stealthy~ hyperuniform~ condition:} ~&
    \abs{\FTJ{p}{\vect{k}}}
    = 0,    \quad \qquad \mathrm{for~ } 0\leq \abs{\vect{k}}<K,    \label{eq:SHU_PT_layered}\\
    \fl \mbox{No~ Bragg~ peaks:} ~&
    \abs{\FTJ{p}{\vect{k}}} < C,  \quad \qquad \mathrm{for~ any~ } \vect{k},    \label{eq:bounded_PT_layered}\\
    \fl \mbox{Large-$\abs{\vect{k}}$~ scaling~ from~ Eq. \eqref{decay}:} ~& \abs{\FTJ{p}{\vect{k}}} 
    \leq   \frac{C'}{\abs{\vect{k}}^{(d+1)/2}},   \qquad \mathrm{for~ }Q \leq \abs{\vect{k}} , \label{eq:decay_PT_layered} 
\end{eqnarray}
where $C'$ and $C$ are some finite positive numbers, and $Q$ is a large wavenumber much greater than $K$.

\subsection{Layered Media}
\label{sec:PT-layered-3pt}

In Sec. \ref{sec:trans}, we analytically show that for SHU layered medium with $\spD{k}=0$ for $k<K$, the finite perfect transparency interval given in Eq. \eqref{eq:trans-interval} conjunction with Eq. \eqref{eq:eps_layered} is exact through the second-order terms.
Here, we provide detailed proof that this transparency interval is exact through the third-order terms $A_3^\perp$.

Without losing generality, one can rewrite the third-order condition for perfect transparency as follows:
\begin{eqnarray}   \label{eq:im-layered-3pt_app}
    \Im[\fn{A_3^\perp}{k;1}] =0 ,\quad (k<K/2).
\end{eqnarray}
Substituting 
$
    \fn{\tilde{\Delta}_3^{(p)}}{q_1,q_2}
        =
        -\phi_p \E{\fn{\tilde{\mathcal{J}}^{(p)}}{q_1} \fn{\tilde{\mathcal{J}}^{(p)}}{q_2 - q_1} \fn{\tilde{\mathcal{J}}^{(p)}}{-q_2} } 
        - {\phi_p}^2 (2\pi) \fn{\delta}{q_1 - q_2}\spD{q_2} + \spD{q_1} \spD{q_2}
$ into the Fourier representation of $A_3^\perp$ given in Eq. \eqref{eq:A3-strat-Fourier} allows us to separate the third-order coefficient into three terms:
\begin{eqnarray}
    \fl\fn{A_3^\perp}{k;1}
    &= \frac{-1}{\phi_p} \left(\frac{k^2}{2\pi} \right)^2
    \int_{-\infty}^\infty \dd{q_1}
    \int_{-\infty}^\infty \dd{q_2}
    \frac{1}{(k+q_1)^2-k^2} \frac{1}{(k+q_2)^2-k^2}
    \nonumber \\
    \fl \qquad&\times  \Bigg[
        \spD{q_{1}} \spD{q_{2}}
        -{\phi_p}^2 (2\pi) \fn{\delta}{q_1-q_2} \spD{q_2} 
    -\phi_p \E{\FTJ{p}{-q_1}\FTJ{p}{q_1-q_2}\FTJ{p}{q_2}}
    \Bigg] 
    \nonumber \\
    \fl&=
    \frac{-1}{\phi_p}\frac{{k}^4}{(2\pi)^2} 
        \Bigg\{ \fn{\CL{1}}{k} - {\phi_p}^2 (2\pi) \fn{\CL{2}}{k} - {\phi_p} \fn{\CL{3}}{k} 
        \Bigg\}, \label{eq:A3-layered-decomposed}
\end{eqnarray}
where $\fn{\CL{1}}{k}$ and $\fn{\CL{2}}{k}$ depend on the two-point statistics, whereas $\fn{\CL{3}}{k}$ depends on the three-point statistics:
\begin{eqnarray}
    \fn{\CL{1}}{k} 
    \equiv &  
    \left[2\pi {k}^{-2} \F{1}{k}\right]^2
    \label{eq:C1_layered},  \\
    \fn{\CL{2}}{k}
    \equiv &
    \int_{-\infty}^\infty \dd{q_{1}} \left[(k+q_{1})^2-{k}^2 \right]^{-2} \spD{q_{1}} 
    \label{eq:C2_layered}, \\
    \fn{\CL{3}}{k}
    \equiv & 
    \int_{-\infty}^\infty \dd{q_{1}} \int_{-\infty}^\infty \dd{q_{2}} \frac{\E{\FTJ{p}{q_{1}} \FTJ{p}{-q_{1}+q_{2}} \FTJ{p}{-q_{2}}}}{[(k+q_{1})^2-{k}^2][(k+q_{2})^2-{k}^2]} 
    \label{eq:C3_layered}.
\end{eqnarray}
Therefore, we can prove Eq. \eqref{eq:im-layered-3pt_app} by showing that for $n=1,2,3,$
\begin{eqnarray}   \label{eq:Cn-cond_layered}
    \Im[\fn{\CL{n}}{k}]=0 \quad (k<K/2),
\end{eqnarray}
where $\fn{\CL{n}}{k}$ are defined in Eqs. \eqref{eq:C1_layered}-\eqref{eq:C3_layered}, respectively.
We prove \eqref{eq:Cn-cond_layered} by using the three general properties of $\FTJ{p}{k}$ for 1D disordered SHU two-phase media, given in Eqs. \eqref{eq:SHU_PT_layered}-\eqref{eq:decay_PT_layered}.

% To prove \eqref{eq:Cn-cond_layered}, we utilize the three general properties of $\FTJ{p}{k}$ for 1D disordered stealthy hyperuniform two-phase media:
% \begin{eqnarray}
%     \fl \mbox{Stealthy~ hyperuniform~ condition:} ~&
%     \abs{\FTJ{p}{k}}
%     = 0,    \quad \qquad \mathrm{for~ } 0\leq \abs{k}<K,    \label{eq:SHU_PT_layered}\\
%     \fl \mbox{No~ Bragg~ peaks:} ~&
%     \abs{\FTJ{p}{k}} < C,  \quad \qquad \mathrm{for~ any~ } k,    \label{eq:bounded_PT_layered}\\
%     \fl \mbox{Large-$k$~ scaling~ from~ \eqref{eq:decay}:} ~& \abs{\FTJ{p}{k}} 
%     \leq   \frac{C'}{\abs{k}},   \qquad \mathrm{for~ }Q \leq \abs{k} , \label{eq:decay_PT_layered} 
% \end{eqnarray}
% where $C'$ and $C$ are some finite positive numbers, and $Q$ is a large wavenumber much greater than $K$.

One can immediately see that \eqref{eq:Cn-cond_layered} is true for $n=1$.
For $k<K/2$, \eqref{eq:A2-strat-Fourier} gives $\Im[\F{1}{k}]=0$, and thus $\Im[\fn{\CL{1}}{k}]=0$.
We then prove \eqref{eq:Cn-cond_layered} for $n=2$.
The integrand in \eqref{eq:C2_layered} is real-valued and nonnegative except at two singular points ($q_1 = 0$ and $-2k$) because it is the product of the spectral density $\spD{q_1}$ and the square of the Green's function [i.e., $(\abs{k+q_{1}}^2-{k}^2)^{-2}$]. 
For $k<K/2$, however, this integrand has no singularity because the stealthy hyperuniform condition [\eqref{eq:SHU_PT_layered}] can completely remove these two singular points that the Green's function can arise.
Furthermore, due to the properties of $\spD{k}$ [\eqref{eq:bounded_PT_layered} and \eqref{eq:decay_PT_layered}], this nonnegative integrand of \eqref{eq:C2_layered} is bounded from above and decays rapidly like $\abs{q_1}^{-6}$ for large $\abs{q_1}$.
Therefore, for $k<K/2$, the real-valued integrand of \eqref{eq:C2_layered} is absolutely integrable, leading to $\Im[\fn{\CL{2}}{k}]=0$.

We then prove \eqref{eq:Cn-cond_layered} for $n=3$.
The crucial part of the proof is to show that for $k<K/2$, the integrand in \eqref{eq:C3_layered} is absolutely integrable (i.e., the integral of its absolute value is well-defined and finite) such that the Fubini-Tonelli theorem \cite{stewart_calculus_2016} enables one to interchange the integration order in \eqref{eq:C3_layered}.
Once we show the integration order is interchangeable, using the general property of the Fourier transform [i.e., $\FTJ{p}{k'}^* = \FTJ{p}{-k'}$], we can immediately show that for $k<K/2$, $\fn{\CL{3}}{k}$ is real-valued because
\begin{eqnarray}
    \fl [\fn{\CL{3}}{k}]^*
    &= 
    \int_{-\infty}^\infty \dd{q_{1}} \int_{-\infty}^\infty \dd{q_{2}} \frac{\E{[\FTJ{p}{q_{1}} \FTJ{p}{-q_{1}+q_{2}} \FTJ{p}{-q_{2}}]^*} }{[(k+q_{1})^2-{k}^2] [(k+q_{2})^2-{k}^2]}  
    \nonumber \\
    \fl &=
    \int_{-\infty}^\infty \dd{q_{1}} \int_{-\infty}^\infty \dd{q_{2}} \frac{\E{\FTJ{p}{-q_{1}} \FTJ{p}{q_{1}-q_{2}} \FTJ{p}{q_{2}}}}{[(k+q_{1})^2-{k}^2] [(k+q_{2})^2-{k}^2]}  
    \nonumber 
    \\
    \fl &=
    \int_{-\infty}^\infty \dd{q_{2}} \int_{-\infty}^\infty \dd{q_{1}} \frac{\E{\FTJ{p}{q_{2}}
    \FTJ{p}{-q_{2}+q_{1}} \FTJ{p}{-q_{1}} }}{[(k+q_{2})^2-{k}^2] [(k+q_{1})^2-{k}^2] }
    =
    \fn{\CL{3}}{k}. \label{eq:C3_layered-step3}
\end{eqnarray}

Now let us show that the integrand in Eq. \eqref{eq:C3_layered} is absolutely integrable if $k<K/2$.
We first reduce the absolute value of this integrand in a simpler form:
\begin{eqnarray}
    &\abs{\frac{1}{(k+q_{1})^2-{k}^2} \frac{1}{(k+q_{2})^2-{k}^2} 
        \E{\FTJ{p}{q_{1}} \FTJ{p}{-q_{1}+q_{2}} \FTJ{p}{-q_{2}}}}
    \nonumber\\
    \leq &
    % \abs{\frac{1}{(k+q_{1})^2-{k}^2} \frac{1}{(k+q_{2})^2-{k}^2} }
    %     \E{
    %         \abs{\FTJ{p}{q_{1}} \FTJ{p}{-q_{1}+q_{2}} \FTJ{p}{-q_{2}}}
    %     }
    % \nonumber
    % \\
    % &=
    \abs{\frac{1}{(k+q_{1})^2-{k}^2}} \abs{\frac{1}{(k+q_{2})^2-{k}^2} }
        \E{
            \abs{\FTJ{p}{q_{1}}}
            \abs{\FTJ{p}{-q_{1}+q_{2}}} 
            \abs{\FTJ{p}{-q_{2}}}
        }    
    \nonumber \\
    \leq &
    \abs{\frac{1}{(k+q_{1})^2-{k}^2}} \abs{\frac{1}{(k+q_{2})^2-{k}^2} }
        C \E{
            \abs{\FTJ{p}{q_{1}}}
            \abs{\FTJ{p}{-q_{2}}}
        } , \label{eq:C3_layered-step3-1}
\end{eqnarray}
where we have used Eq. \eqref{eq:bounded_PT_layered}.
If the integral of \eqref{eq:C3_layered-step3-1} is bounded from above by a finite constant, then we prove that the integrand in \eqref{eq:C3_layered} is absolutely integrable.
\begin{eqnarray}
    \fl &\int_{-\infty}^\infty \dd{q_1}
    \int_{-\infty}^\infty \dd{q_2}
    \abs{\frac{1}{(k+q_{1})^2-{k}^2}} \abs{\frac{1}{(k+q_{2})^2-{k}^2} }
        C \E{
            \abs{\FTJ{p}{q_{1}}}
            \abs{\FTJ{p}{-q_{2}}}
        }
    \nonumber
    \\
    &=
    % C \E{
    % \int_{-\infty}^\infty \dd{q_1}
    % \int_{-\infty}^\infty \dd{q_2}
    % \frac{\abs{\FTJ{p}{q_{1}}}}{\abs{(k+q_{1})^2-{k}^2}} \frac{\abs{\FTJ{p}{-q_{2}}}}{\abs{(k+q_{2})^2-{k}^2} }
    % }
    % =
    C
    \E{
        \left[\int_{-\infty}^\infty \dd{q_1}
        \frac{\abs{\FTJ{p}{q_{1}}}}{\abs{(k+q_{1})^2-{k}^2}} \right]^2
    }   \label{eq:C3_layered-step4}.
\end{eqnarray}
We then compute the integral that appears in the upper bound on Eq. \eqref{eq:C3_layered-step4}:
\begin{eqnarray}
    \fl &\int_{-\infty}^\infty \dd{q_1}
        \frac{\abs{\FTJ{p}{q_{1}}}}{\abs{(k+q_{1})^2-{k}^2}}
    =
    \left[
        \int_{\abs{q_1}<K} 
    +
        \int_{K\leq \abs{q_1}<Q}
    +
        \int_{Q\leq \abs{q_1}}
    \right]\dd{q_1}
    \frac{\abs{\FTJ{p}{q_{1}}}}{\abs{(k+q_{1})^2-{k}^2}}
    \nonumber \\
    \fl & \leq 
    0
    +
    \int_{K\leq \abs{q_1}<Q} \dd{q_1}
    \frac{C}{\abs{(k+q_{1})^2-{k}^2}}
    +
    \int_{Q\leq \abs{q_1}} \dd{q_1}
    \frac{C'/ \abs{q_1}}{\abs{(k+q_{1})^2-{k}^2}},  \label{eq:C3_layered-step5}
\end{eqnarray}
where we have applied Eqs. \eqref{eq:SHU_PT_layered}-\eqref{eq:decay_PT_layered} sequentially in the three integrals.
It is important to note that the first integral over $-K<q_1<K$ is divergent unless the system is stealthy hyperuniform.
The second integral in \eqref{eq:C3_layered-step5} is positive and bounded from above, as shown below:
\begin{eqnarray}
    \fl &
    \int_{K\leq \abs{q_1}<Q} \dd{q_1}
    \frac{C}{\abs{(k+q_{1})^2-{k}^2}}
    \leq 
    2C \int_{-Q}^{-K} \dd{q_1}
    \frac{1}{q_{1}(q_1 + 2k)}
    =
    \frac{C}{k} \fn{\ln}{\frac{K}{K-2k}\frac{Q-2k}{Q}}.
    \label{eq:C3_layered-step5-2}
\end{eqnarray}
The third integral in \eqref{eq:C3_layered-step5} is also positive and bounded from above, as shown below:
\begin{equation}
    \fl\int_{Q\leq \abs{q_1}} \dd{q_1}
    \frac{C'/ \abs{q_1}}{\abs{(k+q_{1})^2-{k}^2}}
    \leq
    2C'\int_{-\infty}^{-Q} \dd{q_1}
    \frac{-1}{{q_1}^2 (q_1+2k)}
    =
    \frac{-C'}{k}
    \left[ 
        \frac{1}{Q} + \frac{1}{2k}\fn{\ln}{1-\frac{2k}{Q}}
    \right].  \label{eq:C3_layered-step5-3}
\end{equation}
Combining Eqs. \eqref{eq:C3_layered-step3}-\eqref{eq:C3_layered-step5-3} shows that the integrand in \eqref{eq:C3_layered} is absolutely integrable, 
%i.e.,
% \begin{eqnarray}
%     \fl&
%     \int_{-\infty}^\infty \dd{q_1}
%     \int_{-\infty}^\infty \dd{q_2}
%     \abs{\frac{1}{(k+q_{1})^2-{k}^2} \frac{1}{(k+q_{2})^2-{k}^2} 
%         \E{\FTJ{p}{q_{1}} \FTJ{p}{-q_{1}+q_{2}} \FTJ{p}{-q_{2}}}}
%     \nonumber \\
%     \fl &\leq 
%     C\left\{\frac{C}{k} \fn{\ln}{\frac{K}{K-2k}\frac{Q-2k}{Q}} + \frac{-C'}{k}
%     \left[ 
%         \frac{1}{Q} + \frac{1}{2k}\fn{\ln}{1-\frac{2k}{Q}}
%     \right]\right\}^2.
% \end{eqnarray}
leading \eqref{eq:Cn-cond_layered} for $n=3$ to be true as noted in \eqref{eq:C3_layered-step3}.

\subsection{Transversely Isotropic Media}
\label{sec:PT-trans-3pt}

In Sec. \ref{sec:trans}, we analytically show that for SHU transversely isotropic systems with $\spD{k}=0$ for $k<K$, the finite perfect transparency interval, Eq. \eqref{eq:trans-interval} conjunction with Eq. \eqref{eq:eps_2D_TM}, is exact through the second-order terms.
Here, we provide detailed proof that this transparency interval is exact through the third-order terms.

Without losing generality, we can rewrite the third-order condition for the perfect transparency as follows:
\begin{eqnarray}   \label{eq:im-trans-3pt}
    \Im[\ATM{3}{k;1}] =0 ,\quad (k<K/2).
\end{eqnarray}
As in Eq. \eqref{eq:A3-layered-decomposed}, we decompose $A_3^{TM}$ given in Eq. \eqref{eq:A3-TM} into three terms:
\begin{eqnarray}
    \ATM{3}{k;1}
    &=
    \frac{-1}{\phi_p}\frac{{k}^4}{(2\pi)^2} 
        \left[ \fn{\Ctm{1}}{k} - {\phi_p}^2 (2\pi)^2 \fn{\Ctm{2}}{k} - {\phi_p} \fn{\Ctm{3}}{k} 
        \right], \label{eq:A3-TM-decomposed}
\end{eqnarray}
where $\fn{\Ctm{1}}{k}$ and $\fn{\Ctm{2}}{k}$ depend on the two-point statistics, whereas $\fn{\Ctm{3}}{k}$ depends on the three-point statistics:
\begin{eqnarray}
    \fl \fn{\Ctm{1}}{k} 
    \equiv &  
    \left[-\frac{2\pi^3}{{k}^2} \F{2}{k} \right]^2
    \label{eq:C1_TM},  \\
    \fl \fn{\Ctm{2}}{k}
    \equiv &
    \int_{\R^2} \dd{\vect{q}_{1}} \left[\frac{1}{\abs{k\uvect{y}+\vect{q}_{1}}^2-{k}^2} \right]^2 \spD{q_{1}} 
    \label{eq:C2_TM}, \\
    \fl \fn{\Ctm{3}}{k}
    \equiv & 
    \int_{\R^2} \dd{\vect{q}_{1}} \int_{\R^2} \dd{\vect{q}_{2}} 
    \frac{\E{\FTJ{p}{\vect{q}_{1}} \FTJ{p}{-\vect{q}_{1}+\vect{q}_{2}} \FTJ{p}{-\vect{q}_{2}}}}{(\abs{k\uvect{y}+\vect{q}_{1}}^2-{k}^2) (\abs{k\uvect{y}+\vect{q}_{2}}^2-{k}^2)}  
    \label{eq:C3_TM},
\end{eqnarray}
which are the 2D counterparts of Eqs. \eqref{eq:C1_layered}-\eqref{eq:C3_layered}.
We now prove \eqref{eq:im-trans-3pt} by showing that for $n=1,2,3,$
\begin{eqnarray}   \label{eq:Cn-conditions}
    \Im[\fn{\Ctm{n}}{k}]=0 \quad (k<K/2).
\end{eqnarray}
Because the proof of Eq. \eqref{eq:Cn-conditions} is analogous to \eqref{eq:Cn-cond_layered}, we outline how to prove it by using the three general properties of $\FTJ{p}{\vect{k}}$ for 2D disordered SHU two-phase media, given in Eqs. \eqref{eq:SHU_PT_layered}-\eqref{eq:decay_PT_layered}.

One can immediately see that \eqref{eq:Cn-conditions} is true for $n=1$ because $\Im[\F{2}{k}]=0$ for $k<K/2$, which is identical to the perfect transparency at the two-point level.
One can also immediately show \eqref{eq:Cn-conditions} for $n=2$ because the integrand in \eqref{eq:C2_TM} is (i) real-valued and (ii) absolutely integrable for $k<K/2$.
The condition (ii) can be confirmed because this integrand has no singularity due to the stealthy hyperuniform condition [\eqref{eq:SHU_PT_layered}], and its absolute value is also bounded from above and decays rapidly as $\abs{\vect{q}_1}$ increases due to Eqs. \eqref{eq:bounded_PT_layered} and \eqref{eq:decay_PT_layered}.

We now outline the proof of \eqref{eq:Cn-conditions} for $n=3$.
Similar to \ref{sec:PT-layered-3pt}, the crucial part of this proof is to show that for $k<K/2$, the absolute value of the integrand in \eqref{eq:C3_TM} is integrable (i.e., absolutely integrable), and thus the value of $\fn{\Ctm{3}}{k}$ is independent of the integration order due to the Fubini-Tonelli theorem \cite{stewart_calculus_2016}.
Once we enable the interchange of the integration order, we can immediately show that $\Im[\CL{3}(k)]=0$ for $k<K/2$, as in Eq. \eqref{eq:C3_layered-step3}.

One can show that the integrand in \eqref{eq:C3_TM} is absolutely integrable, as we did so in \ref{sec:PT-layered-3pt}.
Roughly speaking, the proof is based on Eqs. \eqref{eq:SHU_PT_layered}-\eqref{eq:decay_PT_layered}.
We first show that the integrand in Eq. \eqref{eq:C3_TM} has no singularity because while all possible singularities of the Green's functions in this integrand lie along two circles [i.e., $\abs{\vect{q}_1+k\uvect{y}}=k$ and $\abs{\vect{q}_2+k\uvect{y}}=k$], $\E{\FTJ{p}{\vect{q}_1}\FTJ{p}{-\vect{q}_1+\vect{q}_2}\FTJ{p}{-\vect{q}_2}}=0$ along these singularities because of \eqref{eq:SHU_PT_layered}.
In addition, due to Eq. \eqref{eq:bounded_PT_layered}, the absolute value of the integrand in \eqref{eq:C3_TM} is bounded from above for any $\vect{q}_1$ and $\vect{q}_2$.
Furthermore, because of Eq. \eqref{eq:decay_PT_layered}, the absolute value of the integrand vanishes as fast as or faster than $\abs{\vect{q}_1}^{-7/2} \abs{\vect{q}_2}^{-7/2}$ for large $\abs{\vect{q}_1}$ and $\abs{\vect{q}_2}$.
These three observations directly show that the integrand is absolutely integrable.

\ack The research was sponsored by the U.S. Army Research
Office and was accomplished under Cooperative Agreement No. W911NF-22-2-0103.

\section*{References}
%\bibliography{ref}
%\bibliographystyle{dcu}
% \bibliographystyle{iopart-num-mod}
% \bibliography{new, ref}

\begin{thebibliography}{100}
\expandafter\ifx\csname url\endcsname\relax
  \def\url#1{{\tt #1}}\fi
\expandafter\ifx\csname urlprefix\endcsname\relax\def\urlprefix{URL }\fi
\providecommand{\eprint}[2][]{\url{#2}}
% Bibliography created with iopart-num v2.1
% /biblio/bibtex/contrib/iopart-num

\bibitem{To03a}
Torquato S and Stillinger F~H 2003 Local density fluctuations, hyperuniform
  systems, and order metrics {\em Phys. Rev. E\/}
  \href{http://dx.doi.org/10.1103/PhysRevE.68.041113}{{\bf 68} 041113}

\bibitem{Ga02}
Gabrielli A, Joyce M and Labini F~S 2002 Glass-like universe: {R}eal-space
  correlation properties of standard cosmological models {\em Phys. Rev. D\/}
  \href{http://dx.doi.org/10.1103/PhysRevD.65.083523}{{\bf 65} 083523}

\bibitem{To18a}
Torquato S 2018 Hyperuniform states of matter {\em Phys. Rep.\/}
  \href{http://dx.doi.org/10.1016/j.physrep.2018.03.001}{{\bf 745} 1--95}

\bibitem{Lo18a}
{Lomba} E, {Weis} J~J and {Torquato} S 2018 Disordered multihyperuniformity
  derived from binary plasmas {\em Phys. Rev. E\/}
  \href{http://dx.doi.org/10.1103/PhysRevE.97.010102}{{\bf 97} 010102(R)}

\bibitem{Fl22}
Flack Aand~Majumdar S~N and Schehr G 2022 Gap probability and full counting
  statistics in the one-dimensional one-component plasma {\em J. Stat. Mech.:
  Theory \& Exp.\/} \href{http://dx.doi.org/10.1088/1742-5468/ac6a59}{{\bf
  2022} 053211}

\bibitem{Uc04b}
Uche O~U, Stillinger F~H and Torquato S 2004 Constraints on collective density
  variables: Two dimensions {\em Phys. Rev. E\/}
  \href{http://dx.doi.org/10.1103/PhysRevE.70.046122}{{\bf 70} 046122}

\bibitem{Zh15a}
Zhang G, Stillinger F and Torquato S 2015 Ground states of stealthy
  hyperuniform potentials: {I}. {E}ntropically favored configurations {\em
  Phys. Rev. E\/} \href{http://dx.doi.org/10.1103/PhysRevE.92.022119}{{\bf 92}
  022119}

\bibitem{Zh17a}
Zhang G, Stillinger F~H and Torquato S 2017 Can exotic disordered ``stealthy''
  particle configurations tolerate arbitrarily large holes? {\em Soft Matter\/}
  \href{http://dx.doi.org/10.1039/c7sm01028a}{{\bf 13} 6197--6207}

\bibitem{Do05d}
Donev A, Stillinger F~H and Torquato S 2005 Unexpected density fluctuations in
  disordered jammed hard-sphere packings {\em Phys. Rev. Lett.\/}
  \href{http://dx.doi.org/10.1103/PhysRevLett.95.090604}{{\bf 95} 090604}

\bibitem{Ma23}
Maher C~E, Jiao Y and Torquato S 2023 Hyperuniformity of maximally random
  jammed packings of hyperspheres across spatial dimensions {\em Phys. Rev.
  E\/} \href{http://dx.doi.org/10.1103/PhysRevE.108.064602}{{\bf 108} 064602}

\bibitem{Ri17}
Ricouvier J, Pierrat R, Carminati R, Tabeling P and Yazhgur P 2017 Optimizing
  hyperuniformity in self-assembled bidisperse emulsions {\em Phys. Rev.
  Lett.\/} \href{http://dx.doi.org/10.1103/PhysRevLett.119.208001}{{\bf 119}
  208001}

\bibitem{Ku11}
Kurita R and Weeks E~R 2011 Incompressibility of polydisperse
  random-close-packed colloidal particles {\em Phys. Rev. E\/}
  \href{http://dx.doi.org/10.1103/PhysRevE.84.030401}{{\bf 84} 030401}

\bibitem{Dr15}
{Dreyfus} R, {Xu} Y, {Still} T, {Hough} L~A, {Yodh} A~G and {Torquato} S 2015
  Diagnosing hyperuniformity in two-dimensional, disordered, jammed packings of
  soft spheres {\em Phys. Rev. E\/}
  \href{http://dx.doi.org/10.1103/PhysRevE.91.012302}{{\bf 91} 012302}

\bibitem{Si09}
Silbert L~E and Silbert M 2009 Long-wavelength structural anomalies in jammed
  systems {\em Phys. Rev. E\/}
  \href{http://dx.doi.org/10.1103/PhysRevE.80.041304}{{\bf 80} 041304}

\bibitem{Be11}
Berthier L, Chaudhuri P, Coulais C, Dauchot O and Sollich P 2011 Suppressed
  compressibility at large scale in jammed packings of size-disperse spheres
  {\em Phys. Rev. Lett.\/}
  \href{http://dx.doi.org/10.1103/PhysRevLett.106.120601}{{\bf 106} 120601}

\bibitem{Ja15}
Jack R~L, Thompson I~R and Sollich P 2015 Hyperuniformity and phase separation
  in biased ensembles of trajectories for diffusive systems {\em Phys. Rev.
  Lett.\/} \href{http://dx.doi.org/10.1103/PhysRevLett.114.060601}{{\bf 114}
  060601}

\bibitem{He15}
{Hexner} D and {Levine} D 2015 {Hyperuniformity of critical absorbing states}
  {\em Phys. Rev. Lett.\/}
  \href{http://dx.doi.org/10.1103/PhysRevLett.114.110602}{{\bf 114} 110602}

\bibitem{We15}
Weijs J~H, Jeanneret R, Dreyfus R and Bartolo D 2015 Emergent hyperuniformity
  in periodically driven emulsions {\em Phys. Rev. Lett.\/}
  \href{http://dx.doi.org/10.1103/PhysRevLett.115.108301}{{\bf 115} 108301}

\bibitem{Tj15}
Tjhung E and Berthier L 2015 Hyperuniform density fluctuations and diverging
  dynamic correlations in periodically driven colloidal suspensions {\em Phys.
  Rev. Lett.\/} \href{http://dx.doi.org/10.1103/PhysRevLett.114.148301}{{\bf
  114} 148301}

\bibitem{He17a}
Hexner D and Levine D 2017 {Noise, Diffusion, and Hyperuniformity} {\em Phys.
  Rev. Lett.\/} \href{http://dx.doi.org/10.1103/PhysRevLett.118.020601}{{\bf
  118} 020601}

\bibitem{He17b}
Hexner D, Chaikin P~M and Levine D 2017 Enhanced hyperuniformity from random
  reorganization {\em Proc. Nat. Acad. Sci. U. S. A.\/}
  \href{http://dx.doi.org/10.1073/pnas.1619260114}{{\bf 114} 4294--4299}

\bibitem{We17}
Weijs J~H and Bartolo D 2017 Mixing by unstirring: {H}yperuniform dispersion of
  interacting particles upon chaotic advection {\em Phys. Rev. Lett.\/}
  \href{http://dx.doi.org/10.1103/PhysRevLett.119.048002}{{\bf 119} 048002}

\bibitem{Kw17}
Kwon S and Kim J~M 2017 Hyperuniformity of initial conditions and critical
  decay of a diffusive epidemic process belonging to the {M}anna class {\em
  Phys. Rev. E\/} \href{http://dx.doi.org/10.1103/PhysRevE.96.012146}{{\bf 96}
  012146}

\bibitem{Wi18}
Willis G and Pruessner G 2018 Spatio-temporal correlations in the {M}anna model
  in one, three and five dimensions {\em Int. J. Mod. Phys. B\/}
  \href{http://dx.doi.org/10.1142/S0217979218300025}{{\bf 32} 1830002}

\bibitem{Le14}
{Lesanovsky} I and {Garrahan} J~P 2014 {Out-of-equilibrium structures in
  strongly interacting Rydberg gases with dissipation} {\em Phys. Rev. A\/}
  \href{http://dx.doi.org/10.1103/PhysRevA.90.011603}{{\bf 90} 011603}

\bibitem{Ji14}
Jiao Y, Lau T, Hatzikirou H, Meyer-Hermann M, Corbo J~C and Torquato S 2014
  Avian photoreceptor patterns represent a disordered hyperuniform solution to
  a multiscale packing problem {\em Phys. Rev. E\/}
  \href{http://dx.doi.org/10.1103/PhysRevE.89.022721}{{\bf 89} 022721}

\bibitem{Ma15}
Mayer A, Balasubramanian V, Mora T and Walczak A~M 2015 How a well-adapted
  immune system is organized {\em Proc. Nat. Acad. Sci. U. S. A.\/}
  \href{http://dx.doi.org/10.1073/pnas.1421827112}{{\bf 112} 5950--5955}

\bibitem{Ge23}
Ge Z 2023 The hidden order of {T}uring patterns in arid and semi-arid
  vegetation ecosystems {\em Proc. Nat. Acad. Sci.\/}
  \href{http://dx.doi.org/10.1073/pnas.2306514120}{{\bf 120} e2306514120}

\bibitem{To08b}
Torquato S, Scardicchio A and Zachary C~E 2008 Point processes in arbitrary
  dimension from {F}ermionic gases, random matrix theory, and number theory
  {\em J. Stat. Mech.: Theory Exp.\/}
  \href{http://dx.doi.org/10.1088/1742-5468/2008/11/p11019}{{\bf 2008} P11019}

\bibitem{Fe56}
Feynman R~P and Cohen M 1956 Energy spectrum of the excitations in liquid
  helium {\em Phys. Rev.\/}
  \href{http://dx.doi.org/10.1103/PhysRev.102.1189}{{\bf 102} 1189--1204}

\bibitem{sanchez_disordered_2023}
S{\'a}nchez J~A, Maldonado R~C, Amig{\'o} M~L, Nieva G, Kolton A and Fasano Y
  2023 Disordered hyperuniform vortex matter with rhombic distortions in fese
  at low fields {\em Phys. Rev. B\/}
  \href{http://dx.doi.org/10.1103/PhysRevB.107.094508}{{\bf 107} 094508}

\bibitem{Zh16a}
Zhang G, Stillinger F~H and Torquato S 2016 The perfect glass paradigm:
  Disordered hyperuniform glasses down to absolute zero {\em Sci. Rep.\/}
  \href{http://dx.doi.org/10.1038/srep36963}{{\bf 6} 36963}

\bibitem{Mon73}
Montgomery H~L 1973 The pair correlation of zeros of the zeta function {\em
  Amer. Math. Soc.\/}  181--193

\bibitem{Dy70}
Dyson F~J 1970 Correlations between eigenvalues of a random matrix {\em Comm.
  Math. Phys.\/} \href{http://dx.doi.org/10.1007/bf01646824}{{\bf 19} 235--250}

\bibitem{Me91}
Metha M~L 1991 {\em Random Matrices\/} (New York: Academic Press)

\bibitem{La19}
Lacroix-A-Chez-Toine B, Garz{\'o}n J~A~M, Calva C~S~H, Castillo I, Kundu A,
  Majumdar S~N and Schehr G 2019 Intermediate deviation regime for the full
  eigenvalue statistics in the complex ginibre ensemble {\em Phys. Rev. E\/}
  \href{http://dx.doi.org/10.1103/PhysRevE.100.012137}{{\bf 100} 012137}

\bibitem{Fl09b}
Florescu M, Torquato S and Steinhardt P~J 2009 Designer disordered materials
  with large complete photonic band gaps {\em Proc. Nat. Acad. Sci. U. S. A.\/}
  \href{http://dx.doi.org/10.1073/pnas.0907744106}{{\bf 106} 20658--20663}

\bibitem{Ae22}
Aeby S, Aubry G~J, Froufe-P{\'e}rez L~S and Scheffold F 2022 Fabrication of
  hyperuniform dielectric networks via heat-induced shrinkage reveals a bandgap
  at telecom wavelengths {\em Adv. Opt. Mater.\/}
  \href{http://dx.doi.org/10.1002/adom.202200232}{{\bf 2022} 2200232}

\bibitem{Le16}
{Leseur} O, {Pierrat} R and {Carminati} R 2016 {High-density hyperuniform
  materials can be transparent} {\em Optica\/}
  \href{http://dx.doi.org/10.1364/optica.3.000763}{{\bf 3} 763--767}

\bibitem{Fr17}
{Froufe-P{\'e}rez} L~S, {Engel} M, {Jos{\'e} S{\'a}enz} J and {Scheffold} F
  2017 {Band gap formation and Anderson localization in disordered photonic
  materials with structural correlationss} {\em Proc. Nat. Acad. Sci.\/}
  \href{http://dx.doi.org/10.1073/pnas.1705130114}{{\bf 114} 9570–--9574}

\bibitem{Ki20a}
Kim J and Torquato S 2020 Multifunctional composites for elastic and
  electromagnetic wave propagation {\em Proc. Nat. Acad. Sci.\/}
  \href{http://dx.doi.org/10.1073/pnas.1914086117}{{\bf 117} 8764--8774}

\bibitem{To21a}
Torquato S and Kim J 2021 Nonlocal effective electromagnetic wave
  characteristics of composite media: Beyond the quasistatic regime {\em Phys.
  Rev. X\/} \href{http://dx.doi.org/10.1103/PhysRevX.11.021002}{{\bf 11}
  021002}

\bibitem{Ki23}
Kim J and Torquato S 2023 Effective electromagnetic wave properties of
  disordered stealthy hyperuniform layered media beyond the quasistatic regime
  {\em Optica\/} \href{http://dx.doi.org/10.1364/OPTICA.489797}{{\bf 10}
  965--972}

\bibitem{Fr23}
{Froufe-P{\'e}rez} L~S, Aubry G~J, Scheffold F and Magkiriadou S 2023 Bandgap
  fluctuations and robustness in two-dimensional hyperuniform dielectric
  materials {\em Opt. Express\/}
  \href{http://dx.doi.org/10.1364/OE.484232}{{\bf 31} 18509--18515}

\bibitem{Kl22}
Klatt M~A, Steinhardt P~J and Torquato S 2022 Wave propagation and band tails
  of two-dimensional disordered systems in the thermodynamic limit {\em Proc.
  Nat. Acad. Sci.\/} \href{http://dx.doi.org/10.1073/pnas.2213633119}{{\bf 119}
  e2213633119}

\bibitem{alhaitz_experimental_2023}
Alha{\"i}tz L, Conoir J~M and {Valier-Brasier} T 2023 Experimental evidence of
  isotropic transparency and complete band gap formation for ultrasound
  propagation in stealthy hyperuniform media {\em Phys. Rev. E\/}
  \href{http://dx.doi.org/10.1103/PhysRevE.108.065001}{{\bf 108} 065001}

\bibitem{Sg22}
Sgrignuoli F, Torquato S and Dal~Negro L 2022 Subdiffusive wave transport and
  weak localization transition in three-dimensional stealthy hyperuniform
  disordered systems {\em Phys. Rev. B\/}
  \href{http://dx.doi.org/doi.org/10.1103/PhysRevB.105.064204}{{\bf 105}
  064204}

\bibitem{Sc22}
Scheffold F, Haberko J, Magkiriadou S and {Froufe-P{\'e}rez} L~S 2022 Transport
  through amorphous photonic materials with localization and bandgap regimes
  {\em Phys. Rev. Lett.\/}
  \href{http://dx.doi.org/10.1103/PhysRevLett.129.157402}{{\bf 129} 157402}

\bibitem{Bi19}
Bigourdan F, Pierrat R and Carminati R 2019 Enhanced absorption of waves in
  stealth hyperuniform disordered media {\em Optics Express\/}
  \href{http://dx.doi.org/10.1364/oe.27.008666}{{\bf 27} 8666--8682}

\bibitem{merkel_stealthy_2023}
Merkel M, Stappers M, Ray D, Denz C and Imbrock J 2023 Stealthy hyperuniform
  surface structures for efficiency enhancement of organic solar cells {\em
  Adv. Photonics Res.\/} \href{http://dx.doi.org/10.1002/adpr.202300256}{{\bf
  n/a} 2300256}

\bibitem{Gk17}
Gkantzounis G, Amoah T and Florescu M 2017 Hyperuniform disordered phononic
  structures {\em Phys. Rev. B\/}
  \href{http://dx.doi.org/10.1103/PhysRevB.95.094120}{{\bf 95} 094120}

\bibitem{Ro19}
Romero-Garc{\'\i}a V, Lamothe N, Theocharis G, Richoux O and Garc{\'\i}a-Raffi
  L~M 2019 Stealth acoustic materials {\em Phys. Rev. Appl.\/}
  \href{http://dx.doi.org/10.1103/PhysRevApplied.11.054076}{{\bf 11} 054076}

\bibitem{Roh20}
Rohfritsch A, Conoir J~M, Valier-Brasier T and Marchiano R 2020 Impact of
  particle size and multiple scattering on the propagation of waves in
  stealthy-hyperuniform media {\em Phys. Rev. E\/}
  \href{http://dx.doi.org/10.1103/PhysRevE.102.053001}{{\bf 102} 053001}

\bibitem{Zh19}
Zhang H, Chu H, Giddens H, Wu W and Hao Y 2019 Experimental demonstration of
  luneburg lens based on hyperuniform disordered media {\em Appl. Phys.
  Lett.\/} \href{http://dx.doi.org/10.1063/1.5055295}{{\bf 114} 053507}

\bibitem{Ch21}
Christogeorgos O, Zhang H, Cheng Q and Hao Y 2021 Extraordinary {{Directive
  Emission}} and {{Scanning}} from an {{Array}} of {{Radiation Sources}} with
  {{Hyperuniform Disorder}} {\em Phys. Rev. Appl.\/}
  \href{http://dx.doi.org/10.1103/PhysRevApplied.15.014062}{{\bf 15} 014062}

\bibitem{tang_hyperuniform_2023}
Tang K, Wang Y, Wang S, Gao D, Li H, Liang X, Sebbah P, Zhang J and Shi J 2023
  Hyperuniform disordered parametric loudspeaker array (\textit{Preprint}
  \eprint{2301.00833})

\bibitem{tamraoui_hyperuniform_2023}
Tamraoui M, Roux E and Liebgott H 2023 Hyperuniform disordered sparse array for
  3d ultrasound imaging {\em 2023 IEEE International Ultrasonics Symposium
  (IUS)\/} pp 1--4 \urlprefix\url{www.doi.org/10.1109/IUS51837.2023.10308368}

\bibitem{granchi_nearfield_2023}
Granchi N, Lodde M, Stokkereit K, Spalding R, {van Veldhoven} P~J, Sapienza R,
  Fiore A, Gurioli M, Florescu M and Intonti F 2023 Near-field imaging of
  optical nanocavities in hyperuniform disordered materials {\em Phys. Rev.
  B\/} \href{http://dx.doi.org/10.1103/PhysRevB.107.064204}{{\bf 107} 064204}

\bibitem{To18c}
Torquato S and Chen D 2018 Multifunctional hyperuniform cellular networks:
  optimality, anisotropy and disorder {\em Multifunct. Mater.\/}
  \href{http://dx.doi.org/10.1088/2399-7532/aaca91}{{\bf 1} 015001}

\bibitem{To16a}
Torquato S 2016 Hyperuniformity and its generalizations {\em Phys. Rev. E\/}
  \href{http://dx.doi.org/10.1103/PhysRevE.94.022122}{{\bf 94} 022122}

\bibitem{Za09}
Zachary C~E and Torquato S 2009 Hyperuniformity in point patterns and two-phase
  heterogeneous media {\em J. Stat. Mech.: Theory \& Exp.\/}
  \href{http://dx.doi.org/10.1088/1742-5468/2009/12/P12015}{{\bf 2009} P12015}

\bibitem{To16b}
Torquato S 2016 Disordered hyperuniform heterogeneous materials {\em J. Phys.:
  Cond. Mat\/} \href{http://dx.doi.org/10.1088/0953-8984/28/41/414012}{{\bf 28}
  414012}

\bibitem{To02a}
Torquato S 2002 {\em Random Heterogeneous Materials: Microstructure and
  Macroscopic Properties\/} (New York: Springer-Verlag)

\bibitem{Sa03}
Sahimi M 2003 {\em Heterogeneous Materials {I}: {L}inear Transport and Optical
  Properties\/} (New York: Springer-Verlag)

\bibitem{Zh16b}
Zhang G, Stillinger F~H and Torquato S 2016 Transport, geometrical and
  topological properties of stealthy disordered hyperuniform two-phase systems
  {\em J. Chem. Phys\/} \href{http://dx.doi.org/10.1063/1.4972862}{{\bf 145}
  244109}

\bibitem{To22b}
Torquato S 2022 Extraordinary disordered hyperuniform multifunctional
  composites {\em J. Composite Mater.\/}
  \href{http://dx.doi.org/10.1177/0021998322111643}{{\bf 56} 3635--3649}

\bibitem{To21c}
Torquato S 2021 Structural characterization of many-particle systems on
  approach to hyperuniform states {\em Phys. Rev. E\/}
  \href{http://dx.doi.org/10.1103/PhysRevE.103.052126}{{\bf 103} 052126}

\bibitem{St87b}
Stanley H~E 1987 {\em Introduction to Phase Transitions and Critical
  Phenomena\/} (New York: Oxford University Press)

\bibitem{Bi92}
Binney J~J, Dowrick N~J, Fisher A~J and Newman M~E~J 1992 {\em The Theory of
  Critical Phenomena: An Introduction to the Renormalization Group\/} (Oxford,
  England: Oxford University Press)

\bibitem{Ma82}
Mandelbrot B~B 1982 {\em The fractal geometry of nature\/} (New York: W. H.
  Freeman)

\bibitem{Og19}
O{\u g}uz E~C, Socolar J~E~S, Steinhardt P~J and Torquato S 2019
  Hyperuniformity and anti-hyperuniformity in one-dimensional substitution
  tilings {\em Acta Cryst. Section A: Foundations \& Advances\/}
  \href{http://dx.doi.org/10.1107/s2053273318015528}{{\bf A75} 3--13}

\bibitem{Be78a}
Bergman D~J 1978 The dielectric constant of a composite material-- {A} problem
  in classical physics {\em Phys. Rep. C\/}
  \href{http://dx.doi.org/10.1016/0370-1573(78)90009-1}{{\bf 43} 377--407}

\bibitem{Av88}
Avellaneda M, Cherkaev A~V, Lurie K~A and Milton G~W 1988 On the effective
  conductivity of polycrystals and a three-dimensional phase-interchange
  inequality {\em J. Appl. Phys.\/}
  \href{http://dx.doi.org/10.1063/1.340445}{{\bf 63} 4989--5003}

\bibitem{To90e}
Torquato S 1990 Relationship between permeability and diffusion-controlled
  trapping constant of porous media {\em Phys. Rev. Lett.\/}
  \href{http://dx.doi.org/10.1103/PhysRevLett.64.2644}{{\bf 64} 2644--2646}

\bibitem{To91f}
Torquato S and Avellaneda M 1991 Diffusion and reaction in heterogeneous media:
  {P}ore size distribution, relaxation times, and mean survival time {\em J.
  Chem. Phys.\/} \href{http://dx.doi.org/10.1063/1.461519}{{\bf 95} 6477--6489}

\bibitem{Av91b}
Avellaneda M and Torquato S 1991 Rigorous link between fluid permeability,
  electrical conductivity, and relaxation times for transport in porous media
  {\em Phys. Fluids A\/} \href{http://dx.doi.org/10.1063/1.858194}{{\bf 3}
  2529--2540}

\bibitem{Gi95b}
Gibiansky L~V and Torquato S 1995 Rigorous link between the conductivity and
  elastic moduli of fibre-reinforced composite materials {\em Phil. Trans.
  Royal Soc. Lond. A.\/} \href{http://dx.doi.org/10.1098/rsta.1995.0099}{{\bf
  353} 243--278}

\bibitem{Gi96b}
Gibiansky L~V and Torquato S 1996 Connection between the conductivity and bulk
  modulus of isotropic composite materials {\em Proc. R. Soc. Lond. A\/}
  \href{http://dx.doi.org/10.1098/rspa.1996.0015}{{\bf 452} 253--283}

\bibitem{Gi97c}
Gibiansky L~V and Torquato S 1997 Thermal expansion of isotropic multiphase
  composites and polycrystals {\em J. Mech. Phys. Solids\/}
  \href{http://dx.doi.org/10.1016/S0022-5096(96)00129-9}{{\bf 45} 1223--1252}

\bibitem{To04b}
Torquato S and Donev A 2004 Minimal surfaces and multifunctionality {\em Proc.
  R. Soc. Lond. A\/} \href{http://dx.doi.org/10.1098/rspa.2003.1269}{{\bf 460}
  1849--1856}

\bibitem{Se09}
Sevostianov I and Kachanov M 2009 Connections between elastic and conductive
  properties of heterogeneous materials {\em Advances in Applied Mechanics\/}
  vol~42 (Elsevier) pp 69--252

\bibitem{To21d}
Torquato S 2021 Diffusion spreadability as a probe of the microstructure of
  complex media across length scales {\em Phys. Rev. E\/}
  \href{http://dx.doi.org/10.1103/PhysRevE.104.054102}{{\bf 104} 054102}

\bibitem{kim_accurate_2024}
Kim J and Torquato S 2024 Theoretical prediction of the effective dynamic
  dielectric constant of disordered hyperuniform anisotropic composites beyond
  the long-wavelength regime {\em Opt. Mater. Express\/}
  \href{http://dx.doi.org/10.1364/OME.507918}{{\bf 14} 194}

\bibitem{anderson_absence_1958}
Anderson P~W 1958 Absence of diffusion in certain random lattices {\em Phys.
  Rev.\/} \href{http://dx.doi.org/10.1103/PhysRev.109.1492}{{\bf 109}
  1492--1505}

\bibitem{mcgurn_anderson_1993}
McGurn A~R, Christensen K~T, Mueller F~M and Maradudin A~A 1993 Anderson
  localization in one-dimensional randomly disordered optical systems that are
  periodic on average {\em Phys. Rev. B\/}
  \href{http://dx.doi.org/10.1103/PhysRevB.47.13120}{{\bf 47} 13120--13125}

\bibitem{sheng_introduction_2006}
Sheng P 2006 {\em Introduction to wave scattering, localization, and mesoscopic
  phenomena\/} 2nd ed ({\em Springer series in materials science\/} no~88)
  (Berlin ; New York: Springer) ISBN 978-3-540-29155-8

\bibitem{aegerter_coherent_2009}
Aegerter C~M and Maret G 2009 Coherent backscattering and anderson localization
  of light {\em Progress in Optics\/} vol~52 (Elsevier) pp 1--62

\bibitem{izrailev_anomalous_2012}
Izrailev F~M, Krokhin A~A and Makarov N~M 2012 Anomalous localization in
  low-dimensional systems with correlated disorder {\em Physics Reports\/}
  \href{http://dx.doi.org/10.1016/j.physrep.2011.11.002}{{\bf 512} 125--254}

\bibitem{wiersma_disordered_2013}
Wiersma D~S 2013 Disordered photonics {\em Nature Photonics\/}
  \href{http://dx.doi.org/10.1038/nphoton.2013.29}{{\bf 7} 188--196}

\bibitem{Ki21}
Kim J and Torquato S 2021 Characterizing the hyperuniformity of ordered and
  disordered two-phase media {\em Phys. Rev. E\/}
  \href{http://dx.doi.org/10.1103/PhysRevE.103.012123}{{\bf 103} 012123}

\bibitem{De49}
Debye P and Bueche A~M 1949 Scattering by an inhomogeneous solid {\em J. Appl.
  Phys.\/} \href{http://dx.doi.org/10.1063/1.1698419}{{\bf 20} 518--525}

\bibitem{De57}
Debye P, Anderson H~R and Brumberger H 1957 Scattering by an inhomogeneous
  solid. {II}. {T}he correlation function and its applications {\em J. Appl.
  Phys.\/} \href{http://dx.doi.org/10.1063/1.1722830}{{\bf 28} 679--683}

\bibitem{To85b}
Torquato S and Stell G 1985 Microstructure of two-phase random media: {V}.
  {T}he $n$-point matrix probability functions for impenetrable spheres {\em J.
  Chem. Phys.\/} \href{http://dx.doi.org/10.1063/1.448475}{{\bf 82} 980--987}

\bibitem{Ye98a}
Yeong C~L~Y and Torquato S 1998 Reconstructing random media {\em Phys. Rev.
  E\/} \href{http://dx.doi.org/10.1103/PhysRevE.57.495}{{\bf 57} 495--506}

\bibitem{To20}
Torquato S 2020 Predicting transport characteristics of hyperuniform porous
  media via rigorous microstructure-property relations {\em Adv. Water
  Resour.\/} \href{http://dx.doi.org/10.1016/j.advwatres.2020.103565}{{\bf 140}
  103565}

\bibitem{Sk21}
Skolnick M and Torquato S 2021 Understanding degeneracy of two-point
  correlation functions via debye random media {\em Phys. Rev. E\/}
  \href{http://dx.doi.org/10.1103/PhysRevE.104.045306}{{\bf 104} 045306}

\bibitem{Ha86}
Hansen J~P and McDonald I~R 1986 {\em Theory of Simple Liquids\/} (New York:
  Academic Press)

\bibitem{Pe64}
Percus J 1964 Pair distribution function in classical statistical mechanics
  {\em The Equilibrium Theory of Classical Fluids\/} ed Frisch H~L and Lebowitz
  J~L (Benjamin)

\bibitem{Ze27}
Zernike F and Prins J~A 1927 Die {B}eugung von {R}{\"o}ntgenstrahlen in
  {F}l{\"u}ssigkeiten als {E}ffekt der {M}olek{\"u}lanordnung {\em Z. Phys.\/}
  {\bf 41} 184--194

\bibitem{kim_new_2019}
Kim J and Torquato S 2019 New tessellation-based procedure to design perfectly
  hyperuniform disordered dispersions for materials discovery {\em Acta
  Mater.\/} \href{http://dx.doi.org/10.1016/j.actamat.2019.01.026}{{\bf 168}
  143--151}

\bibitem{bruggeman_berechnung_1935}
Bruggeman D~A~G 1935 Berechnung verschiedener physikalischer konstanten von
  heterogenen substanzen. i. dielektrizit\"atskonstanten und leitf\"ahigkeiten
  der mischk\"orper aus isotropen substanzen {\em Annalen der Physik\/}
  \href{http://dx.doi.org/10.1002/andp.19354160705}{{\bf 416} 636--664}

\bibitem{yu_engineered_2021}
Yu S, Qiu C~W, Chong Y, Torquato S and Park N 2021 Engineered disorder in
  photonics {\em Nat Rev Mater\/}
  \href{http://dx.doi.org/10.1038/s41578-020-00263-y}{{\bf 6} 226--243}

\bibitem{jackson_classical_1999}
Jackson J~D 1999 {\em Classical Electrodynamics\/} 3rd ed (New York: John Wiley
  \& Sons, Inc.)

\bibitem{To22a}
Torquato S, Skolnick M and Kim J 2022 Local order metrics for two-phase media
  across length scales {\em J. Phys. A: Math. \& Theor.\/}
  \href{http://dx.doi.org/10.1088/1751-8121/ac72d7}{{\bf 55} 274003}

\bibitem{Kl_local_2023}
Klatt M~A, Steinhardt P~J and Torquato S 2023 ~~ {\em preprint\/} In
  preparation

\bibitem{yaghjian_electric_1980}
Yaghjian A 1980 Electric dyadic green's functions in the source region {\em
  Proc. IEEE\/} \href{http://dx.doi.org/10.1109/PROC.1980.11620}{{\bf 68}
  248--263}

\bibitem{tsang_scattering_1981}
Tsang L and Kong J~A 1981 Scattering of electromagnetic waves from random media
  with strong permittivity fluctuations {\em Radio Sci.\/}
  \href{http://dx.doi.org/10.1029/RS016i003p00303}{{\bf 16} 303--320}

\bibitem{mackay_strongpropertyfluctuation_2000}
Mackay T~G, Lakhtakia A and Weiglhofer W~S 2000 Strong-property-fluctuation
  theory for homogenization of bianisotropic composites: Formulation {\em Phys.
  Rev. E\/} \href{http://dx.doi.org/10.1103/PhysRevE.62.6052}{{\bf 62}
  6052--6064}

\bibitem{stewart_calculus_2016}
Stewart J 2016 {\em Calculus: early transcendentals\/} 8th ed (Boston, MA, USA:
  Cengage Learning) ISBN 978-1-305-27237-8

\end{thebibliography}

\providecommand{\newblock}{}

\end{document}